\documentclass[a4paper]{article}

    \usepackage{amssymb,amsmath,amsfonts,eurosym,geometry,ulem,graphicx,color,setspace,sectsty,comment,footmisc,caption,natbib,pdflscape,subfigure,array,url}
    \usepackage{hyperref}
    \hypersetup{
        colorlinks=true,        
        linkcolor=blue,         
        citecolor=blue,         
        urlcolor=blue,          
        filecolor=blue          
    }
    \usepackage{chngcntr}
    \usepackage{booktabs}
    \usepackage[figuresright]{rotating}
    \usepackage{adjustbox}
    \usepackage{multirow}
    \usepackage{float}
    \usepackage{arydshln}
    \usepackage{pifont}
    \usepackage{enumitem}
    \usepackage{mathabx}

    \usepackage{thmtools}
    \usepackage{thm-restate}
    \usepackage{cleveref}

    \usepackage{algorithm}
    \usepackage{algorithmic}

    \newenvironment{proofprop}[1]
    {\noindent\textbf{Proof of \textbf{Proposition \ref{#1}}.} }
    {\ \rule{0.5em}{0.5em} \vspace{\baselineskip}}
    \newenvironment{proofthm}[1]
    {\noindent\textbf{Proof of \textbf{Theorem \ref{#1}}.} }
    {\ \rule{0.5em}{0.5em} \vspace{\baselineskip}}
    \newenvironment{prooflmm}[1]
    {\noindent\textbf{Proof of \textbf{Lemma \ref{#1}}.} }
    {\ \rule{0.5em}{0.5em} \vspace{\baselineskip}}

    \newtheorem{proposition}{Proposition}
    \newtheorem{lemma}[proposition]{Lemma}
    \newtheorem{theorem}[proposition]{Theorem}
    \newtheorem{corollary}[proposition]{Corollary}
    
    \newtheorem{assumption}{Assumption}
    \newtheorem{example}{Example}
    
    \newenvironment{remark*}{\vspace{0.5em}\noindent \textbf{{Remark.}} \itshape}{\vspace{0.5em}}

    \DeclareMathOperator*{\argmin}{argmin}

    \newcommand{\bs}{\boldsymbol}
    \newcommand{\mc}{\mathcal}
    
    \newcommand{\mb}{\mathbb}
    \newcommand{\mr}{\mathrm}

    \newcommand{\cp}{\stackrel{p}{\longrightarrow}}

    \newcommand{\cd}{\stackrel{d}{\longrightarrow}}

    \newcommand{\leqtext}[1]{\stackrel{\text{#1}}{\leq}}    
      
    \newcommand{\geqtext}[1]{\stackrel{\text{#1}}{\geq}}    
    \newcommand{\eqtext}[1]{\stackrel{\text{#1}}{=}}

    \makeatletter
    \newsavebox{\@brx}
    \newcommand{\llangle}[1][]{\savebox{\@brx}{\(\m@th{#1\langle}\)}%
    \mathopen{\copy\@brx\kern-0.5\wd\@brx\usebox{\@brx}}}
    \newcommand{\rrangle}[1][]{\savebox{\@brx}{\(\m@th{#1\rangle}\)}%
    \mathclose{\copy\@brx\kern-0.5\wd\@brx\usebox{\@brx}}}
    \makeatother

    \newcommand{\vt}[1]{{\vert\kern-0.25ex\vert #1 
    \vert\kern-0.25ex\vert}}

    \numberwithin{equation}{section}

    \title{\textsc{Flexible Imputation of Incomplete Network Data} \thanks{We thank Aureo de Paula, Chen-Wei Hsiang, Andrei Zeleneev, and   the participants of the UCL Econometrics Brownbag Seminar for their valuable comments.}} 
    
    \author{
        Ge \textsc{Sun}\thanks{University of Notre Dame: \textsf{gsun4@nd.edu}.} \quad \quad Weisheng \textsc{Zhang} \thanks{%
        University College London: \textsf{weisheng.zhang.21@ucl.ac.uk}.}
    }

\begin{document}

    \maketitle

    \begin{abstract}

        Sampled network data are widely used in empirical research because collecting complete network information is costly. However, empirical analyses based on sampled networks may lead to biased estimators. We propose a nonparametric imputation method for sampled networks and show that empirical analyses based on imputed networks yield consistent estimates.
        Our approach imputes missing network links by combining a projection onto covariates with a local two-way fixed-effects regression. The method avoids parametric assumptions, does not rely on low-rank restrictions, and flexibly accommodates both observed covariates and unobserved heterogeneity. We establish entrywise convergence rates for the imputed matrix and prove the consistency of generalized method of moments (GMM) estimators based on imputed networks. We further derive the convergence rate of the corresponding estimator in the linear-in-means peer-effects model. Simulations show strong performance of our method both in terms of imputation accuracy and in downstream empirical analysis. We illustrate our method with an application to the microfinance network data of \citet{banerjee2013diffusion}.

    \end{abstract}


    \thispagestyle{empty}
    \clearpage

\setcounter{page}{1}

\clearpage

    \section{Introduction}

    There is a large and growing literature studying how social networks shape individuals' behavior and economic outcomes. Examples include how social interactions  affect information diffusion \citep{banerjee2013diffusion, beaman2018diffusion}, test scores \citep{sacerdote2001peer}, and demand for financial assets \citep{conley2010learning, cai2015social}. While network structure is central to these analyses, collecting complete network data is often expensive \citep{breza2020using}, so applied work typically relies on sampled or partially observed networks instead. 
    However, it has been noted that using sampled networks in place of the full network may lead to biased regression coefficients and generalized method of moments (GMM) estimators; see, for example, \citet{chandrasekhar2011econometrics}. As a result, many empirical conclusions may be undermined by incomplete network data.

    To overcome this limitation, we propose a flexible imputation method for sampled networks and show that downstream empirical analysis using the imputed network delivers consistent parameter estimators.  Our  method is designed for networks generated by dyadic formation processes and observed through egocentric sampling, a widely used design in which researchers randomly sample\footnote{
        Our method also applies to more general sampling schemes in which individuals are randomly sampled with heterogeneous probabilities that depend on both observed and unobserved characteristics.
    } a subset of individuals and record all of their links in the network, not only those within the sample. The proposed procedure imputes missing links by combining two parts: (i) one explained by observable covariates, recovered via projection onto the observed covariates, and (ii) a residual part estimated using a local two-way fixed-effects regression. The method has several advantages over existing approaches, making it particularly well suited to empirical studies of social networks in economics.

    First, the proposed method avoids parametric assumptions. Many imputation approaches, such as \citet{chandrasekhar2011econometrics}, \cite{breza2020using}, and \cite{boucher2020estimating}, require specifying the functional form of the network formation model. In practice, this is a demanding requirement, as applied researchers rarely have reliable prior information about the correct functional form. Our approach circumvents this issue by using a fully nonparametric procedure and therefore does not require specifying the network formation process.

    Second, our imputation method does not rely on a low-rank assumption. Assuming a low-rank network formation model and exploiting this structure for imputation is a common approach in the statistical network literature \citep{li2023link, cai2016structured}, but this assumption may be violated when the network formation process is nonlinear, as in the specification in \citet{graham2017econometric}. Building on ideas from \citet{freeman2023linear} and \citet{feng2023optimal}, our method instead approximates network formation using \emph{local} two-way fixed-effects regressions, making it applicable to a broad class of \emph{nonlinear} dyadic formation processes.

    Third, our procedure jointly exploits observed covariates and accommodates unobserved heterogeneity in link formation. Social networks typically exhibit homophily---meaning that individuals with similar characteristics are more likely to form links---and degree heterogeneity---individuals differ systematically in the number of links.  An effective imputation method should therefore exploit information from observed covariates and accommodate latent features. The method proposed in this paper achieves both objectives without imposing additional structure on the joint distribution of latent factors and observed covariates.

    The general nonparametric network formation setting leads to a technical challenge: because the underlying formation model is unknown and potentially nonlinear, we cannot directly estimate latent factors, as is common in linear factor models, and then use them to recover the formation process and impute missing links. To address this problem, we draw on the insight of \citet{zhang2017estimating} that similarity in observed connections can reveal similarity in latent factors, and use the sampled subnetwork to construct a pseudo-distance that serves as a proxy for latent distance. For each unsampled individual, we then use this pseudo-distance to locate sampled individuals who are “close" to her and use their observed connections to impute her missing links.

    We establish a convergence rate for the imputed link probabilities as the network size and the number of sampled individuals go to infinity. The convergence is uniform over all missing links, which implies that the imputation error is controlled at the link level and is particularly useful for applications that require accurate imputation of each link probability. 
    We further provide a bias-variance decomposition of the imputation error, which, to the best of our knowledge, is new in this context.  The decomposition shows that the approximation error introduced by using pseudo-distance has an asymmetric effect: it is asymptotically negligible for the bias, but contributes to the variance.

    We then establish the consistency of the GMM estimator constructed from the imputed network. This result is particularly relevant for empirical applications, where interest typically centers on the consistency of estimators based on the imputed network rather than on the accuracy of the imputed links themselves. In addition, because our imputation procedure does not require specifying a parametric network formation model and does not rely on low-rank assumptions, the resulting GMM estimator is correspondingly more robust than those obtained using existing approaches.

    It is technically challenging to characterize how imputation error affects the asymptotic distribution of the GMM estimator. This difficulty arises because (i) imputation error is non-classical and correlated across links, and (ii) the moment conditions rely on network statistics, which aggregate these errors and thereby further complicate the analysis. 
    In this paper, we focus on the widely used linear-in-means peer-effects model and derive the asymptotic orders of the bias and variance of the resulting GMM estimator. We show that the optimal bandwidth for link-level imputation is generally not optimal for downstream estimation. We also provide guidance to help applied researchers assess the valid inference of parameter estimates obtained using our imputation approach.

    We provide extensive simulation evidence to evaluate the finite-sample performance of our imputation method. First, we compare its imputation accuracy with alternative approaches and show that our method outperforms them across a wide range of sampling rates and network sparsity levels.  We then evaluate the performance of our method in empirical analysis, including regressions based on network statistics and the linear-in-means peer-effects model. These simulations demonstrate that the estimates of regression coefficients and GMM estimators based on our imputed networks exhibit strong finite-sample performance. Finally, we illustrate the practical usefulness of the proposed method through an empirical application to the study of how social networks affect microfinance adoption \citep{banerjee2013diffusion}.

    \paragraph{Related literature} We contribute to the literature on estimating model parameters with sampled network data by developing an imputation method that allows for a fully nonparametric network formation model. \citet{chandrasekhar2011econometrics} show that models estimated using sampled networks are generally biased. They propose an analytical correction for certain special cases and,  for more general settings, they estimate a parametric network formation model and conduct GMM using simulated moments. \citet{boucher2020estimating} study the estimation of the linear-in-means peer-effects model when the network data are corrupted, taking egocentrically sampled networks as a special case. Their approach relies on obtaining a consistent estimator of the distribution of the network, which is typically available only under parametric specifications. \citet{herstad2023estimating} also examine peer-effects estimation with missing links, focusing on settings with large networks and relying on the network formation model of \citet{graham2017econometric}. A key distinction relative to this line of work is that our method does not rely on parametric specifications of the network formation process and does not require specifying the distribution of unobserved heterogeneity. Therefore, parameter estimation based on our imputation method is more robust. \citet{hsieh2024non} propose weighted estimators to recover several network-level statistics without assuming any network formation model. In comparison, our approach accommodates more general sampling mechanisms and supports a broader class of downstream regression analyses. In addition, our approach to GMM estimation differs from that of \citet{chandrasekhar2011econometrics}, who construct moments by integrating over the missing network data. Instead, we directly plug the imputed network into the moment conditions to avoid heavy computational burdens.

    Another related line of research studies the recovery of missing links under partial network data or even in the absence of network data. \citet{mccormick2015latent} and \citet{breza2020using} demonstrate that aggregated relational data can help recover network structure without observing any links. \citet{de2025identifying} study how network ties can be identified from panel data. \citet{thirkettle2019identification} show that network statistics are set identified under partially sampled networks, although their analysis relies on a strategic network formation model, whereas ours is based on a dyadic formation framework. Other sampling designs have also been studied—for example, induced subgraphs in \citet{chatterjee2015matrix} and censored networks in \citet{griffith2022name}.

    Our work also relates to the growing literature on estimation with mismeasured network data, as empirical analysis based on imputed networks inevitably inherits the errors introduced in the imputation stage. For peer-effects models, \citet{lewbel2024ignoring} show that when the number of misclassified links grows at a rate slower than the network size, the 2SLS estimator of \citet{bramoulle2009identification} remains consistent and standard inference methods remain valid. \citet{lewbel2025estimating} propose an adjusted 2SLS estimator for peer-effects models under i.i.d. link misclassification. \citet{hardy2019estimating} also study link misclassification under i.i.d. errors and obtain consistent treatment effects using an expectation-maximization algorithm.  \citet{cai2022linear} examine linear regression on centrality measures under link-level i.i.d. Bernoulli errors, and propose bias-correction and inference methods for OLS estimators. The setting we consider is technically more challenging than those in the existing literature. Our imputed networks contain errors that are dependent across links and may have bias of the same order as their standard deviation. Our paper contributes to the literature by characterizing the convergence rate of linear-in-means peer-effects estimators in settings that allow for far more general forms of network measurement error.

    Methodologically, the studies most directly connected to ours are in the recent literature on nonlinear factor models in panel-data causal inference, particularly the methods of \citet{feng2020causal}, \citet{feng2023optimal}, and \citet{deaner2025inferring}, which also allow for nonparametric and nonlinear factor structures. Our local two-way fixed-effects regression is analogous to a local linear regression and therefore achieves a second-order approximation error. In contrast, the estimator in \citet{deaner2025inferring} corresponds to a local constant regression and attains only a first-order approximation error. The local PCA methods of \citet{feng2020causal} and \citet{feng2023optimal} can be viewed as local polynomial regressions that fully exploit smoothness of nonlinear functions. Our approach differs from local PCA in that it does not require specifying the latent factor dimension. Another important difference is that, unlike most methods in this strand of the literature, our approach flexibly incorporates observed covariates to improve prediction accuracy.

    Finally, our analysis is connected to several strands of the econometrics and statistics literature. The use of pseudo-distance as a measure of latent similarity was first developed by \citet{zhang2017estimating} and has since been applied in a range of contexts, including nonparametric graphon estimation \citep{zhang2017estimating, zeleneev2020identification}, controlling for unobservables using network data \citep{auerbach2022identification}, and nonlinear factor models \citep{feng2020causal, feng2023optimal, beyhum2024inference, mugnier2025simple, deaner2025inferring}. Our analysis is also related to the dyadic network formation literature \citep{graham2017econometric, dzemski2019empirical,  chen2021nonlinear, ma2022detecting} and the graphon estimation literature \citep{gao2015rate, xu2018rates}. \citet{kitamura2024estimating} consider covariate-assisted graphon estimation similar to ours, but they focus on stochastic block models, whereas we focus on more general nonlinear link-formation process.   Finally, extracting the component of the network explained by covariates relies on the dyadic nonparametric regression developed by \citet{graham2021minimax}.

    \vspace{1em}

    The rest of the paper is organized as follows. Section~\ref{sec:method} introduces the framework and estimation procedure. Section~\ref{sec:asymtotics} provides statistical guarantees for the imputation. Section~\ref{sec:empirics} presents theoretical properties of the GMM estimators based on imputed networks. Section~\ref{sec:simulation} presents simulation results for both the imputation and the downstream empirical analysis. The empirical application to \citet{banerjee2013diffusion} is given in Section~\ref{sec:application}. 
    Section~\ref{sec:conclusion} concludes.

    \section{Method}\label{sec:method}

\subsection{Network Formation Model}

    Let $A\in\{0,1\}^{N\times N}$ denote the adjacency matrix of an  unweighted network, where $A_{ij} = 1$ if there is a link between $i$ and $j$ and $A_{ij} = 0$ otherwise. We impose $A_{ij} = A_{ji}$ for all $i, j$ and $A_{ii} = 0$, so that the network is symmetric and has no self-loops. For each individual $i$, let $X_i \in \mathbb{R}^{d_X}$ denote the observed covariates (e.g., gender, education, and other demographic characteristics), and let $\xi_i \in \mathbb{R}^{d_\xi}$ denote the latent factors. 
    The network $A$ is generated according to the following process. Conditional on observable and unobservable characteristics $\{X_i,\xi_i\}_{i=1}^{N}$,  links $\{A_{ij}\}_{1\leq i<j\leq N}$ are generated independently with probability $\mb{P}\left(A_{ij} = 1 \mid \{X_i,\xi_i\}_{i=1}^{N} \right) = f(X_i, \xi_i, X_j, \xi_j)$, where $f: \mb{R}^{d_X + d_{\xi}}\times \mb{R}^{d_X + d_{\xi}} \rightarrow [0, 1] $ is a symmetric function, referred to as the \emph{graphon}, whose functional form is unknown to econometricians. It is therefore convenient to write 
    \begin{equation}\label{eq:graphon}
        A_{ij} = f(X_i, \xi_i, X_j, \xi_j) + \epsilon_{ij}, \quad \forall 1\leq i< j \leq N, 
    \end{equation}
    where $\epsilon_{ij}$ are realization errors, which is conditionally independent across pairs $\{(i,j) \mid  1\leq i< j \leq N\}$ given $\{(X_i,\xi_i)\}_{i=1}^{N}$. 

    The network formation model in~\eqref{eq:graphon} is a conditionally independent dyadic model. Conditional on the observable and unobservable characteristics $\{X_i,\xi_i\}_{i=1}^{N}$, the graphon $f$ does not depend on other individuals or on other links in the network. In other words, while $A_{ij}$ and $A_{ik}$ ($k \neq j$) may be correlated unconditionally, they are independent given $\{X_i,\xi_i\}_{i=1}^{N}$. This framework is flexible and encompasses a wide range of commonly used network formation models. Below, we provide several examples. 

    \begin{example}[Stochastic block model]\label{example:SBM}
        A widely used example consistent with our network formation framework is the stochastic block model (SBM). The SBM assumes that each individual belongs to one of $G$ groups, indexed by $g \in \{1, \ldots, G\}$, and that the probability of link formation depends solely on group membership. In particular, the probability of a link between two individuals is determined entirely by the groups to which they belong, implying that individuals within the same group share identical connection probabilities with other individuals. The stochastic block model can be viewed as a special case of our network formation process in which the support of $(X_i, \xi_i)$ is finite and consists of only $G$ distinct values. Under this restriction, the graphon $f$ reduces to a piecewise constant function defined over the $G^2$ possible group pairs.
    \end{example}
    \begin{example}[Network formation with transferable utility]\label{example:graham}
        Suppose that individuals $i$ and $j$ form a link if the total surplus from doing so is positive, 
        \begin{equation}\label{eq:transfer_degree_heterogeneity}
            A_{ij} = \bs{1}\!\left( \omega(X_i, X_j)'\beta + g(\xi_i, \xi_j) - U_{ij} \ge 0 \right). 
        \end{equation}
        The total surplus consists of three components:
        \begin{itemize}
            \item \textbf{(Homophily in observables)}  $\omega(X_i, X_j)$ measures the distance between observable characteristics of $i$ and $j$.  For instance, one may define  $\omega(X_i, X_j) = (|X_{i1} - X_{j1}|,\ldots, |X_{id_X} - X_{jd_X}| )'$  for absolute differences, or  $\omega(X_i, X_j) = \big((X_{i1} - X_{j1})^2,\ldots, (X_{id_X} - X_{jd_X})^2 \big)'$  for squared differences.  A negative coefficient $\beta < 0$ captures homophily, meaning that individuals with similar observables are more likely to form a link. 
            \item \textbf{(Individual heterogeneity)}  The function $g(\xi_i, \xi_j)$ captures how unobserved individual heterogeneity affects the probability of link formation. For example, consider the specification $g(\xi_i, \xi_j) = \xi_{i1} + \xi_{j1} - (\xi_{i2} - \xi_{j2})^2$. Here, $\xi_{i1}$ represents degree heterogeneity, under which individuals with larger values of $\xi_{i1}$ tend to form more connections. The quadratic term $-(\xi_{i2} - \xi_{j2})^2$ captures homophily in unobservables, with link probabilities decreasing in the distance between individuals' latent characteristics. 
            \item \textbf{(Idiosyncratic shocks)} $\{U_{ij}\mid 1\leq i<j\leq N \}$ captures the idiosyncratic preference shock that is independent of $(X, \xi)$. Let $F(\cdot)$ be the cumulative distribution of $U_{ij}$, then $\mb{P}(A_{ij} = 1\mid \{X_i,\xi_i\}_{i=1}^{N}) = F(\omega(X_i, X_j)'\beta +  g(\xi_i, \xi_j))$. Specifically, when $U_{ij}$ follows standard logistic distribution, 
            \begin{align*}
                f(X_i, \xi_i,X_j, \xi_j) = \frac{\exp\left(\omega(X_i, X_j)'\beta +  g(\xi_i, \xi_j)\right)}{1 + \exp\left(\omega(X_i, X_j)'\beta +  g(\xi_i, \xi_j) \right)}. 
            \end{align*}
            When $d_{\xi}=1$ and $g(\xi_i, \xi_j)$ includes only the additive component $\xi_i + \xi_j$, \eqref{eq:transfer_degree_heterogeneity} reduces to the model of \citet{graham2017econometric}, which is common in empirical applications.
        \end{itemize}

    \end{example}
    
    Beyond these examples, our network formation framework in \eqref{eq:graphon} can flexibly accommodate richer forms of heterogeneity.  For example, it allows for heterogeneous slope coefficients when the homophily effect on observables, captured by $\beta$ in \eqref{eq:transfer_degree_heterogeneity}, depends on the latent factors $(\xi_i, \xi_j)$.  Similarly, the unobserved heterogeneity term $g(\xi_i, \xi_j)$ can capture more complex interaction structures, such as interactive fixed effects of the form $\xi_i'\xi_j$. However, our dyadic model excludes strategic network formation, where the formation of one link may depend on the presence of other links (see \citet{de2020econometric}).

\subsection{Data} 
    
    Economists observe individual covariates $X_i$ (e.g., gender, age) for \emph{all} $i$ in the network but do not observe the full network $A$; instead, only a sampled network is available. This setting is common in applications because collecting covariates is typically less costly than collecting network information. 
    
    In this paper, we focus on egocentric sampling,  which is one of the most commonly used sampling designs in empirical work. Specifically, we randomly sample $n < N$ nodes in the network and ask all their friends not only in the sample but in the whole network. The sampling probabilities can depend on both covariates $X_i$ and unobserved characteristics $\xi_i$.  
    Let $\mc{S} \subset \{1,2, \ldots, N\}$ denote the set of sampled individuals with $|\mc{S}| = n$. Then a link $A_{ij}$ is observed if  $i\in \mc{S}$ or $j\in \mc{S}$, whereas links between two non-sampled nodes (i.e., $i\notin\mc{S}$ and $j\notin\mc{S}$) are unobserved. A simple example follows to illustrate the setup. 

    \begin{figure}[H]
        \centering
        \begin{minipage}[t]{0.48\textwidth}
        \centering
        \includegraphics[width=0.4\textwidth]{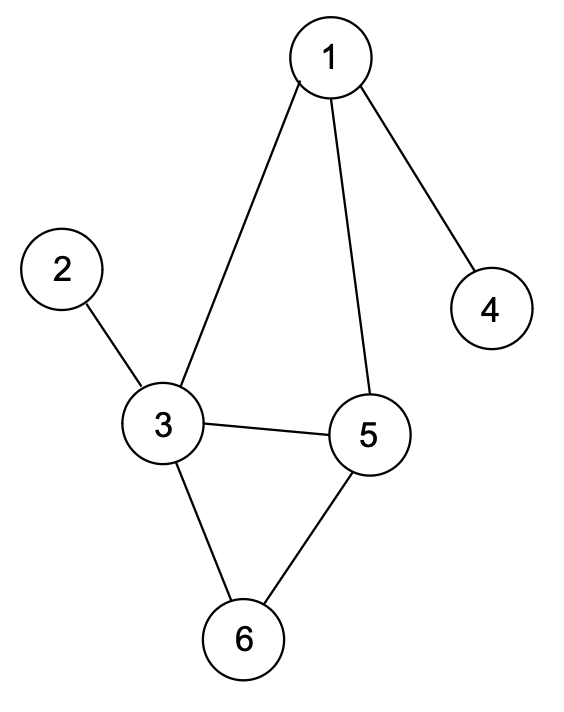}
        \caption*{(a) Full network}
        \end{minipage}
        \begin{minipage}[t]{0.48\textwidth}
        \centering
        \includegraphics[width=0.4\textwidth]{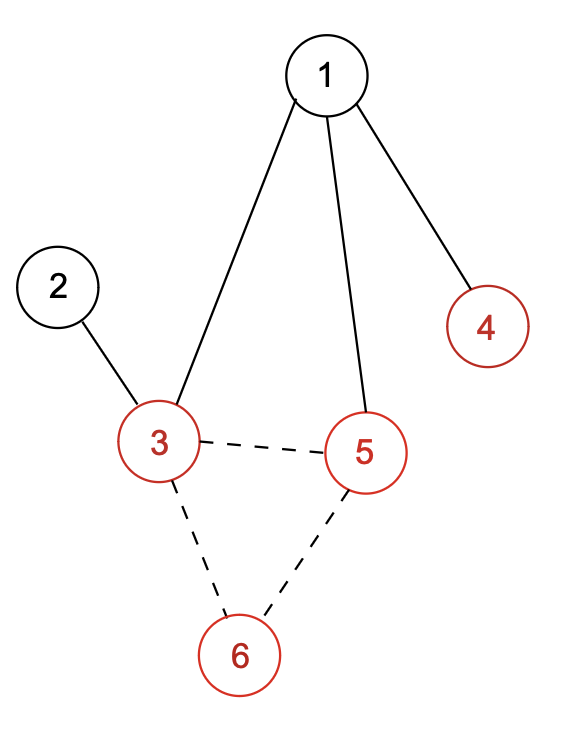}
        \caption*{(b) Sampled network by sampling $\left\{1,2\right\}$, red nodes are not in the sample, dashed edges are not observed} 
        \end{minipage}
        \begin{minipage}[t]{0.48\textwidth}
            \begin{align*}
                \left(\begin{array}{cccccc}
                    0 & 0 & 1 & 1 & 1 & 0 \\
                    0 & 0 & 1 & 0 & 0 & 0 \\
                    1 & 1 & 0 & 0 & 1 & 1 \\
                    1 & 0 & 0 & 0 & 0 & 0 \\
                    1 & 0 & 1 & 0 & 0 & 1 \\
                    0 & 0 & 1 & 0 & 1 & 0 
                \end{array}\right)
                \Rightarrow 
                \left(\begin{array}{cc:cccc}
                    0 & 0 & 1 & 1 & 1 & 0 \\
                    0 & 0 & 1 & 0 & 0 & 0 \\
                    \hdashline
                    1 & 1 & \times & \times & \times & \times \\
                    1 & 0 & \times & \times & \times & \times \\
                    1 & 0 & \times & \times & \times & \times\\
                    0 & 0 & \times & \times & \times & \times 
                \end{array}\right)
           \end{align*}
           \caption*{(c) From full adjacency matrix to partially observed adjacency matrix, $\times$ denotes missing entry }  
        \end{minipage}
        \caption{An example of how egocentric sampling works}
        \label{fig:sample}
    \end{figure}
     
    Since it causes no harm to relabel the agents, we can do so such that $\mathcal{S} = \{1,2,\ldots,n\}$, in which case the observed adjacency matrix $A^{\mathrm{obs}}$ admits the following block structure: 
    \begin{equation}\label{eq:block_structure}
        A^{\mr{obs}} = 
        \begin{pmatrix}
            A_{\mc{S}\mc{S}} & A_{\mc{S}\mc{S}^c}\\
            A_{\mc{S}^c\mc{S}} & \textcolor{red}{A_{\mc{S}^c\mc{S}^c} = \times}
        \end{pmatrix}. 
    \end{equation}
    Here, $A_{\mc{S}\mc{S}}$ consists of links between two sampled individuals, $A_{\mc{S}^c\mc{S}} = A_{\mc{S}\mc{S}^c}' $ contains links between sampled and unsampled individuals, and the bottom right block  $A_{\mc{S}^c\mc{S}^c}$ is unobserved  because both individuals in each dyad are unsampled; we highlight this block in red.

\subsection{Imputation}
    
    To illustrate the main idea of our imputation approach, let us start with a simple case in which  both $X_i$ and $\xi_i$ are observable for every individual in the network.  In this case, imputing missing links  would be straightforward: we could  estimate $f(\cdot, \cdot)$ nonparametrically from the observed adjacency matrix to obtain estimates $\hat{f}(\cdot, \cdot)$, and then use these estimates to impute the missing links in the submatrix $A_{\mc{S}^c \mc{S}^c}$.
  
    The main difficulty, however, is that the latent factors $\xi$ are unobserved for all individuals. To make this challenge more transparent, let us first abstract from the observed covariates $X$ and focus on the case where link formation depends solely on the latent factors. In this setting, the imputation problem reduces to recovering information about the latent factors $\xi$ from the observed subnetwork and using them to predict the missing  entries of the adjacency matrix $A_{\mc{S}^c \mc{S}^c}$.

    Our approach is motivated by the following idea. When network formation is dyadic, as in~\eqref{eq:graphon}, individuals with similar $\xi$ are expected to have similar linking probabilities. This observation provides a useful heuristic: similarity in observed connections reflect similarity in latent positions. Building on this intuition, for any unsampled individual $i$, we use their observed links to sampled individuals to measure their similarity to sampled nodes. We then use the observed links of sampled individuals to other unsampled individuals, weighted by their similarity to individual $i$, to impute $i$'s missing links to other unsampled individuals.

    Because the information about $\xi$ is estimated rather than directly observed, it is inappropriate to ignore the observed covariates or absorb them into the latent factors, as doing so would discard the noiseless information contained in $X$. Therefore, when $X$ is available, our method first extracts the component of link formation explained by the observed covariates and then applies the latent-factor-based procedure described above to the residuals. These two components are subsequently combined to obtain the final imputed network.

\subsubsection{Extract information explained by $X$} 

    As discussed previously, when individual covariates $X$ are available, it is appropriate to first extract the component that can be explained by $X$. Consider the decomposition:
    \begin{align*}
        f(X_i, \xi_i, X_j, \xi_j) = \underbrace{\mb{E}\left(f(X_i, \xi_i, X_j, \xi_j)\mid X_i, X_j\right)}_{\Pi_{ij}} + \underbrace{f(X_i, \xi_i, X_j, \xi_j) - \mb{E}\left(f(X_i, \xi_i, X_j, \xi_j)\mid X_i, X_j\right)}_{\Pi_{ij}^{\bot} }. 
    \end{align*}
    where $\Pi_{ij}$ represents the part of the graphon that can be explained by the observed covariates, and $\Pi_{ij}^{\bot}$ captures the residual component of the graphon that cannot be explained solely by the observed covariates. Note that 
    the observed link $A_{ij}$ can then be written as $A_{ij} = \Pi_{ij} + (\Pi_{ij}^{\bot} + \epsilon_{ij})$,  and the composite error term satisfies $\mb{E}(\Pi_{ij}^{\bot} + \epsilon_{ij}\mid X_i, X_j) = 0$, which implies that dyadic nonparametric regression of $A_{ij}$ on $(X_i, X_j)$ consistently estimates $\Pi_{ij}$. 
    \citet{graham2021minimax} derive the pointwise and uniform minimax risks for estimating such conditional expectations and show that the Nadaraya-Watson (NW) estimator attains the optimal convergence rate\footnote{
        In practice, we implement local linear regression rather than using NW estimator. We discuss this in  Appendix~\ref{appendix:extra_dyadic}. 
    }. We implement the dyadic nonparametric regression using the observed subset of the adjacency matrix. Since the covariates $X$ are also observed for unsampled individuals, the nonparametric estimator can then be evaluated directly for missing links, that is, the estimator $\hat{\Pi}_{ij}$ can be computed even when both $i$ and $j$ are unsampled. The implementation details are discussed in Appendix~\ref{appendix:extra_dyadic}. 

    In the next two steps, we focus on estimating this residual component ${\Pi}^{\bot}$ using pseudo-distance and two-way fixed-effect imputation procedure described below.

\subsubsection{Calculate similarity}

    We now describe how to infer similarity between latent factors using observed network links.  
    
    For notational simplicity, let $\zeta_{i} := (X_{i}, \xi_{i})$ for each $i$, and let $f_{ij}=f(\zeta_i, \zeta_j)$ for all $1\leq i, j\leq N$. The adjacency matrix $A$ can be expressed as  $A = P + E$. Here, $P$ collects the conditional probabilities of link formation, 
    i.e., $P_{ij} = f_{ij}$ when $i\neq j$, and $P_{ii} = 0$ (because self-loops are ruled out), and $E$ collects error terms $\epsilon_{ij}$\footnote{
        Likewise, we impose $\epsilon_{ii} = 0$ for each $i$.
    }.  
    Our approach relies on the following idea. For any unsampled individual $i$ and sampled individual $i'$, if their linking probabilities to other sampled individuals (or equivalently, the corresponding rows in $P$) appear sufficiently similar, then under suitable regularity conditions, $\zeta_i$ and $\zeta_{i'}$ are also similar in the latent space.  

    A natural measure of the difference between linking probabilities is the squared $L_2$ distance between the functions $f(\zeta_i, \cdot)$ and $f(\zeta_{i'}, \cdot)$, defined as 
    \begin{align*}
        L^2_2(\zeta_i, \zeta_{i'}) := \int (f(\zeta_i, \tilde{\zeta}) - f(\zeta_{i'}, \tilde{\zeta}))^2 \mr{d}\mb{P}(\tilde{\zeta}). 
    \end{align*}
    The squared $L_2$ distance equals zero if and only if the two individuals $i$ and $i'$ have identical link formation probabilities with all other individuals in the network, that is, $f(\zeta_i, \cdot) = f(\zeta_{i'}, \cdot)$ almost everywhere. Therefore, if one could construct a consistent estimator of $L^2_2(\zeta_i, \zeta_{i'}) $ from the observed network, it would provide a natural proxy for latent similarity. However, the squared $L_2$ distance is not directly estimable. Its sample analogue based on observed network links is given by
    \begin{align*}
        \frac{1}{n}\sum_{\ell = 1}^{n} (A_{i\ell} - A_{i'\ell})^2 \cp L^2_2(\zeta_i, \zeta_{i'}) + \mb{E}(\epsilon_{i\ell}^2\mid \zeta_i) + \mb{E}(\epsilon^2_{i'\ell}\mid \zeta_{i'}),
    \end{align*}
    where the additional variance terms do not vanish asymptotically and therefore contaminate the estimation. A formal derivation and further discussion are provided in Appendix~\ref{appendix:extra_informativeness}.  

    To address this issue, we follow \cite{zhang2017estimating} and define a new metric between the functions $f(\zeta_i, \cdot)$ and $f(\zeta_{i'}, \cdot)$ as 
    \begin{align*}
        d(\zeta_i, \zeta_{i'}): = \sup_{\zeta \in \mr{supp}(\zeta)} \left| \int f(\zeta, \tilde{\zeta})(f(\zeta_i, \tilde{\zeta}) - f(\zeta_{i'}, \tilde{\zeta})) \mr{d}\mb{P}(\tilde{\zeta})\right|. 
    \end{align*} 
    The new metric $d(\cdot, \cdot)$ has two useful properties. First, it is straightforward to verify that $d(\zeta_i, \zeta_{i'}) \geq \frac{1}{2} L^2_2(\zeta_i, \zeta_{i'})$ and that $d(\zeta_i, \zeta_{i'}) = 0$ if $\zeta_i = \zeta_{i'}$. Therefore, controlling $d(\cdot,\cdot)$ is sufficient to control $L^2_2(\zeta_i, \zeta_{i'})$. Second, unlike the squared $L_2$ distance, $d(\zeta_i, \zeta_{i'})$ admits a consistent estimator based on the observed network. Specifically, define 
    \begin{align*}
        \hat{d}_{ii'} := \max_{\substack{k=1, \ldots, N, \\ k\neq  i, i'}} \left|\frac{1}{n} \sum_{\ell = 1}^{n} A_{k\ell}(A_{i\ell } - A_{i'\ell})\right| \cp d(\zeta_i, 
        \zeta_{i'}), \quad i, i' = 1, \ldots, N. 
    \end{align*} 
    We show that $\hat{d}_{ii'} \cp d(\zeta_i, \zeta_{i'})$. We refer to $\hat{d}_{ii'}$ as the pseudo-distance and $d(\zeta_i, \zeta_{i}')$ as the population pseudo-distance  throughout the paper. Calculating $\hat{d}_{ii'}$ relies only on the friendships of sampled individuals and can therefore be computed from the observed subnetwork. 
    In addition, under suitable regularity conditions, the pseudo-distance $\hat{d}_{ii'}$ is informative about latent similarity, in the sense that small values of $\hat{d}_{ii'}$ imply that $\zeta_i$ and $\zeta_{i'}$ are close.

    It is important to note that the discussion here concerns the composite characteristic $\zeta_i$, rather than the unobserved heterogeneity $\xi_i$ alone. 
    This is because (i) $X_i$ and $\xi_i$ enter the network formation process in a non-separable way, making it impossible to isolate their effects without additional assumptions; and (ii) the information in $X_i$ that can be directly explained has already been extracted in the dyadic nonparametric regression, so there is no need to distinguish between $X_i$ and $\xi_i$ in this step.

\subsubsection{Local two-way fixed-effects imputation}

    For the projection residuals, we employ the pseudo-distance $\hat{d}$ obtained in the previous step to construct a local estimator for $\Pi_{ij}^{\bot}$, referred to as the two-way fixed-effect imputation. Specifically, for unsampled individuals $i$ and $j$,  we estimate the local two-way fixed-effects regression to obtain $\hat{a}_i$ and $\hat{b}_j$: 
    \begin{equation}\label{eq:TWFE}
        (\hat{\bs{a}}, \hat{\bs{b}}) \in \argmin_{\bs{a}, \bs{b} \in \mb{R}^{n+1}} \sum_{ \substack{i'\in \mc{S}\cup\{i\},  j'\in \mc{S}\cup\{j\} \\ (i', j') \neq (i, j)}}  \left(A_{i'j'} - \hat{\Pi}_{i'j'} - a_{i'} - b_{j'} \right)^2 K_h\left(\hat{d}_{ii'}\right)K_h\left(\hat{d}_{jj'}\right), 
    \end{equation}
    where $K_h(\cdot) := h^{-1}K(\cdot / h)$ is a kernel function with bandwidth $h$, and $\hat{d}$ denotes the pseudo-distance.  Then the imputed $\Pi_{ij}^{\bot}$ is obtained via $\hat{\Pi}_{ij}^{\bot} = \hat{a}_i + \hat{b}_j$, and the imputed network link is given by
    \begin{align*}
        \hat{A}_{ij} = \hat{\Pi}_{ij} + \hat{\Pi}_{ij}^{\bot} = \hat{\Pi}_{ij} + \hat{a}_i + \hat{b}_j. 
    \end{align*}
    
    The imputation is “local” in the sense that, by using a kernel estimator, we rely only on information from individuals $i'$ and $j'$ whose pseudo-distances $\hat{d}_{ii'}$ and $\hat{d}_{jj'}$ (or, equivalently, the distances between their latent positions) are sufficiently small when estimating fixed effects. Note that, although our notations omit the indices $(i, j)$ for simplicity, the estimator $(\hat{\bs{a}}, \hat{\bs{b}})$ in \eqref{eq:TWFE} is pair-specific and therefore must be re-estimated for each $(i, j)$. 

    The procedure is referred to as a “two-way fixed-effects imputation"  because \eqref{eq:TWFE} takes the form of the two-way fixed-effects regression rather than a local constant regression. This is analogous to replacing a local constant estimator with a local polynomial estimator in nonparametric regression: incorporating the two-way fixed-effects provides a higher-order local approximation and thus improves estimation accuracy. We will formally justify this result later using  a Taylor expansion. 

    It is important to note that  $\hat{A}_{ij}$  can only approximate the conditional link-formation probability $f_{ij}$ rather than the missing link $A_{ij}$. This is because the missing link contains an unpredictable idiosyncratic error $\epsilon_{ij}$, which cannot be predicted from the available data.

    We impute all missing links based on the previous steps and merge these imputed entries with the observed ones to obtain a complete imputed matrix $\hat{A}$.  To facilitate understanding of the main idea of our algorithm, we summarize below a simplified version of our method that excludes cross-fitting and cross-validation, which we would discuss later. 
    \begin{algorithm}[h]
        \caption{Imputation (Basic Idea)}\label{alg:simple}
        \begin{algorithmic}
            \REQUIRE Kernel $K(\cdot)$, bandwidths $h$, incomplete network $A^{\mr{obs}}$, and covariate matrix $X$. 
            \ENSURE Imputed network $\hat{A}$.
                \STATE \textbf{Step 1: Extract information explained by $X$}  
                \STATE Implement dyadic nonparametric regression to obtain $\hat{\Pi}(\cdot, \cdot)$.  
                \STATE \textbf{Step 2: Calculate similarity}  
                \STATE For each $i \in \mc{S}^c$ and $i'\in \mc{S}$, calculate the pseudo-distance using observed adjacency matrix: 
                \begin{align}\label{eq:alg:simple:d}
                    \hat{d}_{ii'} = \max_{\substack{k=1, \ldots, N, \\ k\neq  i, i'}} \left|\frac{1}{n} \sum_{\ell = 1}^{n} A_{k\ell}(A_{i\ell } - A_{i'\ell})\right|. 
                \end{align} 
                \STATE \textbf{Step 3: Local imputation}   
                \STATE For each missing entry $(i, j)$, i.e., $i\in \mc{S}^c$ and  $j\in \mc{S}^c $, estimate the local two-way fixed-effects regression
                \begin{align}\label{eq:alg:simple:Aij}
                    (\hat{\bs{a}}, \hat{\bs{b}}) \in \argmin_{\bs{a}, \bs{b} \in \mb{R}^{n+1}} \sum_{\substack{i'\in \mc{S}\cup\{i\},  j'\in \mc{S}\cup\{j\}, \\ (i', j') \neq (i, j)}}   \left(A_{i'j'} - \hat{\Pi}_{i'j'} - a_{i'} - b_{j'} \right)^2 K_h\left(\hat{d}_{ii'}\right)K_h\left(\hat{d}_{jj'}\right)
                \end{align}
                to obtain $\tilde{A}_{ij} = \hat{a}_i + \hat{b}_j + \hat{\Pi}_{ ij}$.  By combining the imputed entries with the observed entries of the adjacency matrix, we obtain the (preliminary) imputed adjacency matrix $\tilde{A}$. Truncate $\tilde{A}_{ij}$ to get 
                \begin{align*}
                    \hat{A}_{ij} = \left\{
                        \begin{array}{ll}
                           \tilde{A}_{ij}, & \text{if } \tilde{A}_{ij}\in [0, 1]\\
                            1 & \text{if } \tilde{A}_{ij} > 1 \\
                            0 & \text{if } \tilde{A}_{ij} < 0
                        \end{array}
                    \right., 
                \end{align*}
                and zero the diagonal of $\hat{A}$ such that $\hat{A}_{ii} = 0$ for all $i$. 
        \end{algorithmic}
    \end{algorithm}

\subsubsection{Cross-validation and sample-splitting}

    Although the proposed algorithm is straightforward to implement, deriving the asymptotic properties of $\hat{A}_{ij}$ can be technically challenging because of the potential dependence between the estimated pseudo-distance $\hat{d}$ and the error terms $\epsilon$ in the two-way fixed-effects regression. 
    
    A standard approach to eliminate this dependence is to employ cross-fitting, in which the data are partitioned into folds so that the first-step and second-step estimations are conducted on separate subsets of the data. Specifically, we randomly split the sampled individuals $\mc{S}$ into two parts, $\mc{S}_1$ and $\mc{S}_2$, of equal size. We then use only the friendships of $\mc{S}_1$ to calculate the similarity $\hat{d}(\cdot,\cdot)$:
    \begin{align*}
        \hat{d}_{ii'} = \max_{\substack{k=1, \ldots, N, \\ k\neq  i, i'}} \left|\frac{1}{|\mc{S}_1 |} \sum_{\ell \in \mc{S}_1 } A_{k\ell}(A_{i\ell } - A_{i'\ell})\right|,  \quad i \in  \mc{S}^c, i'\in \mc{S}_2. 
    \end{align*}
    In the imputation stage, we use no link information from $\mc{S}_1$, and all calculations rely on the friendships of $\mc{S}_2$: 
    \begin{gather*}
        (\hat{\bs{a}}, \hat{\bs{b}}) \in \argmin_{\bs{a}, \bs{b} \in \mb{R}^{|\mc{S}_2| + 1}} \sum_{ \substack{i'\in \mc{S}_2 \cup\{i\} , j'\in \mc{S}_2 \cup\{j\}, \\ (i', j') \neq (i, j)}   }  \left(A_{i'j'} - \hat{\Pi}_{i'j'}  - z_{i'} - b_{j'} \right)^2 K_{h}\left(\hat{d}_{ii'}\right)K_{h}\left(\hat{d}_{jj'}\right), \\
        \hat{A}_{ij} = \hat{a}_i + \hat{b}_j + \hat{\Pi}_{ ij}. 
    \end{gather*}
    
    In addition, a practical concern is the choice of the tuning parameter $h$. We recommend selecting $h$ using a leave-one-out cross-validation procedure, and  the details of cross-validation, together with the sample-splitting,  are provided in  Appendix~\ref{appendix:extra_algorithm}.

    \section{Theory for Imputation}\label{sec:asymtotics}
     
    We now introduce the regularity conditions.

    \begin{assumption}[Regularity Conditions]\label{assumption:regularity}
        Suppose that 
        \begin{enumerate}[label=(\roman*)]
            \item \label{item:regularity_assumption_asymptotics} \textbf{(Asymptotics)} Large-$N$, large-$n$ asymptotics with $N, n\rightarrow \infty$ and $\frac{\log N}{n} \rightarrow 0$. 
            \item \label{item:regularity_assumption_independence} \textbf{(Independence)} $\{(X_i, \xi_i)\}_{i=1}^{N}$ are i.i.d., and conditional on $\{(X_i, \xi_i)\}_{i=1}^{N}$, $\{A_{ij}\}_{1\leq i < j \leq N}$ are generated independently  according to \eqref{eq:graphon}. 
            \item \label{item:regularity_assumption_sampling} \textbf{(Sampling)} Let $D_i\in\{0,1\}$ indicate whether $i$ is sampled. Assume $D_i \bot (A,\{(X_i,\xi_i)\}_{i=1}^{N})$ for all $i$, and $\{D_i\}_{i=1}^N$ are independent across $i$.
            \item \label{item:regularity_assumption_compact_support} \textbf{(Compact support)} Let $\zeta_i = (X_i', \xi_i')'$ and $d_{\zeta} = d_X + d_{\xi}$, assume that the support of $\zeta_i$ is compact on $\mb{R}^{d_{\zeta}}$. In addition, there exist constants $0 < \underline{c} < \bar{c}   < \infty$ such that for any $\zeta \in \mr{supp}(\zeta)$ and any $h>0$, $\underline{c} h^{d_{\zeta}} \leq \mb{P} (\|\zeta_i - \zeta\|\leq h) \leq  \bar{c} h^{d_{\zeta}}$.
            \item \label{item:regularity_assumption_smoothness} \textbf{(Smoothness)} For every fixed $\tilde{\zeta}\in \mr{supp}(\zeta)$, the mapping $\zeta \mapsto f(\zeta ,\tilde{\zeta})$ is twice continuously differentiable, and its second derivatives are uniformly bounded for all $\zeta, \tilde{\zeta} \in \mr{supp}(\zeta)$. 
        \end{enumerate}
    \end{assumption}

    Assumption~\ref{assumption:regularity}\ref{item:regularity_assumption_asymptotics} indicates that our method is designed for large-network settings in which both the network size $N$ and the number of sampled individuals $n$ grow large.  Our analysis does not rely on $n$ and $N$ growing at the same rate and continues to hold when $n$ grows slowly, e.g., $n = O(N^{\alpha})$ for any $\alpha \in (0,1)$.  
    
    Assumption~\ref{assumption:regularity}\ref{item:regularity_assumption_independence} imposes that $\{(X_i, \xi_i)\}_{i=1}^{N}$ are i.i.d., and the data-generating process of $A$ follows a conditionally independent dyadic model \eqref{eq:graphon}. This model also follows from the Aldous-Hoover Theorem \citep{aldous1981representations, hoover1979relations}, which shows that any infinitely exchangeable random graph admits such a representation. As discussed earlier, although \eqref{eq:graphon} is flexible enough to encompass a wide range of network formation models commonly used in economics and statistics, it rules out strategic network formation, in which individuals linking decisions depend on those of others; see \citet{de2020econometric} for a review.

    Assumption~\ref{assumption:regularity}\ref{item:regularity_assumption_sampling} is standard in egocentric sampling designs, which assume that individuals are randomly drawn from the population. Our method also applies to more general sampling schemes in which individuals are sampled with heterogeneous probabilities that depend on both observed and unobserved characteristics, i.e., $D_i \bot A \mid \{(X_i, \xi_i)\}_{i=1}^{N}$. In this case, the asymptotic analysis can be extended with only minor modifications, as long as the sampling probabilities $\mb{P}(D_i = 1\mid \{(X_i, \xi_i)\}_{i=1}^{N})$ are uniformly bounded away from zero across individuals.

    The first part of Assumption~\ref{assumption:regularity}\ref{item:regularity_assumption_compact_support} imposes the boundedness of $\zeta$, and the second part  introduces a standard bounded-density condition. Intuitively, it ensures that the distribution of $\zeta_i$ is neither locally too sparse nor too concentrated by requiring that,   for any point $\zeta$ and a small radius $h>0$, the probability that $\zeta_i$ falls within a ball of radius $h$ centered at $\zeta$ is of the same order as the volume of that ball. This requirement is automatically satisfied when (i) $\mr{supp}(\zeta)$ is compact and convex, (ii) $\mr{affine}(\mr{supp}(\zeta)) = \mb{R}^{d_{\zeta}}$,  and (iii) the probability density function of $\zeta$ is uniformly bounded above and bounded away from zero. This assumption is common for establishing the asymptotic properties of nonparametric kernel estimators.

    Assumption~\ref{assumption:regularity}\ref{item:regularity_assumption_smoothness} requires the graphon $f$ to be twice continuously differentiable with uniformly bounded second derivatives. This smoothness condition is needed because our two-way fixed-effects imputation method relies on a second-order Taylor expansion of $\Pi^{\bot}_{ij}$ within a local neighborhood to eliminate the first-order approximation bias, thereby improving accuracy. Methods that rely on higher-order Taylor expansions, such as the local PCA approach proposed by \citet{feng2023optimal}, correspondingly require the existence of higher-order derivatives. We discuss this point in detail later.  

    It should be noted that the requirement of second-order differentiability of $f$ is not a mild condition. For instance, consider the network formation model with degree heterogeneity in Example~\ref{example:graham}. If the distance function $\omega(\cdot,\cdot)$ is defined by the absolute distance, i.e., $\omega(X_i, X_j) = |X_i - X_j|$, which is common in applied research, then the corresponding graphon $f$ is at most Lipschitz continuous rather than twice differentiable. Nevertheless, our method remains valid and can consistently impute the conditional probability of link formation, although the approximation error no longer attains the second-order rate.

\subsection{Asymptotic analysis of pseudo-distance}
    
   We now develop the asymptotic properties of the pseudo-distance $\hat{d}_{ii'}$ and show that a sufficiently small $\hat{d}_{ii'}$ implies that $\zeta_i$ and $\zeta_{i'}$ are close in the latent space. In the following lemma, we establish the convergence of $\hat{d}_{ii'}$ to its population counterpart $d(\zeta_i, \zeta_{i'})$. 
    \begin{lemma}\label{lemma:consistency_hat_d}
        Under Assumption~\ref{assumption:regularity}, there exist constants $\gamma_1, \gamma_2  >0$ such that
         \begin{align*}
            \mb{P}\left(\max_{  i \in \mc{S}^c,  i' \in \mc{S}_2 } \left| \hat{d}_{ii'} - d(\zeta_i, \zeta_{i'}) \right| \geq \gamma_1 \frac{\log N }{N^{1/d_{\zeta}}} + \gamma_2 \frac{\log N }{\sqrt{n}}   \right) \leq n^{-1/2}. 
        \end{align*}
    \end{lemma}
    Lemma~\ref{lemma:consistency_hat_d} shows that the approximation error of $\hat{d}_{ii'}$ relative to its population counterpart $d(\zeta_i, \zeta_{i'})$ can be decomposed into two components. The first term, $\gamma_1 \frac{\log N}{N^{1/d_{\zeta}}}$, captures the matching discrepancy in the latent characteristics $\zeta_i$. Its rate deteriorates with the dimension of the latent space $d_{\zeta}$, which is consistent with the well-known “curse of dimensionality” in the matching literature. The second term, $\gamma_2 \frac{\log N}{\sqrt{n}}$, reflects the effect of the idiosyncratic error $\epsilon_{ij}$. It shows that although the true graphon is unobservable, the pseudo-distance can asymptotically denoise the observed network. The matching discrepancy error dominates the effect of idiosyncratic errors when $N\sim n$ and $d_{\xi} \geq 3$, which is the common case in practice.  Finally, for notational simplicity, let $$\delta_{N, n} := \gamma_1 \frac{\log N }{N^{1/d_{\zeta}}} + \gamma_2 \frac{\log N }{\sqrt{n}}, $$ and focus on the event under which $|\hat{d}_{ii'} - d(\zeta_i, \zeta_{i'})|$ is uniformly bounded by $\delta_{N, n}$, as implied by Lemma~\ref{lemma:consistency_hat_d}.

    The following assumption is critical and requires that the population pseudo-distance $d(\zeta_i, \zeta_{i'})$ be sufficiently informative to reveal the latent distance $\|\zeta_i - \zeta_{i'}\|$. 
    \begin{assumption}[Informativeness]\label{assumption:informativeness}
        There exists a constant $\lambda >0$ such that for all $\zeta_i, \zeta_{i'} \in \mr{supp}(\zeta)$, $d(\zeta_i, \zeta_{i'}) \geq \lambda \|\zeta_i - \zeta_{i'}\|$. 
    \end{assumption}
    This assumption ensures that when the population pseudo-distance $d(\zeta_i, \zeta_{i'})$ is sufficiently small, the latent positions $\zeta_i$ and $\zeta_{i'}$ are close to each other. Hence, $d(\zeta_i, \zeta_{i'})$ can be used to identify individuals with similar latent characteristics. This assumption is crucial for establishing link-level imputation error bounds. In  Appendix~\ref{appendix:extra_informativeness}, we provide a set of primitive sufficient conditions for Assumption~\ref{assumption:informativeness} and verify that it holds under commonly used network formation models. 
    
    \setcounter{example}{0}

    \begin{example}[Continued]
        When each individual belongs to one of $G$ groups, $g\in\{1, \ldots, G\}$, Assumption~\ref{assumption:informativeness} is  guaranteed as long as there is heterogeneity across different groups, that is, no two groups share identical linking patterns. A formal illustration is provided in Lemma~\ref{lemma:SBM_informativeness} and the proof is provided  in Appendix~\ref{appendix:supplementary}. When the graphon $f$ takes a more general low-rank form, the informativeness condition still holds under similarly mild assumptions. We refer the reader to the discussions in \citet{feng2023optimal} and \citet{deaner2025inferring} for further details.
    \end{example}

    \begin{example}[Continued]
        In Example~\ref{example:graham}, the latent type $\zeta$ may be continuously distributed, and the graphon $f$ is not necessarily low rank. Verifying Assumption~\ref{assumption:informativeness} in this setting is generally challenging. Nevertheless, we show that the informativeness assumption holds in the following model which serves as a benchmark in empirical applications. Specifically, Assumption~\ref{assumption:informativeness} holds when (i) $U_{ij}$ in~\eqref{eq:transfer_degree_heterogeneity} follows a standard logistic distribution or normal distribution, and (ii) the model allows for degree heterogeneity and quadratic-distance homophily effects on both observable and unobservable characteristics (see Lemma~\ref{lemma:verification_sigle_index_quadratic} and Corollary~\ref{corollary:logit_informativeness} in  Appendix~\ref{appendix:extra_informativeness}). This finding implies that the informativeness assumption is empirically plausible and unlikely to impose restrictive constraints in applied network analysis. By contrast, we also show that under absolute-distance homophily, Assumption~\ref{assumption:informativeness} may fail in multidimensional settings (see Lemma~\ref{lemma:verification_absolute_logit} and Lemma~\ref{lemma:verification_impossible} in Appendix~\ref{appendix:extra_informativeness}). 
    \end{example}

    \paragraph{Remark}  An alternative to the pseudo-distance is the difference in average degrees, defined as $\hat{\rho}_{ii'} = \frac{1}{|\mc{S}_1|}\sum_{\ell \in \mc{S}_1}A_{i\ell} - \frac{1}{|\mc{S}_1|}\sum_{\ell \in \mc{S}_1} A_{i'\ell}$.  However, using the average-degree distance requires a stronger informativeness assumption than Assumption~\ref{assumption:informativeness}. In particular, even when $\zeta$ is one-dimensional, using average-degree distance requires that the population average degree, $\zeta \mapsto \int  f(\zeta, \tilde{\zeta})\,d\mb{P}(\tilde{\zeta})$, is injective in $\zeta$. (In the case  where $\zeta$ is a scalar, this is equivalent to requiring the population average degree to be strictly monotonic in $\zeta$.) This condition does not hold in either Example~\ref{example:SBM} or Example~\ref{example:graham} and is stronger than Assumption~\ref{assumption:informativeness} in our paper. 

    \paragraph{Remark} While Assumption~\ref{assumption:informativeness} is convenient for theoretical analysis, it should not be interpreted as necessary for good imputation performance. Even under the cases where the informativeness condition may fail, our simulation results in Appendix~\ref{appendix:extra_simulation} show that our proposed method still performs well in practice and outperforms alternative approaches.

\subsection{Asymptotic analysis of imputation}

    The error of our proposed imputation method can be decomposed into two parts corresponding to the dyadic nonparametric regression step and the local two-way fixed-effects regression step.  The first-stage error is typically asymptotically negligible relative to the second-stage error for two reasons. First, the dyadic nonparametric regression only extracts variation explained by the \emph{observed} covariates $X$, whereas the local two-way fixed-effects regression involves unobserved heterogeneity $\xi$. Because $\xi$ must be approximated through the pseudo-distance, there is an additional source of estimation error in the second stage. Second, the dimension of the regressors in the first-stage nonparametric regression is $2d_X$. By contrast, because the graphon $f$ is nonseparable in $X$ and $\xi$, the dimension in the second-stage regression is $d_{\zeta} = d_X + d_{\xi}$, leading to a slower convergence rate. Therefore,  our asymptotic analysis primarily focuses on the second-stage estimation error. 
    
    The asymptotic properties of dyadic nonparametric regression are studied in the literature (see \citep{graham2021minimax}). We therefore summarize the required  conditions in the following assumption.
    \begin{assumption}\label{assumption:dyadic_nonparametric} 
        The first-stage dyadic nonparametric regression estimator $\hat{\Pi}_{ij}$ satisfies the following uniform bounds on its bias and variance: 
    \begin{align*}
        \sup_{x_i, x_j \in \mc{X}}  \left|\mb{E}\left(\hat{\Pi}(x_i, x_j) - \Pi(x_i, x_j)\right)\right| =   O\left(n^{\frac{-2}{4 + d_{X}}} \log (N)\right) , \quad \sup_{x_i, x_j \in \mc{X}} \mr{Var}\left(\hat{\Pi}(x_i, x_j) \right) = O \left(n^{\frac{-4}{4 + d_{X}}} \log  (N)\right). 
    \end{align*}
    \end{assumption}

    Assumption~\ref{assumption:dyadic_nonparametric} imposes uniform convergence rates on the bias and variance of the first-stage dyadic nonparametric regression estimator. The convergence rates depend only on $d_X$ and the number of sampled individuals $n$ (up to a logarithmic factor).  Under Assumption~\ref{assumption:regularity}, these rates can be achieved by the Nadaraya-Watson estimator, as shown in \citet{graham2021minimax}. 

    \paragraph{Remark} The convergence rates in Assumption~\ref{assumption:dyadic_nonparametric} are the MSE-optimal rates for twice continuously differentiable target functions with $n$ i.i.d. observations. As noted by \citet{graham2021minimax}, dyadic nonparametric regression differs from standard nonparametric regression with i.i.d. data in two respects: (i) although the dyadic conditional expectation $\Pi(\cdot,\cdot)$ takes $2d_X$ arguments, the convergence rates behave as if $\Pi$ depended on only $d_X$ arguments; and (ii) when the entire network is observed, although the regression involves $N^2$ dyads, the effective sample size is just $N$ because the estimator is a U-statistic. The convergence rates in Assumption~\ref{assumption:dyadic_nonparametric} are consistent with the first feature, but differ from the second because the network is incomplete in our setting. One use  Hoeffding decomposition to show that the effective sample size in our setting is the number of sampled individuals $n$ rather than $N$.

    \begin{assumption}\label{assumption:kernel}
        Suppose that the kernel $K: \mb{R}\mapsto \mb{R}_{+}$ is bounded and supported on $[-1, 1]$.   In addition, $K(0)>0$ and  $K(\cdot)$ is Lipschitz continuous with constant $\bar{K}>0$. 
    \end{assumption}
    This assumption imposes regularity conditions on kernel $K$ used in the second step. The requirement is standard in the literature, and  many commonly used kernels, including the Epanechnikov kernel, satisfy Assumption~\ref{assumption:kernel}.

    \bigskip 
    
    We now briefly explain why $\hat{d}_{ii'}$ can be viewed as a noisy measure of $\|\zeta_i - \zeta_{i'}\|$ and can thus serve as its proxy in the kernel function. First, let $h$ denote the bandwidth of the kernel function. When imputing $\Pi^{\bot}_{ij}$, Assumption~\ref{assumption:kernel} allows us to restrict attention to links $(i', j')$ for which both $\hat{d}_{ii'}\leq h$ and $\hat{d}_{jj'}\leq h$. Second, by Lemma~\ref{lemma:consistency_hat_d}, we have $|\hat{d}_{ii'} - d(\zeta_i, \zeta_{i'})| \leq \delta_{N, n}$ with high probability. When $\delta_{N, n} / h \to 0$, the approximation error in $\hat{d}_{ii'}$ is asymptotically negligible relative to the bandwidth $h$, so $d(\zeta_i, \zeta_{i'})$ can be replaced by $\hat{d}_{ii'}$ without loss of first-order accuracy. Furthermore, under Assumption~\ref{assumption:informativeness}, $d(\zeta_i, \zeta_{i'})$ is informative about the latent distance in the sense that $\|\zeta_i - \zeta_{i'}\|\leq \lambda^{-1} d(\zeta_i, \zeta_{i'})$. Combining these results yields, with probability approaching one, $\|\zeta_i - \zeta_{i'}\|\leq \lambda^{-1} h$, which implies that using pseudo-distances in the kernel function selects pairs whose latent distances  lie within a $\lambda^{-1} h$-neighborhood when imputing missing links.

    The following theorem provides the theoretical guarantee for our imputation method.
    \begin{theorem}[Imputation errors]\label{thm:bias_variance}
        Under Assumption~\ref{assumption:regularity}, \ref{assumption:informativeness}, \ref{assumption:dyadic_nonparametric}, \ref{assumption:kernel},  if (i) $h\rightarrow 0$, (ii) $\delta_{N, n} / h \rightarrow 0$, and (iii) $nh^{d_{\zeta}}/\log N \rightarrow \infty$, then there exists constant $\gamma_3>0$ such that
        \begin{align}\label{eq:thm:bias_variance:pbound}
            \mb{P}\left(\max_{i, j \in \mc{S}^c} \left|\hat{A}_{ij} - P_{ij}\right| \leq \gamma_3 \left(h^2 + \frac{\log N}{\sqrt{nh^{d_{\zeta}}}} \right)\right) \geq 1-n^{-1/2}. 
        \end{align}
        Moreover, the imputation error admits the following bias--variance decomposition.  Specifically, with probability at least $1-n^{-1/2}$,
        \begin{equation}\label{eq:thm:bias_variance:bvbound}
            \begin{gathered}
            \max_{i, j \in \mc{S}^c}  \left| \mb{E}\left(\hat{A}_{ij} - P_{ij} \mid \{\zeta_i \}_{i=1}^{N} \right)\right|    =  O\left(h^2 + \frac{1}{nh^{d_{\zeta}}} \right), \\
            \max_{i, j \in \mc{S}^c} \mr{Var}\left(\hat{A}_{ij}\mid \{\zeta_i \}_{i=1}^{N} \right) =  O\left(  \frac{\log N}{nh^{d_{\zeta}}} + \delta^2_{N, n} h^2 \right). 
        \end{gathered}
        \end{equation} 
    \end{theorem}

    Theorem~\ref{thm:bias_variance} establishes link-level error bounds, rather than a Frobenius-norm bound for a submatrix (e.g.,  \cite{gao2015rate} and \cite{xu2018rates}). The convergence is uniform over all missing links, which implies that the imputation error is controlled at the link level and is particularly useful for applications that require accurate imputation of each link probability.  

    Following the classical nonparametric regression literature, we decompose the estimation error into bias and variance. The bias term is of order $\left(h^2 + 1/(n h^{d_{\zeta}})\right)$, where (i) $h^2$ arises because our local two-way fixed-effects regression removes the first-order bias, leaving only the second-order approximation error, and (ii) $1/(n h^{d_{\zeta}})$ reflects the contamination from diagonal entries of the matrix $A$ (which are $0$ instead of $f_{ii} + \epsilon_{ii}$ because we rule out self-loops). In most empirically relevant cases, the first term, $h^2$, is dominant. In particular, under the MSE-optimal bandwidth choice, it can be directly verified that $1/(n h^{d_{\zeta}}) = o(h^2) $. 
    
    The variance consists of two terms. The first term, $\log N /(n h^{d_{\zeta}})$, matches the rate in the standard nonparametric literature. The second term, $\delta^2_{N,n} h^2$, is specific to our estimator and arises from the use of the pseudo-distance as a substitute for the unobserved latent factors $\zeta_i$ and $\zeta_j$. When the latent factors are observable, we have $\delta_{N,n}=0$, and the second term disappears, reducing our result to classic conclusions in the nonparametric regression literature (see \citet{Tsybakov2008introduction} and, for dyadic regression, \citet{graham2021minimax}). 

    The bandwidth $h$ governs the bias-variance trade-off. However, because the bias and variance each consist of two components, the trade-off deviates from the classical setting. As $h$ decreases, $h^2$ and $h^2 \delta_{N, n}^2$ diminish, while $1/(n h^{d_{\zeta}})$ increases. Therefore, the optimal choice of $h$ depends jointly on the relative magnitudes of $n$ and $N$ and on $d_{\zeta}$. Nevertheless, it is straightforward to verify that, when $n$ and $N$ are of the same order, setting $h \asymp n^{-1/(4 + d_{\zeta})}$ attains the optimal MSE rate (up to logarithmic factors).

    \paragraph{Remark} 
    Our method employs two-way fixed-effects regression for imputation, which effectively removes the first-order approximation error. The intuition behind this approximation can be illustrated by a Taylor expansion. Heuristically, to impute $A_{ij}$, if we ignore the first-step nonparametric dyadic regression and consider the Taylor expansion of $f(\zeta_{i'}, \zeta_{j'})$ around the fixed point $(\zeta_i, \zeta_j)$: 
    \begin{align*}
        f(\zeta_{i'}, \zeta_{j'}) = &  \underbrace{f(\zeta_{i}, \zeta_{j}) + \left(\nabla_{\zeta_i} f(\zeta_{i}, \zeta_{j}) \right)'(\zeta_{i'} - \zeta_{i})}_{a_{i'}} + \underbrace{ \left(\nabla_{\zeta_j} f(\zeta_{i}, \zeta_{j}) \right)'(\zeta_{j'} - \zeta_{j})}_{b_{j'}} + O(\|\zeta_{i'} - \zeta_{i}\|^2 + \|\zeta_{j'} - \zeta_{j}\|^2 ) \\
        = &   a_{i'} + b_{j'} + O(\|\zeta_{i'} - \zeta_{i}\|^2 + \|\zeta_{j'} - \zeta_{j}\|^2 ). 
    \end{align*}
    Because we employ  a kernel-weighted estimator based on the pseudo-distances $\hat{d}_{ii'}, \hat{d}_{jj'}$ with bandwidth $h$, and since, as discussed above, $\hat{d}_{ii'}$ and $\hat{d}_{jj'}$ differ from the latent-space distances $\|\zeta_{i'} - \zeta_{i}\|$ and $\|\zeta_{j'} - \zeta_{j}\|$ only up to a bounded factor, the contribution to imputation comes from links $(i', j')$ with $\hat{d}_{ii'}, \hat{d}_{jj'}\leq h$. For those links, we obtain
    \begin{align*}
        f(\zeta_{i'}, \zeta_{j'}) = a_{i'} + b_{j'} +  O(h^2) . 
    \end{align*}
    Hence, the two-way fixed-effects regression absorbs  first-order  approximation errors, leaving only the second-order remainder $O(h^2)$.  
    
    The idea of using two-way fixed-effects imputation is closely related to  the approaches developed in \cite{freeman2023linear}  and \cite{beyhum2024inference}. Our approach differs from theirs in two respects. First, their methods do not involve missing data and are primarily designed to estimate structural parameters in panel data models, whereas our method aims to predict the formation probability of missing links. Second, they discretize the data into groups using  k-means clustering, while we employ a kernel estimator. 

    The methods most conceptually related to ours are those of \citet{feng2023optimal} and \citet{deaner2025inferring}. \citet{deaner2025inferring} propose a local-constant estimator to impute potential outcomes. Because they assume only that the potential outcome function is Lipschitz continuous, their approximation error is of order $O(h)$. Building on the ideas of \citet{freeman2023linear}  and \citet{beyhum2024inference}, we assume that the graphon is twice continuously differentiable and, by using a two-way fixed-effects imputation, we reduce the approximation error to $O(h^2)$. In comparison, the local PCA method proposed by \citet{feng2023optimal}  leverages factor models together with higher-order Taylor expansions to achieve more accurate approximations. In particular, \citet{feng2023optimal} show that the local PCA estimator attains Stone's minimax-optimal rate for nonparametric regression. We refer interested readers to that paper for further details. 

    We adopt the two-way fixed-effects model instead of  local PCA approach for two reasons.
    First,  local PCA method requires prior knowledge of the dimension $d_{\zeta}$, which is typically unknown in practice. In contrast, our approach does not require prior knowledge of $d_{\zeta}$. 
    Second, networks in most economic applications are relatively small.  For instance, in development economics, village networks typically contain around $200$ households, and with a sampling rate of roughly $40\%$, the effective sample size available for imputing a specific missing link can be fewer than $20$. Under such small-sample conditions, achieving theoretical higher-order accuracy is difficult. Our numerical simulations also show that the finite-sample performance of our method is better to that of the local PCA.

    \section{Theory for Downstream Estimation}\label{sec:empirics}

    We consider the application of our method to estimating parameters in downstream economic models. The data $\{(A_m, X_m, W_m, Y_m)\}_{m=1}^{M}$ consist of $M$ large i.i.d. networks indexed by $m = 1,2,\ldots, M$, and each network has size $N_m$.  Here, $A_m\in \{0, 1\}^{N_m\times N_m}$ is the adjacency matrix,  $Y_m \in \mb{R}^{N_m}$ is the outcome variable, $X_m \in \mb{R}^{N_m\times d_{X}}$ denotes the covariates that directly enter the network formation model as in~\eqref{eq:graphon}, whereas $W_m \in \mb{R}^{N_m\times d_{W}}$ contains variables that do not necessarily appear in the network formation model but affect the outcome $Y_m$. Note that $W_m$ and $X_m$ may overlap, i.e., $W_m\bigcap X_m \neq\emptyset$.

    The adjacency matrix $A_m$ is not fully observed because of egocentric sampling, but the outcome $Y_m$ and the covariates $(X_m, W_m)$ are observed for all individuals in the network.  Let $n_m$ be the number of sampled  individuals in network $m$. We mainly focus on the asymptotics in which $M, N_m, n_m\rightarrow \infty$.  
    The large-network asymptotics $N_m, n_m \rightarrow \infty$ are required for consistent imputation of the adjacency matrix. The large-$M$ asymptotics are introduced to accommodate within-network dependence of error terms in the downstream regression. If one is willing to impose independence of these errors within a network (see Examples~\ref{example:regression} and \ref{example:linear_in_mean_peer_effect} in the following text), then the consistency analysis can also be extended to the single-large-network setting.

    We allow for heterogeneous network formation process $f_m$ for each $m = 1, \ldots, M$, so that information from other networks does not help in imputing a given network. Therefore, we implement imputation method separately for each incomplete network, obtain the imputed adjacency matrix $\hat{A}_m$, and use it, rather than the observed network $A^{\mr{obs}}$, in the moment conditions to contruct the GMM estimator.

\subsection{Consistency}
    
    \paragraph{GMM} We work with the moment function $$\psi\left(A_m, W_m, Y_m, \alpha \right): [0, 1]^{N_m\times N_m}   \times \mb{R}^{N_m\times d_{W}} \times  \mb{R}^{d_{\alpha}} \mapsto \mb{R}^{q}, $$ where $q$ is the number of moment conditions and $\alpha\in \mb{R}^{d_{\alpha}}$ is the structural parameter of interest in the downstream analysis. We assume the following moment condition 
    \begin{align}\label{eq:moment_condition}
        \mb{E}\left(\psi\left(A_m, W_m, Y_m, \alpha_0 \right)\right) = 0, 
    \end{align}
    where $\alpha_0$ is the true parameter. Since  $A_m$ is partially observed due to sampling, we impute each sampled network separately to obtain $\{\hat{A}_m\}_{ m=1, \ldots, M}$, and then use the imputed  networks to solve 
    \begin{align}\label{eq:GMM}
        \hat{\alpha} = \arg\min_{\alpha \in \mr{supp}(\alpha) } \left(\frac{1}{M}\sum_{m=1}^{M} \psi \left(\hat{A}_m, W_m, Y_m, \alpha \right)\right)'\Sigma_M \left(\frac{1}{M}\sum_{m=1}^{M} \psi \left(\hat{A}_m, W_m, Y_m, \alpha \right)\right), 
    \end{align}
    where $\Sigma_{M} \in \mb{R}^{q\times q}$ is the weight matrix. 

    \begin{example}[Regression on centrality]\label{example:regression} 
        A large literature  studies the relationship between individuals' network positions (centrality) and their outcomes, for example, \citet{hochberg2007whom}, \citet{cruz2017politician},  and \citet{banerjee2013diffusion}.  The regressions in such studies is 
        \begin{align}\label{eq:regression}
            Y_{mi} = \alpha_C + \alpha_1 \phi_{mi}(A_m) + e_{mi}, \quad i=1, 2, \ldots, N_m, \text{ } m = 1,2, \ldots, M ,  
        \end{align} 
        where $\alpha = (\alpha_C, \alpha_1)'$ is the parameter of interest, $\phi_{mi} : [0,1]^{N_m \times N_m} \to \mathbb{R}$ computes the centrality of individuals $i$. For example:  
        \begin{enumerate}[label=(\roman*)]
            \item \textbf{(Degree centrality)} For matrix $\tilde{A}_m \in [0,1]^{N_m \times N_m}$ and $i=1, \ldots, N_m$, the degree centrality is defined as $\phi_{mi}(\tilde{A}_m) := \frac{1}{N_m} \sum_{j=1}^{N_m} \tilde{A}_{m, ij}$. It is straightforward to verify that $\phi_{mi}(\cdot)$ represents the normalized avreage degree of individual $i$ when $\tilde{A}$ is an adjacency matrix. 
            \item \textbf{(Eigenvector centrality)}  For any symmetric and non-negative matrix $\tilde{A}_m \in [0,1]^{N_m \times N_m}$, let $\tilde{A}_m = U_m D_m U_m'$ be the eigenvalue decomposition of $\tilde{A}$, where $U_m \in \mb{R}^{N_m\times N_m}$ is an orthonormal matrix containing the eigenvectors of $\tilde{A}_m$ and $D_m \in \mb{R}^{N_m\times N_m}$ is a diagonal matrix with eigenvalues arranged in decreasing order.  Define $\phi_{mi}(\tilde{A}_m) = \sqrt{N_m} U_{m, i1}$, where $U_{m,i1}$ denotes the $i$-th entry of the leading eigenvector\footnote{
                Because eigenvectors are defined only up to sign, we normalize the leading eigenvector so that all its entries are non-negative. This normalization is justified by the Perron-Frobenius theorem. 
            }. It is straightforward to verify that, when $\tilde{A}_m$ is a symmetric adjacency matrix, $\phi_{mi}(\cdot)$ corresponds to the eigenvector centrality of individual $i$.
        \end{enumerate}
        Under the exogeneity of $\{e_{mi}\}_{m=1, \ldots, M,  i = 1, \ldots, N_m}$, the moment function can be constructed as
        \begin{align*}
            \psi\left(A_m, W_m, Y_m, \alpha \right):= \frac{1}{N_m}\sum_{i=1}^{N_m} (1, \phi_{mi}(A_m))' (Y_{mi}  - \alpha_C - \alpha_1 \phi_{mi}(A_m)). 
        \end{align*}
        Since the network $A_m$ is incomplete, we replace $\phi_{mi}(A_m)$ with $\phi_{mi}(\hat{A}_m)$ to obtain $\psi\left(\hat{A}_m, W_m, Y_m, \alpha \right)$. We then use~\eqref{eq:GMM} (with $\Sigma_M$ equal to the identity matrix) to obtain the OLS estimator $\hat{\alpha}$.  
        In addition, since the error terms $\{e_{mi}\}_{m=1, \ldots, M,  i = 1, \ldots, N_m}$ may be correlated within networks, inference should be clustered at the network level. 
    \end{example}

    \begin{example}[Linear-in-means peer-effects model]\label{example:linear_in_mean_peer_effect}
        The linear-in-means peer-effects model is widely used to study peer influence in social networks. The model is 
        \begin{align}\label{eq:linear_in_mean_peer_effect}
            Y_{m} = \alpha_C +  \alpha_{\bar{Y}} G_m Y_m +   W_m \alpha_W  +  G_m W_m \alpha_{\bar{W}}  + e_m,  
        \end{align}
        where $G_m$ is the row-normalized adjacency matrix \footnote{
            For any $i, j \in 1, \ldots, N_m, N_m$, $G_{m, ij}$ is obtained through 
            \begin{align*}
                G_{m, ij} = \frac{A_{m, ij}}{\sum_{j' = 1}^{N_m}A_{m, ij' }}. 
            \end{align*}
        }, and $\alpha_{\bar{Y}}$ captures the endogenous effect. Since $G_m Y_m$ is endogenous due to the reflection problem (\citet{manski1993identification}), we estimate the model using instrumental variables. Let $\alpha = (\alpha_C, \alpha_{\bar{Y}}, \alpha_W', \alpha_{\bar{W}}')'$ be the collection of parameters. Under the identification conditions in \citet{bramoulle2009identification}, the following moment function can be constructed as 
        \begin{align*}
            \psi(A_m, W_m, Y_m, \alpha) = \frac{1}{N_m} Z_m' (Y_{m} - \alpha_C  -  \alpha_{\bar{Y}} G_m Y_m -    W_m \alpha_W -  G_m W_m\alpha_{\bar{W}}), 
        \end{align*}
        where $Z_m = [1, W_m, G_m W_m, G^2_m W_m]$ is the instrument.  When the network $A_m$ is incomplete, we compute  $\hat{G}_m$ and $\hat{Z}_m$ using imputed network $\hat{A} $ \footnote{
            For any $i, j \in 1, \ldots, N_m$, $\hat{G}_{m, ij}$ is obtained through
            \begin{align*}
                \hat{G}_{m, ij} = \frac{\hat{A}_{m, ij}}{\sum_{j' = 1}^{N_m}\hat{A}_{m, ij' }}, 
            \end{align*}
            and $\hat{Z}_m:= [1, W_m, \hat{G}_mW_m, \hat{G}^2_mW_m ]$
        }, the corresponding moment function becomes  
        \begin{align*}
            \psi(\hat{A}_m, W_m, Y_m, \alpha) = \frac{1}{N_m}\hat{Z}_m' (Y_{m} - \alpha_C  -  \alpha_{\bar{Y}} \hat{G}_m Y_m -  \alpha_W  W_m  - \alpha_{\bar{W}} \hat{G}_m W_m). 
        \end{align*}
        We then use~\eqref{eq:GMM} to obtain the IV estimator $\hat{\alpha}$.   Since the errors $\{e_{mi}\}_{m=1,\ldots,M, i=1,\ldots,N_m}$ may be correlated within networks, inference should be clustered at the network level. 
    \end{example}

    \paragraph{Remark} Our approach for GMM estimation differs from that of \citet{chandrasekhar2011econometrics}. Specifically, 
    \citet{chandrasekhar2011econometrics} construct moments by integrating over the missing network data,  using simulation to approximate (conditional)  expectations.
    As a consequence, their implementation is computationally demanding in GMM settings and relies on parametric assumptions about the distribution of the error terms (see the discussion in \citet{chandrasekhar2011econometrics} and \citet{boucher2020estimating}). By contrast, our  approach  directly plugs the imputed network into the moment function $\psi(\cdot)$, avoiding simulation and distributional assumptions on the errors. A  key advantage of \citet{chandrasekhar2011econometrics}'s approach is that it can accommodate complex network statistics, such as average path length, that cannot be computed using a simple plug-in strategy.  However, even in such cases, our imputed adjacency matrix can still serve as the basis for the simulation-based GMM procedure, and their analytical framework (developed in Section~4.2 of \citet{chandrasekhar2011econometrics}) can then be used to establish consistency of the downstream estimator.
    
    \bigskip

    We now introduce a set of regularity conditions on the sampling process and the moment function $\psi(\cdot)$ used to establish consistency of the downstream estimator.

    \begin{assumption}[GMM]\label{assumption:consistency_empirics}
        Suppose that 
        \begin{enumerate}[label=(\roman*)]
             \item \label{item:consistency_empirics_assumption_asymptotics}\textbf{(Asymptotics)} Large-$M$ asymptotics with $M\rightarrow \infty$. 
            \item \label{item:consistency_empirics_assumption_independence}\textbf{(Independence)} $\{(A_m, X_m, W_m, Y_m)\}_{m=1}^{M}$ are i.i.d.. 
            \item  \label{item:consistency_empirics_assumption_identification} \textbf{(Identification)} There is a unique $\alpha_0$ such that $\mb{E}\left(\psi\left(A_m, W_m, Y_m, \alpha_0  \right)\right) = 0$. 
            \item \label{item:consistency_empirics_assumption_W_n} $\Sigma_M\cp  \Sigma$ and $\Sigma$ is positive definite. 
            \item \label{item:consistency_empirics_assumption_compactness_and_integrability} \textbf{(Compactness and     integrability)} The support of $\alpha$ is compact, $\alpha \mapsto \psi(A_m, W_m, Y_m, \alpha)$ is continuous almost everywhere, and $\mb{E}\left(\sup_{\alpha \in \mr{supp}(\alpha)}\|\psi(A_m, W_m, Y_m, \alpha)\|^2\right)<\infty$. 
            \item \label{item:consistency_empirics_assumption_ULLN} \textbf{(Lipschitz)} Let $\|\cdot\|$ denote the Frobenius norm of a matrix. For any $\tilde{P}_{m} \in [0, 1]^{N_m\times N_m}$,  
            \begin{align*}
                \sup_{\alpha \in \mr{supp}(\alpha) } \left\| \psi\left(\tilde{P}_{m}, W_m, Y_m, \alpha \right) - \psi\left(P_{m}, W_m, Y_m, \alpha \right) \right\| \leq L( W_m, Y_m) N_m^{-1} \|\tilde{P}_{m}- P_{m}\|_{\mr{F}}
            \end{align*}
            and $\mb{E}L^2(W_m, Y_m) <\infty$.
            \item  \label{item:consistency_empirics_assumption_graphon} \textbf{(Idiosyncrasy-robustness)} Let $P_m$ denote the matrix of conditional link-formation probabilities. Then, as $N \to \infty$,   
            \begin{align*}
                \sup_{1\leq m\leq M} \sup_{\alpha \in \mr{supp}(\alpha) } \left\|   \psi(P_m, W_m, Y_m, \alpha ) -   \psi(A_m, W_m, Y_m, \alpha ) \right\|= o_P(1). 
            \end{align*}   
        \end{enumerate}
    \end{assumption}

    Assumptions~\ref{assumption:consistency_empirics}\ref{item:consistency_empirics_assumption_asymptotics}--\ref{item:consistency_empirics_assumption_ULLN} are standard conditions to ensure the consistency of M-estimator. Assumption~\ref{item:consistency_empirics_assumption_asymptotics} considers the asymptotic regime in which the number of networks $M$ grows large. Assumption~\ref{item:consistency_empirics_assumption_independence} requires each network to be independent and identically distributed across $m$. Assumption~\ref{assumption:consistency_empirics}\ref{item:consistency_empirics_assumption_identification} is an identification condition and ensures that the moment condition does not hold at $\alpha\neq\alpha_0$.  Assumption~\ref{item:consistency_empirics_assumption_W_n} requires that the weighting matrix $\Sigma_M$ converges to a positive definite matrix as $M\rightarrow\infty$.   Assumption~\ref{assumption:consistency_empirics}\ref{item:consistency_empirics_assumption_compactness_and_integrability} requires the parameter space to be compact, the moment function to vary continuously with $\alpha$, and the second moments of $\psi$ to remain uniformly bounded over the parameter space. 
    Assumption~\ref{assumption:consistency_empirics}\ref{item:consistency_empirics_assumption_ULLN} imposes a Lipschitz condition requiring that  perturbations in $P_m$ lead to changes in $\psi$ up to a factor $L(W_m, Y_m)$ with a finite second moment, uniformly over $\alpha$.
    Since our parameter of interest is $\alpha$ and $P_m$ serves as a nuisance parameter,  this assumption can be interpreted as a smoothness condition on the nuisance parameters within the framework of semi-parametric M-estimation.  

    Assumption~\ref{assumption:consistency_empirics}\ref{item:consistency_empirics_assumption_graphon} requires that replacing the adjacency matrix $A$ with the probability matrix $P$ in the moment function becomes asymptotically negligible as the network size $N$ grows. Unlike the previous assumptions which are standard in the M-estimation and GMM literature, Assumption~\ref{assumption:consistency_empirics}\ref{item:consistency_empirics_assumption_graphon} is specific to our setting and is the key condition for establishing consistency of the plug-in estimator. This is because our imputation procedure can only recover the conditional link-formation probabilities, rather than the realized missing links themselves. Intuitively, the assumption requires that the idiosyncratic link-formation shocks average out asymptotically in the moment conditions. This condition is satisfied in many empirical applications based on network statistics, such as linear-in-means peer-effects models and regressions on network centralities.

    \setcounter{example}{2}

    \begin{example}[Continued]
        Since verifying Assumptions~\ref{assumption:consistency_empirics}\ref{item:consistency_empirics_assumption_asymptotics}--\ref{item:consistency_empirics_assumption_compactness_and_integrability} under linear models is standard, we focus on Assumptions~\ref{assumption:consistency_empirics}\ref{item:consistency_empirics_assumption_graphon} and \ref{item:consistency_empirics_assumption_ULLN} for linear regressions on degree centrality and eigenvector centrality. 

        For degree centrality, since it is simply the row average of the matrix, Assumption~\ref{assumption:consistency_empirics}\ref{item:consistency_empirics_assumption_ULLN} follows directly from the Cauchy-Schwarz inequality. In addition, by~\eqref{eq:graphon}, 
        we have $\psi_{m, i}(A_m) - \psi_{m, i}(P_m) = \frac{1}{N_m}\sum_{j=1}^{N_m} \epsilon_{m, ij}$. Since $\{\epsilon_{m, i}\}_{1\leq j\leq N_m}$ are uniformly bounded and independent (Assumption~\ref{assumption:regularity}\ref{item:regularity_assumption_independence}), we obtain $\psi_{m, i}(A_m) - \psi_{m, i}(P_m) = O_p(1/\sqrt{N_m})$. Therefore, Assumption~\ref{assumption:consistency_empirics}\ref{item:consistency_empirics_assumption_graphon} also follows.
        
        For eigenvector centrality, we impose a standard eigen-gap condition requiring that the largest eigenvalue of $P_m$ is separated from the second-largest eigenvalue. Such conditions are common in spectral analysis and ensure that the eigenvector-centrality is well-defined. Following Example~\ref{example:regression}, let $\phi(P_m), \phi(\tilde{P}_m), \phi(\tilde{A}_m) \in \mb{R}^{N_m}$ denote the vectors of eigenvector centralities of  $P_m$, $\tilde{P}_m$ and $A_m$, respectively. Let $\|\cdot\|_{\mr{op}}$ and $\|\cdot\|_{\mr{F}}$ denote the operator norm and the Frobenius norm of a matrix, respectively. Then, the Davis-Kahan theorem \citep{yu2015useful} implies that
        \begin{align*}
            \|\phi(\tilde{P}_m) - \phi(P_m)\| = O(\|\tilde{P}_m - P_m\|_{\mr{op}}) = O(\|\tilde{P}_m - P_m\|_{\mr{F}}). 
        \end{align*}
        Assumption~\ref{assumption:consistency_empirics}\ref{item:consistency_empirics_assumption_ULLN} follows immediately.

        Using a similar argument, $\|\phi(A_m) - \phi(P_m)\| = O(\|E_m \|_{\mr{op}})$ where $E_m:=A_m-P_m$ containing the errors in network formation models. By \citet{bandeira2016sharp}, $\|E_m\|_{\mr{op}}/N_m =O_p(1/\sqrt{N_m})$, which establishes Assumption~\ref{assumption:consistency_empirics}\ref{item:consistency_empirics_assumption_graphon}.
    \end{example}

    \begin{example}[Continued]
        For the linear-in-means peer-effects model, Assumptions~\ref{assumption:consistency_empirics}\ref{item:consistency_empirics_assumption_asymptotics}--\ref{item:consistency_empirics_assumption_W_n} follow directly under the identification conditions of \citet{bramoulle2009identification}. In addition, if $W$ is uniformly bounded, then Assumptions~\ref{assumption:consistency_empirics}\ref{item:consistency_empirics_assumption_compactness_and_integrability}--\ref{item:consistency_empirics_assumption_ULLN}  can be verified directly. 
    \end{example}

    \bigskip

    \begin{theorem}[Consistency of GMM estimator]\label{thm:consistency_plug_in} 
        Under conditions in Theorem~\ref{thm:bias_variance} and Assumption~\ref{assumption:consistency_empirics}, the  GMM estimator $\hat{\alpha}$ defined as in \eqref{eq:GMM} is consistent for $\alpha_0$.  
    \end{theorem}

    Theorem~\ref{thm:consistency_plug_in} shows that, in the downstream regression, the plug-in estimator $\hat{\alpha}$ based on our imputed networks is consistent. It is particularly relevant in empirical applications, where interest typically focuses on the consistency of parameter estimates constructed from the imputed network rather than on the accuracy of the imputed links themselves. The key advantage of our approach is that the consistency of $\hat{\alpha}$ does not depend on (i) specification of the network formation model and (ii) the dimension of latent heterogeneity. In addition, even when a researcher prefers to impose a specific parametric network formation model for computational simplicity or because of strong prior knowledge, our method still provides a benchmark for assessing the extent to which downstream estimators depend on parametric assumptions about the network formation process. 

    \paragraph{Remark} The analysis above focuses on large-$N$, large-$M$ asymptotics. However, under certain conditions, the downstream regression analysis can also be extended to the single-large-network setting (i.e., $M=1$). For example, in Examples~\ref{example:regression} and \ref{example:linear_in_mean_peer_effect}, consistency of the downstream GMM estimator continues to hold when the error terms $\{e_{i}\}_{i=1,\ldots,N}$ are uncorrelated across individuals.

    \subsection{Analysis of linear-in-means peer-effect model} 
    
    Although Theorem~\ref{thm:consistency_plug_in} establishes the consistency of the plug-in estimator based on our imputed adjacency matrices, analyzing its convergence rate and inference is challenging. First, as discussed in \citet{cai2022linear}, even when each link contains only i.i.d. classical measurement error, network statistics used in moment conditions typically aggregate errors across links, thereby inducing non-classical measurement error that complicates the analysis. Second, the measurement error generated by our imputation procedure is correlated across links, and because the imputation is nonparametric, its bias and variance may be of the same order, further complicating the analysis of convergence. 
    
    Since network links typically enter the moment conditions through network statistics, the asymptotic properties of imputed links established in Theorem~\ref{thm:bias_variance} could be different from those of the GMM estimators that rely on the imputed network. Therefore, the bandwidth optimal for imputing links may not be optimal for downstream analysis. 
    In the following text, we focus on the linear-in-means peer-effects regression and study the asymptotic behavior of the corresponding GMM estimator.  We also provide guidance for downstream regression analysis using our imputed networks.

    Consider the linear-in-means peer-effects model~\eqref{eq:linear_in_mean_peer_effect} in Example~\ref{example:linear_in_mean_peer_effect}.  The weight matrix is $\Sigma_M$. Let $\hat{V}_m := [\bs{1}, \hat{G}_m Y_m, W_m, \hat{G}_m W_m]$, $\hat{Z}_m := [\bs{1},  W_m, \hat{G}_m W_m, \hat{G}^2_m W_m]$,  and $\alpha := (\alpha_C, \alpha_{\bar{Y}}, \alpha'_{W}, \alpha'_{\bar{W}})'$. The GMM estimator is given by 
    \begin{align}\label{eq:GMM_estimator}
        \hat{\alpha} = \left(\sum_{m=1}^{M} (\hat{V}_m' \hat{Z}_m/N_m) \Sigma_M (\hat{Z}_m'\hat{V}_m/ N_m) \right)^{-1} \left(\sum_{m=1}^{M} (\hat{V}_m' \hat{Z}_m/N_m) \Sigma_M (\hat{Z}_m'Y_m / N_m) \right). 
    \end{align}
    \begin{assumption}\label{assumption:linear_in_mean_peer_effect} Suppose that 
        \begin{enumerate}[label=(\roman*)]
            \item \label{item:linear_in_mean_peer_effect_sampling} \textbf{(Sampling)} $\{(A_m, X_m, W_m, Y_m)\}_{m=1}^{M}$ are i.i.d.,  and the data is generated according to Example~\ref{example:linear_in_mean_peer_effect}. 
            \item \label{item:linear_in_mean_peer_effect_exogeneity}\textbf{(Exogeneity)} The error vector $e_m$ satisfies $\mb{E}\left(e_m \mid \{X_{m, i}, \xi_{m, i}\}_{i=1}^{N_m}, W_m\right) = 0$. In addition, we assume that $e_m\bot \epsilon_m \mid \left(\{X_{m, i}, \xi_{m, i}\}_{i=1}^{N_m}, W_m\right)$, where $\epsilon_m:= \left\{\epsilon_{m,ij}\mid 1\leq i <j\leq N_m \right\}$ are error terms in the network generating process~\eqref{eq:graphon}.  
            \item \label{item:linear_in_mean_peer_effect_asymptotics} \textbf{(Asymptotics)} The network sizes $\{N_m\mid m=1, \ldots, N\}$ are of the same order, and sample sizes $\{n_m\mid m=1, \ldots, M\}$ are of the same order as well, i.e.,  
            \begin{align*}
                \sup_{M\rightarrow \infty} \inf_{1\leq m_1, m_2\leq M} (N_{m_1}/ N_{m_2}) >0, \quad \sup_{M\rightarrow \infty} \inf_{1\leq m_1, m_2\leq M} (n_{m_1}/ n_{m_2}) >0.
            \end{align*} 
            \item \label{item:linear_in_mean_peer_effect_compact} \textbf{(Compactness)} The support of $W$ is uniformly bounded over $i$ and $m$.
            \item \label{item:linear_in_mean_peer_effect_Sigma} $\Sigma_M\cp  \Sigma$ and $\Sigma$ is positive definite. 
            \item \label{item:linear_in_mean_peer_effect_non_degenerate} \textbf{(Non-degeneracy)} Let $\lambda_{\min}(\cdot)$ denote the smallest singular value of a matrix.  Then,  
            \begin{align*}
                \lambda_{\min}\left(\mb{E}\left(\frac{1}{N_m} Z_m'V_m \right)' \mb{E}\left(\frac{1}{N_m} Z_m'V_m \right)\right) >0. 
            \end{align*} 
        \end{enumerate}
    \end{assumption}

    Assumption~\ref{assumption:linear_in_mean_peer_effect}\ref{item:linear_in_mean_peer_effect_sampling} requires the networks to be independent and identically distributed across $m$. However, we allow the error terms $e_{mi}$ to be correlated across individuals within the same network. Assumption~\ref{assumption:linear_in_mean_peer_effect}\ref{item:linear_in_mean_peer_effect_exogeneity} imposes the strict exogeneity condition and rules out correlated effects (\citealp{bramoulle2020peer}). Note that Assumption~\ref{assumption:linear_in_mean_peer_effect}\ref{item:linear_in_mean_peer_effect_exogeneity} is stronger than the standard assumption in the literature, i.e.,  $\mb{E}\left(e_m \mid G_m, W_m\right) = 0$. This is because the network data are incomplete in our setting, requiring the error term in the downstream regression to be orthogonal to the information used for network imputation. We further assume that the error terms in the linear-in-means peer-effects model, $e_m$,  are conditionally independent of the errors in the network formation process, $\epsilon_m$, which ensures that the imputation error is independent of $Y_m$.

    Assumption~\ref{assumption:linear_in_mean_peer_effect}\ref{item:linear_in_mean_peer_effect_asymptotics} requires that all networks have sizes of the same order and that the sample sizes within networks are also of the same order. In other words, no network or sample becomes disproportionately large or disproportionately small.
    
    Assumption~\ref{assumption:linear_in_mean_peer_effect}\ref{item:linear_in_mean_peer_effect_non_degenerate} 
    imposes a non-degeneracy condition requiring that the smallest singular value of $\mb{E}\left(\frac{1}{N_m} Z_m'V_m \right)' \mb{E}\left(\frac{1}{N_m} Z_m'V_m \right)$ is bounded away from zero. This requirement is analogous to the rank condition in instrumental-variable estimation, and is particularly important under large-network asymptotics.  For example, if $W$ is i.i.d. within each network and independent of the network (e.g., random treatment assignment), the instrument $Z_m$ becomes asymptotically collinear, causing the smallest singular value of the above matrix to shrink toward zero and leading to an inconsistent IV estimator. In practice, one can let  $W_m$ to be correlated with $X_m$ to satisfy this assumption (see \cite{hayes2024peer} for further discussion).

    \begin{theorem}\label{thm:linear_in_mean}
        Under Assumption~\ref{assumption:linear_in_mean_peer_effect} and conditions in Theorem~\ref{thm:bias_variance}, let $\bar{N}: = \frac{1}{M}\sum_{m=1, \ldots, M} N_m$ and $\bar{n}: = \frac{1}{M}\sum_{m=1, \ldots, M} n_m$. Then, the estimator $\hat{\alpha}$ defined as in~\eqref{eq:GMM_estimator}, follows that 
        \begin{align*}
            \sqrt{M}\left(\hat{\alpha} - \alpha_0 - B\right) \cd \mc{N}\left(0, \Omega + V \right), 
        \end{align*}
        where $B = O\left(h^2 + \frac{\log \bar{N} }{\bar{n}^2h^{2d_{\zeta}}}\right)$, $ V  = O\left( \delta_{\bar{N}\bar{n}}^2 h^2 +   \frac{\log \bar{N}}{\bar{n}^2h^{2d_{\zeta}}}\right)$. 
    \end{theorem}
    The theorem states that  $\hat{\alpha}-\alpha_0$  is asymptotically normally distributed, but its limiting distribution is centered away from zero by an $O\left(h^2 + \frac{\log \bar{N} }{\bar{n}^2h^{2d_{\zeta}}}\right)$ bias term. The first component $h^2$ arises from the bias of the imputed links and is typically the dominant term. By contrast, the second component is asymptotically negligible relative to $h^2$ in most empirically relevant settings\footnote{
        To see this, note that the bandwidth minimizing the mean squared error of the imputed links is of order $h^*\sim n^{\frac{-1}{4 + d_{\zeta}}}$. For the second bias component to dominate the first, that is, $h^2 = o\left(\frac{1}{n^2h^{2d_{\zeta}}}\right)$, we would require $h = o\left(n^{-\frac{1}{1 + d_{\zeta}}}\right)$.  This bandwidth shrinks much faster than the MSE-optimal bandwidth $h^*$ and therefore requires substantial undersmoothing, which is uncommon in practice.
    }. 
    The asymptotic variance consists of two components: $\Omega/M$ corresponds to the asymptotic variance that would arise if the network $A$ were perfectly observed, while $V/M$ captures the additional variance introduced by nonparametric imputation of the adjacency matrix. While the bias does not necessarily shrink as $M \to \infty$, the variance component induced by network imputation decreases at rate $1/M$, this highlights an important difference between the asymptotic behavior of the downstream estimator and that of the link-level imputation error.

    This result has important implications for downstream inference. Since the bias term $B$ dominates $V^{1/2}/\sqrt{M}$ in most empirically relevant settings, valid inference for $\alpha_0$ requires choosing the bandwidth in the imputation step so that $B$ is asymptotically negligible relative to the sampling variation, $\Omega^{1/2}/\sqrt{M}$. Consequently, bandwidth selection for valid downstream inference should prioritize reducing the bias term rather than balancing the bias and variance of the link-level imputation error, which requires undersmoothing in the first-stage imputation step.

    In practice, we recommend that applied researchers start from the optimal bandwidth $h$ selected by cross-validation in the imputation step, then gradually decrease $h$ and examine the corresponding estimates $\hat{\alpha}$.  Once the estimates become unstable as $h$ decreases, we should stop undersmoothing at that point.

    \section{Simulation Study}\label{sec:simulation}

    In this section, we examine the finite-sample properties of the proposed method using a series of numerical experiments. We compare its performance with several alternative imputation approaches, including the low-rank imputation methods proposed by \citet{bai2021matrix} and \citet{li2023link}, and the local PCA method developed by \citet{feng2023optimal}. We also use numerical experiments to illustrate (i) the importance of incorporating observed covariates for improving prediction accuracy, and (ii) the influence of sample splitting on finite-sample performance. We begin with introducing alternative method.

    \subsection{Alternative Methods}

    \paragraph{Covariate-only method} This method is a dyadic nonparametric regression-based imputation approach that relies solely on the observed covariates $X$. When unobserved heterogeneity is present, the prediction error of this approach does not diminish as the sample size increases. We compare its numerical performance with that of our proposed method to highlight the importance of accounting for unobserved heterogeneity. We refer to this method as \textbf{X} in the following text. 

    \paragraph{Low-rank imputation} When the observed matrix follows a strong-factor structure with an additive noise component and the missing entries exhibit a block-missing pattern\footnote{
        \citet{bai2021matrix} refer to this pattern as a tall-wide structure in the context of panel data, while \citet{li2023link} study an analogous setting in network data and refer to it as egocentric sampling.
    } (see \eqref{eq:block_structure}), \citet{bai2021matrix} and \citet{li2023link} propose using low-rank estimation to impute the missing data. The key idea is to first recover the latent factors using the observed submatrices $A_{\mc{S}\mc{S}}$, $A_{\mc{S}\mc{S}^c}$, and $A_{\mc{S}^c\mc{S}}$, and then reconstruct the missing bottom-right submatrix $A_{\mc{S}^c\mc{S}^c}$ through the product of the estimated factors. 

    When the graphon $f$ is nonlinear as in Example~\ref{example:graham}, the graphon matrix is typically high-rank and thus violates the low-rank assumption required by \citet{bai2021matrix} and \citet{li2023link}. Although their methods are not theoretically valid in such settings, we include them in our numerical experiments for comparison, given their simplicity and computational efficiency, to assess how our proposed method performs in practice relative to these benchmarks. It is worth noting that both \citet{bai2021matrix} and \citet{li2023link} assume that the true number of factors is known a priori. In practice, this information is rarely available.  Hence, in our numerical experiments, we select the number of factors via cross-validation.

    \paragraph{Local PCA} The local PCA method proposed by \citet{feng2023optimal} is suitable for high-rank graphon. Specifically, it first finds the neighbors of each node $i$ based on the pseudo-distance using $k$-nearest neighbors (kNN), and then performs low-rank imputation on the submatrix formed by node $i$'s neighbors to impute the missing entries. As discussed earlier, when the underlying graphon is smoother than twice continuously differentiable, the local PCA method can, in theory, achieve a higher-order approximation error. We compare our method with local PCA to assess whether this theoretical advantage persists in finite samples. 
    
    A practical concern is that the local PCA method requires prior knowledge of the latent dimension $d_{\zeta}$, since this determines the rank when performing PCA on each submatrix. However, $d_{\zeta}$ is typically unknown and rarely supported by prior knowledge. While \citet{feng2023optimal} propose an ad hoc procedure in their numerical simulations to address this issue, we instead use cross-validation to jointly select both the number of neighbors $k$ in the kNN step and the rank used in PCA.

    It is worth noting that  local PCA does not incorporate observed covariates\footnote{
        \citet{feng2023optimal} briefly discuss how to include covariates in a linear form in their appendix.
    }. In our numerical experiments, we therefore consider two implementations of their method: (1) \textbf{LPCA}, which ignores covariates $X$ and applies local PCA directly to the adjacency matrix, and (2) \textbf{X-LPCA}, which first removes the variation explained by $X$ and then applies local PCA to the residual matrix. 

    \paragraph{Local TWFE without covariates} We also consider an alternative method that directly applies our local TWFE approach to the adjacency matrix without extracting the variation explained by $X$. As discussed earlier, because the information about $\zeta$ recovered from observed data is noisy, it is unwise to ignore the observed covariates and absorb them into the latent factors. We include the numerical simulation results of this method to evaluate the importance of incorporating covariate information. We refer to this method as \textbf{LTWFE} and to our proposed method as  \textbf{X-LTWFE} in the following text.

    \paragraph{Sample splitting} While our asymptotic analysis relies on sample splitting for theoretical tractability, this requirement is primarily technical. In practice, sample splitting reduces the effective sample size, and since network sizes are relatively small in most economics applications, its finite-sample performance can be worse than that of the full-sample implementation. Accordingly, we implement all alternative methods without sample splitting. 
    
    For our proposed method, we consider both versions, with and without sample splitting, to examine how this choice affects finite-sample performance. We refer to our method with sample splitting as \textbf{X-LTWFE-SP} in the following text.

    \paragraph{Summary} We summarize all imputation methods used in the numerical experiments below:
    \begin{enumerate}[label=(\roman*)]
        \item \textbf{(X)} Covariate-only nonparametric method that imputes missing links using $X$ alone. 
        \item \textbf{(LR)} The low-rank imputation methods of \citet{bai2021matrix} and \citet{li2023link}. 
        \item \textbf{(LPCA)} The local PCA method that ignores covariates. 
        \item  \textbf{(X-LPCA)} First removes the variation explained by $X$ and then applies local PCA to the residual matrix. 
        \item  \textbf{(LTWFE)} The local two-way fixed-effects method that ignores covariates.
        \item  \textbf{(X-LTWFE)} Our proposed method. It first removes the variation explained by $X$ and then applies the local two-way fixed-effects regression to the residual matrix. 
        \item  \textbf{(X-LTWFE-SP)} Our proposed method with sample-splitting. 
    \end{enumerate}

\subsection{Simulation for Imputation Accuracy} 

    The network in the simulation is generated as 
    \begin{equation}\label{eq:MC_formation}
    \begin{gathered}
        A_{ij} = \bs{1}(\omega(X_i, X_j)'\beta + g(\xi_i, \xi_j) - U_{ij} \geq 0),  \\
        \omega(X_i, X_j)' = \left((X_{i1} - X_{j1})^2, (X_{i2} - X_{j2})^2\right), \quad g(\xi_i, \xi_j) = \xi_{i1} +  \xi_{j1} - \frac{1}{8} ( \xi_{i2}  -  \xi_{j2} )^2.
    \end{gathered}   
    \end{equation}
    The dimensions of the observable and latent characteristics are $d_{X} = d_{\xi} = 2$, and $\{(X_i, \xi_i)\mid i=1,\ldots, N\}$ is i.i.d. across individuals. Covariates and latent factors are generated according to
    \begin{align*}
        X_{id} = \frac{1}{2}(\xi_{i1} + \xi_{i2}) + \epsilon_{X, id}, \quad \forall d=1,2, 
    \end{align*}
    and 
    \begin{align*}
        \xi_i \sim \mc{N}\left( 0, 
            \begin{pmatrix}
                1 & 0 \\
                0 & 1
            \end{pmatrix}\right), 
            \quad
        \epsilon_{X, i1}, \epsilon_{X, i2} \sim \mr{Uniform}([-1, 1]), 
        \quad
        \epsilon_{X, i1}\perp \epsilon_{X, i2}. 
    \end{align*}
    The idiosyncratic shocks $\{U_{ij}\}_{1 \leq i < j \leq N }$ are independently drawn from a standard logistic distribution.  The coefficient vector $\beta$ captures homophily effects on observables and takes values $\beta\in \left\{(-0.5, -0.5), (-2, -2)\right\}$. We vary the magnitude of $\beta$ to control the sparsity of the network: for example, when $\beta = (-0.5, -0.5)$, the resulting network tends to be denser, with a larger number of links being formed. 
    
    The network size is $N = 200$, and in the simulations we consider sampling rates $\varphi := n/N \in \{0.2, 0.3, 0.4, 0.5, 0.6, 0.8\}$ to compare the performance of the proposed and alternative methods under different sampling scenarios.   
    We conduct $S = 1,000$ Monte Carlo replications in the numerical experiment, and for each replication $s$, we construct the imputed network $\hat A_s$ and compute the mean squared error (MSE) between $\hat A_s$ and the underlying probability $P_{s}$ on the missing subnetwork, $  \frac{1}{(N-n)(N-n-1) } \sum_{i, j\in \mc{S}^c} \left(\hat{A}_{s, ij} - P_{s, ij} \right)^2$. We report the root mean-squared error (RMSE)
    \begin{equation}\label{eq:MC_RMSE}
    \begin{aligned}
        \mr{RMSE} = \sqrt{\frac{1}{S}\sum_{s=1}^{S} \frac{1}{(N-n)(N-n-1) } \sum_{i, j\in \mc{S}^c} \left(\hat{A}_{s, ij} - P_{s, ij} \right)^2 }. 
    \end{aligned}  
    \end{equation}

    \begin{table}[H]
    \centering
    \caption{Simulation: Imputation}\label{tab:simulation_imputation_basic}%
        \begin{tabular}{cccccccc}
        \toprule
            & \multicolumn{1}{c}{X} & \multicolumn{1}{c}{LR} & \multicolumn{1}{c}{LPCA} & \multicolumn{1}{c}{LTWFE} & \multicolumn{1}{c}{X-LPCA} & \multicolumn{1}{c}{X-LTWFE} & \multicolumn{1}{c}{X-LTWFE-SP} \\
            &   $(\times 0.01)$    &  $(\times 0.01)$     &   $(\times 0.01)$    &   $(\times 0.01)$    &  $(\times 0.01)$     &   $(\times 0.01)$    & $(\times 0.01)$ \\
        \midrule
        $\beta = (-.5, -.5)$ &       &       &       &       &       &       &  \\
        $\varphi = 0.2$ & 21.0    & 15.8  & 15.9  & 15.3  & 14.9  & 10.9  & 13.1 \\
        $\varphi = 0.3$ & 20.9  & 13.9  & 14.0    & 13.9  & 12.3  & 9.7   & 11.3 \\
        $\varphi = 0.4$ & 20.8  & 12.4  & 12.5  & 13.0    & 10.8  & 8.9   & 10.4 \\
        $\varphi = 0.5$ & 20.8  & 11.3  & 11.3  & 12.3  & 9.8   & 8.4   & 9.7 \\
        $\varphi = 0.6$ & 20.8  & 10.3  & 10.4  & 11.9  & 9.2   & 8.1   & 9.3 \\
        $\varphi = 0.8$ & 20.6  & 9.2   & 9.2   & 11.0    & 8.3   & 7.5   & 8.6 \\ 
        $\beta = (-2, -2)$ &       &       &       &       &       &       &  \\
        $\varphi = 0.2$ & 13.8  & 16.3  & 16.4  & 15.3  & 12.7  & 10.2  & 11.3 \\
        $\varphi = 0.3$ & 13.7  & 14.8  & 14.9  & 14.0    & 11.5  & 9.4   & 10.3 \\
        $\varphi = 0.4$ & 13.7  & 13.3  & 13.5  & 13.1  & 10.5  & 8.9   & 9.7 \\
        $\varphi = 0.5$ & 13.6  & 11.7  & 11.8  & 12.5  & 9.8   & 8.6   & 9.3 \\
        $\varphi = 0.6$ & 13.6  & 10.5  & 10.6  & 11.9  & 9.3   & 8.3   & 9.0 \\
        $\varphi = 0.8$ & 13.5  & 9.2   & 9.2   & 11.0    & 8.4   & 7.8   & 8.5 \\
        \bottomrule
        \end{tabular}
        \vspace{0.5cm}
        \begin{minipage}{\textwidth}
            \footnotesize
            \textbf{Note:} We conduct Monte Carlo simulations based on $1,000$ repetitions of the network formation model~\eqref{eq:MC_formation}. The network size is $N = 200$. We set $\beta\in \left\{(-0.5, -0.5), (-2, -2)\right\}$ and sampling rates $\varphi  \in \{0.2, 0.3, 0.4, 0.5, 0.6, 0.8\}$.  We report RMSE as in~\eqref{eq:MC_RMSE} for the following imputation methods: (1)  covariate-only nonparametric imputation that imputes missing outcomes using $X$ alone (\textbf{X}), (2) 
            the low-rank imputation methods of \citet{bai2021matrix} and \citet{li2023link} (\textbf{LR}), (3)  the local PCA method that ignores covariates (\textbf{LPCA}), (4) the local PCA method using covariates (\textbf{X-LPCA}), (5)  the local two-way fixed-effects regression that ignores covariates (\textbf{LTWFE}), (6) the local two-way fixed-effects regression using covariates (\textbf{X-LTWFE}), and (7) X-LTWFE method with sample-splitting (\textbf{X-LTWFE-SP}). It is important to note that LR, LPCA, LTWFE, X-LPCA, and X-LTWFE are implemented without sample splitting. RMSEs are presented in units of $0.01$ for ease of reading.   
        \end{minipage}
    \end{table}

    We evaluate the performance of our imputation method and alternative approaches in both dense networks (e.g., $\beta = (-0.5, -0.5)$, for which the average probability of forming a link is about $34\%$) and sparse networks (e.g., $\beta = (-2, -2)$, for which the average probability of forming a link is about $10\%$). The simulation results are reported in Table~\ref{tab:simulation_imputation_basic}, with RMSEs presented in units of $0.01$ for ease of reading.  

    The first column (X) reports the RMSE of covariate-only  method, which imputes missing links using $X$ alone. Because this method ignores unobserved heterogeneity, its imputations are systematically biased, and the error does not vanish as the network size or sampling rate increases. Consistent with the theoretical prediction, the numerical results show that, as the sampling rate $\varphi$ increases, the RMSE of the covariate-only method exhibits very limited improvement (from $0.210$ to $0.206$ in the dense scenario, and from $0.138$ to $0.135$ in the sparse scenario). 

    The second column (LR) reports the RMSE of the low-rank  imputation, and the third column (LPCA) reports the RMSE of the local PCA. Since the underlying graphon in \eqref{eq:MC_formation} is high-rank, local PCA should, in theory, outperform the low-rank approach. However, the numerical results indicate that the RMSEs of the two methods are nearly the same across all scenarios. We conjecture that this pattern is driven by  two reasons:  (1) Although the graphon is not exactly low-rank, it is approximately low-rank, so the low-rank approach still provides a good approximation to the graphon matrix, especially when performance is measured using Frobenius norm. (2) Local PCA relies only on the subnetwork formed by neighbors, which leads to substantial sample loss and limits its performance in finite samples. In line with our theory, the local PCA estimator outperforms the covariate-only (X) method in most cases, and its prediction error decreases as the  sampling rate increases. 
  
    The fourth column (LTWFE) reports the RMSE of the local two-way fixed0effects imputation that ignores covariates. In line with the theory, its prediction error decreases as $n$ increases. Table~\ref{tab:simulation_imputation_basic} shows that at low sampling rates, the LTWFE method performs better than local PCA, whereas once the sampling rate becomes large (e.g., when $\varphi$ exceeds $40\%$), local PCA outperforms LTWFE. This suggests that although local PCA has better asymptotic properties, in the sense that it achieves higher-order approximation error, it requires a larger sample size to realize this advantage, whereas LTWFE is more robust in small samples. In addition, the LTWFE method is more robust to network sparsity: as shown in Table~\ref{tab:simulation_imputation_basic}, its performance is nearly unaffected by changes in the sparsity of the adjacency matrix.

    The fifth column (X-LPCA) and the sixth column (X-LTWFE) report the performance of LPCA and LTWFE when incorporating covariate information. These methods first extract the variation of the missing links that can be explained by $X$, and then apply LPCA and LTWFE to the residual matrix, respectively. Compared with LPCA and LTWFE, incorporating covariate information leads to substantial improvements in prediction accuracy, particularly for X-LTWFE. The gains are especially large when the sampling rate is low or when the network is sparse. X-LPCA exhibits a similar pattern. In addition, Table~\ref{tab:simulation_imputation_basic} shows that the proposed  method outperforms all alternatives in our simulations, and its performance remains robust even under sparse networks and low sampling rates, which are common in empirical applications.

    The last column (X-LTWFE-SP) reports the performance of our proposed estimator with sample splitting. As discussed earlier, although sample-splitting facilitates asymptotic analysis, it reduces the effective sample size, and this issue can be severe in small-sample settings. In addition, it is reasonable to expect (though we do not formally prove it) that the correlation between the pseudo-distance and the error terms diminishes, making sample splitting unnecessary in theory.  Table~\ref{tab:simulation_imputation_basic} shows that performance of proposed imputation under sample splitting is always worse than the performance without splitting the sample, but the difference between the two methods becomes smaller as the sample size increases.

    In summary, the simulations show that our proposed method outperforms the alternative methods and is more robust across different levels of network sparsity and sampling rates. In addition, for finite samples, we recommend using the version without sample splitting in order to achieve higher prediction accuracy. Further numerical results are provided in Appendix~\ref{appendix:extra_simulation}.

\subsection{Simulation for GMM estimators}

    In this subsection, we evaluate the performance of GMM estimators in downstream analyses that rely on the imputed networks. We compare the performance of our method with the alternative approaches introduced previously. We focus on two widely used classes of empirical exercises in economics that are based on network data. 
    \paragraph{Regression on network statistics} Suppose for each network $m=1, \ldots, M$, the outcome $Y_{mi}$ is generated according to: 
    \begin{align}\label{eq:MC_regression}
        Y_{mi} = \alpha_0 + \alpha_1 \phi_{mi}(A_m) + u_m + e_{mi}, \quad u_m \sim \mc{N}(0, 1/4), \quad e_{mi} \sim \mc{N}(0, 1/4), 
    \end{align} 
    where $\phi_{mi} : [0,1]^{N_m \times N_m} \to \mathbb{R}$ computes a network statistic for individual $i$ in network $m$, $\{u_m\}_{1\leq m\leq M}$ are i.i.d. network-level random effects independent of $A_m$, and $e_{mi}$ is i.i.d. idiosyncratic error. We consider two specifically, normalized average degree and eigenvector centrality, as defined in Example~\ref{example:regression}. The networks $\{A_m\}_{1\leq m\leq M}$ according to~\eqref{eq:MC_formation} with $\beta = (-0.5, -0.5)$, which yields relatively dense networks with a larger number of links.   
    
    We set the number of networks to $M = 40$ and fix network size at $N_m = 200$ for all $m$. The simulations consider sampling rates $\varphi := n/N \in \{0.2, 0.3, 0.4, 0.5\}$ to compare the performance of the proposed method and alternative methods under different levels of sampling. For reference, we also report the performance of regressions using the complete networks (\textbf{CD}). We set $\alpha_0 = 0$ and $\alpha_1 = 0.5$.
 
    We conduct $S = 1,000$ Monte Carlo replications. For each replication $s$,  we generate $M$ independent network, impute each network, and estimate the linear regression. The bias and standard deviation of $\hat{\alpha}_1$ are reported in Table~\ref{tab:simulation_centrality_basic}.
    
    Table~\ref{tab:simulation_centrality_basic} displays patterns similar to those in Table~\ref{tab:simulation_imputation_basic}. As the sampling rate increases, both the bias and the standard deviation of the estimator decline in most cases. This improvement is not only due to more accurate imputation at higher sampling rates (Theorem~\ref{thm:bias_variance}), but also because the aggregate statistics themselves are computed more precisely when fewer observations are missing. The latter explains why the bias and standard deviation in the \textbf{X} column also decrease. 
    In most cases, the X-LTWFE estimator delivers the best performance and remains stable even when the sampling rate is small. The only exception arises when eigenvector centrality is used and the sampling rate is high. In this case, LTWFE yields an extremely small bias—almost identical to that obtained using the complete data. This pattern appears to be driven by features of our numerical design.

    It is also worth noting that the bias-std ratio of the linear regression estimator is not close to zero in these simulations. This arises from the use of nonparametric imputation and is particularly evident when eigenvector centrality is employed. Even if researchers choose to rely on a parametric model for imputation, the robustness of our estimator implies that our method remains valuable as a robustness check in applied work.

    \begin{table}[h]
    \centering
    \caption{Simulation: Linear regression on centrality}\label{tab:simulation_centrality_basic}%
        \begin{tabular}{ccccccccc}
        \toprule
            & CD    & X     & LR    & LPCA  & LTWFE & X-LPCA & X-LTWFE & X-LTWFE-SP \\
            &  $(\times 0.01)$     &     $(\times 0.01)$  &   $(\times 0.01)$    &    $(\times 0.01)$   &    $(\times 0.01)$   &  $(\times 0.01)$    &   $(\times 0.01)$    &  $(\times 0.01)$  \\
        \midrule 
        \multicolumn{9}{l}{\textbf{Panel A: Degree centrality}}\\ 
        $\varphi = 0.2$ &       &       &       &       &       &       &       &  \\
        BIAS  & -0.05 & 9.62  & -3.54 & -1.51 & -1.49 & 3.69  & -0.69 & -5.11 \\
        STD   & 4.07  & 12.65 & 4.51  & 4.49  & 4.67  & 7.31  & 4.68  & 4.66 \\
        $\varphi = 0.3$ &       &       &       &       &       &       &       &  \\
        BIAS  & -0.05 & 10.48 & -1.6  & -1.74 & -0.65 & 2.51  & -0.03 & -2.16 \\
        STD   & 4.07  & 9.03  & 4.24  & 4.20   & 4.31  & 5.03  & 4.36  & 4.29 \\
        $\varphi = 0.4$ &       &       &       &       &       &       &       &  \\
        BIAS  & -0.05 & 9.00     & -0.74 & -0.81 & -0.33 & 1.53  & 0.14  & -0.91 \\
        STD   & 4.07  & 7.08  & 4.16  & 4.15  & 4.22  & 4.50   & 4.27  & 4.23 \\
        $\varphi = 0.5$ &       &       &       &       &       &       &       &  \\
        BIAS  & -0.05 & 6.75  & -0.31 & -0.34 & -0.18 & 0.91  & 0.15  & -0.32 \\
        STD   & 4.07  & 5.90   & 4.12  & 4.11  & 4.14  & 4.30   & 4.18  & 4.17 \\
            &       &       &       &       &       &       &       &  \\
    \multicolumn{9}{l}{\textbf{Panel B: Eigenvector centrality}}\\ 
        $\varphi = 0.2$ &       &       &       &       &       &       &       &  \\
        BIAS  & -0.01 & 7.02  & -3.94 & -4.19 & -0.40  & 2.27  & 0.39  & -0.61 \\
        STD   & 0.61  & 1.00     & 0.66  & 0.67  & 0.68  & 0.86  & 0.67  & 0.70 \\
        $\varphi = 0.3$ &       &       &       &       &       &       &       &  \\
        BIAS  & -0.01 & 6.36  & -1.93 & -2.05 & -0.07 & 1.10   & 0.54  & 0.43 \\
        STD   & 0.61  & 0.84  & 0.62  & 0.62  & 0.64  & 0.68  & 0.64  & 0.66 \\
        $\varphi = 0.4$ &       &       &       &       &       &       &       &  \\
        BIAS  & -0.01 & 4.83  & -0.99 & -1.05 & -0.01 & 0.67  & 0.43  & 0.60 \\
        STD   & 0.61  & 0.75  & 0.62  & 0.62  & 0.63  & 0.64  & 0.63  & 0.64 \\
        $\varphi = 0.5$ &       &       &       &       &       &       &       &  \\
        BIAS  & -0.01 & 3.38  & -0.50  & -0.54 & 0.01  & 0.42  & 0.30   & 0.50 \\
        STD   & 0.61  & 0.67  & 0.61  & 0.61  & 0.61  & 0.62  & 0.62  & 0.62 \\
    \bottomrule    
    \end{tabular}
    \vspace{0.5cm}
        \begin{minipage}{\textwidth}
            \footnotesize
            \textbf{Note:} We conduct Monte Carlo simulations based on $1,000$ repetitions of the network formation model~\eqref{eq:MC_formation}. The network size is $N = 200$. The number of networks is $M = 40$.  We set $\beta = (-0.5, -0.5)$ and sampling rates $\varphi  \in \{0.2, 0.3, 0.4, 0.5\}$.  We report bias and standard deviation $\hat{\alpha}_1$  for the following imputation methods: 
            (1) estimates based on complete network data (\textbf{CD}) (2)  covariate-only nonparametric imputation that imputes missing outcomes using $X$ alone (\textbf{X}), (3) 
            the low-rank imputation methods of \citet{bai2021matrix} and \citet{li2023link} (\textbf{LR}), (4)  the local PCA method that ignores covariates (\textbf{LPCA}), (5) the local PCA method using covariates (\textbf{X-LPCA}), (6)  the local two-way fixed-effects regression that ignores covariates (\textbf{LTWFE}), (7) the local two-way fixed-effects regression using covariates (\textbf{X-LTWFE}), and (8) X-LTWFE method with sample-splitting (\textbf{X-LTWFE-SP}). It is important to note that LR, LPCA, LTWFE, X-LPCA, and X-LTWFE are implemented without sample splitting. Bias and standard deviation are presented in units of $0.01$ for ease of reading.   
        \end{minipage}
    \end{table}

\paragraph{Linear-in-mean peer-effects model}
    Suppose for each network $m=1, \ldots, M$, the outcome $Y_{mi}$ is generated according to: 
    \begin{equation}
    \begin{gathered}\label{eq:MC_linear_in_means}
        Y_{m} = \alpha_C +  \alpha_{\bar{Y}} G_m Y_m +  \sum_{d=1}^{2}\alpha_{W, d}  W_{m, d}  + \sum_{d=1}^{2}\alpha_{\bar{W}, d} G_m W_{m, d}  + u_m  + e_m,  \\
        W_{mi, d} = \xi_{mi, d} + \frac{1}{2} X_{mi, d}  \xi_{mi, d}, \quad d = 1, 2,  \\
        u_m \sim   \mc{N}(0,  1/25), \quad e_{mi}  \sim \mc{N}(0, 1), 
    \end{gathered}   
    \end{equation} 
    where $\{u_m\}_{1\leq m\leq M}$ are i.i.d. network-level random effects independent of $A_m$, and $e_{mi}$ is i.i.d. idiosyncratic error.  The networks $\{A_m\}_{1\leq m\leq M}$ according to~\eqref{eq:MC_formation} with $\beta = (-0.5, -0.5)$, which yields relatively dense networks with a larger number of links.   
    
    We set the number of networks to $M = 40$ and fix network size at $N_m = 200$ for all $m$. The simulations consider sampling rates $\varphi := n/N \in \{0.2, 0.3, 0.4, 0.5\}$ to compare the performance of the proposed method and alternative methods under different levels of sampling. For reference, we also report the performance of regressions using the complete networks (\textbf{CD}). We set $\alpha_C = 0$, $\alpha_{\bar{Y}} = 0.5$, $\alpha_W = \alpha_{\bar{W}} = (1, 1)$. We conduct $S = 1,000$ Monte Carlo replications. For each replication $s$,  we generate $M$ independent network, impute each network, and estimate the GMM estimator using~\eqref{eq:GMM_estimator} with $\Sigma_M = \bs{I}_{q}$. The bias and standard deviation of estimate of endogenous effect $\hat{\alpha}_{\bar{Y}}$ are reported in Table~\ref{tab:simulation_linear_in_means}.

    \begin{table}[h]
    \centering
    \caption{Simulation: Linear-in-means peer-effects model}\label{tab:simulation_linear_in_means}
        \begin{tabular}{ccccccc}
        \toprule
            & CD    & X     & LTWFE & X-LPCA & X-LTWFE & X-LTWFE-SP \\
            &       $(\times 0.01)$  &    $(\times 0.01)$   &    $(\times 0.01)$   &   $(\times 0.01)$    &    $(\times 0.01)$   &  $(\times 0.01)$ \\
        \midrule
        \multicolumn{7}{l}{\textbf{Panel A: Estimates $\alpha_{\bar{Y}}$}}\\
        $\varphi = 0.2$ &       &       &       &       &       &  \\
        BIAS  & 0.70   & 28.32 & 16.57 & 11.59 & 10.16 & 13.22 \\
        STD   & 7.50   & 13.49 & 13.31 & 12.63 & 9.23  & 11.38 \\
        $\varphi = 0.3$ &       &       &       &       &       &  \\
        BIAS  & 0.70   & 23.85 & 11.48 & 8.50   & 10.82 & 11.83 \\
        STD   & 7.50   & 11.11 & 9.70   & 9.53  & 8.03  & 9.01 \\
        $\varphi = 0.4$ &       &       &       &       &       &  \\
        BIAS  & 0.70   & 18.07 & 6.89  & 5.66  & 9.73  & 9.69 \\
        STD   & 7.50   & 9.18  & 8.73  & 8.40   & 7.65  & 8.12 \\
        $\varphi = 0.5$ &       &       &       &       &       &  \\
        BIAS  & 0.70   & 12.53 & 3.57  & 3.18  & 7.34  & 7.75 \\
        STD   & 7.50   & 8.12  & 8.44  & 8.07  & 7.54  & 7.78 \\
            &       &       &       &       &       &  \\
        \multicolumn{7}{l}{\textbf{Panel B: Estimates $\alpha_{W,1}$}}\\
        $\varphi = 0.2$ &       &       &       &       &       &  \\
        BIAS  & 0.03  & -11.14 & 18.28 & -3.30  & 3.03  & 5.66 \\
        STD   & 1.49  & 1.63  & 2.11  & 1.73  & 1.58  & 1.79 \\
        $\varphi = 0.3$ &       &       &       &       &       &  \\
        BIAS  & 0.03  & 9.84  & 11.57 & -1.84 & 2.04  & 4.06 \\
        STD   & 1.49  & 1.52  & 1.87  & 1.57  & 1.51  & 1.63 \\
        $\varphi = 0.4$ &       &       &       &       &       &  \\
        BIAS  & 0.03  & -8.55 & 7.02  & -1.29 & 1.19  & 2.68 \\
        STD   & 1.49  & 1.47  & 1.73  & 1.49  & 1.50   & 1.56 \\
        $\varphi = 0.5$ &       &       &       &       &       &  \\
        BIAS  & 0.03  & -7.13 & 4.03  & -0.98 & 0.59  & 1.58 \\
        STD   & 1.49  & 1.45  & 1.64  & 1.48  & 1.49  & 1.51 \\
        \bottomrule
        \end{tabular}\vspace{0.5cm}
        \begin{minipage}{\textwidth}
            \footnotesize
            \textbf{Note:} We conduct Monte Carlo simulations based on $1,000$ repetitions of the network formation model~\eqref{eq:MC_formation}. The network size is $N = 200$. The number of networks is $M = 40$.  We set $\beta = (-0.5, -0.5)$ and sampling rates $\varphi  \in \{0.2, 0.3, 0.4, 0.5\}$.  We report bias and standard deviation of $\hat{\alpha}_{\bar{Y}}$  and $\hat{\alpha}_{W, 1}$ for the following imputation methods: 
                (1) estimates based on complete network data (\textbf{CD}) (2)  covariate-only nonparametric imputation that imputes missing outcomes using $X$ alone (\textbf{X}), (3) the local two-way fixed-effects regression that ignores covariates (\textbf{LTWFE}), (4)  the local PCA method using covariates (\textbf{X-LPCA}), (5)  the local two-way fixed-effects regression using covariates (\textbf{X-LTWFE}), and (6) X-LTWFE method with sample-splitting (\textbf{X-LTWFE-SP}). It is important to note that LTWFE, X-LPCA, and X-LTWFE are implemented without sample splitting. Bias and standard deviation are presented in units of $0.01$ for ease of reading. 
        \end{minipage}
    \end{table}

    Table~\ref{tab:simulation_linear_in_means} reports the bias and standard deviation of different estimators of $\alpha_{\bar{Y}}$ and $\alpha_{W,1}$ obtained using X, LTWFE, X-PCA, X-LTWFE, and sample-splitting X-LTWFE. The full set of simulation results is presented in Table~\ref{tab:simulation_linear_in_means_s1} and Table~\ref{tab:simulation_linear_in_means_s2} in Appendix~\ref{appendix:extra_simulation}. Similar to the results in Table~\ref{tab:simulation_centrality_basic}, as the sampling rate increases, both the bias and the standard deviation of the estimator decline. In most cases, the estimator based on our proposed method (X-LTWFE) has strong performance, especially when the sampling rate is small. However, we also observe that, when the sampling rate exceeds $30\%$, X-LPCA delivers strong performance and attains a smaller bias in estimating the endogenous effect $\alpha_{\bar{Y}}$ than our proposed estimator. In addition, although LTWFE delivers a smaller bias in estimating the endogenous effect than X-LTWFE when $\varphi \geq 40\%$,  its performance deteriorates substantially when the sampling rate becomes small.
    
    The numerical patterns align with the implications of Theorem~\ref{thm:linear_in_mean}: the bias-to-std ratio of the estimators does not vanish to $0$. This phenomenon is inevitable when the graphon is a nonparametric function of latent factors and poses challenges for inference. Nevertheless, by Theorem~\ref{thm:consistency_plug_in}, our proposed estimator is still valuable as a robustness benchmark in empirical applications.

    \section{Empirical Application}\label{sec:application}

    This section presents an empirical application using the network data of 43 villages in rural India from \citet{banerjee2013diffusion}. For each village, the researchers informed certain households (injection nodes or leaders) about the information of microfinance program and asked them to encourage other villagers to join. The researchers want study how households' microfinance take-up decisions are influenced by network characteristics. 
    The data consist of egocentrically sampled social network information and demographic variables (for both sampled and unsampled households). There are $M = 43$ villages, with an average of approximately $N \approx 200$ households per village, and a sampling rate of about $\varphi = 45\%$.

    We need to impute the missing links in the adjacency matrices due to egocentric sampling. Since the dataset includes rich demographic information with $d_X > 10$ and the number of households within each village is relatively small, fully nonparametric regression becomes infeasible. Thus, we replace the first-step nonparametric regression with a linear projection on $\omega(X_i, X_j): = ((X_{i1} - X_{j1})^2, \ldots, (X_{id_{X}} - X_{jd_{X}})^2)$ (another option is to select a subset of important variables and then conduct a nonparametric regression, but this would rely heavily on researchers' discretion). We then apply the local two-way fixed-effects regression to the residuals from the projection. 

    We focus on the \emph{household-level} regressions in \citet{banerjee2013diffusion}. The outcome variable is each household's microfinance take-up decision ($Y_i = 1$ if the household decides to take part in the program). We consider three specifications: (i) how households' normalized average degrees (defined as in Example~\ref{example:regression}) affect their take-up decisions; (ii) how households' eigenvector centralities (defined as in Example~\ref{example:regression}) affect their take-up decisions; and (iii) how peers' decisions affect individual take-up through a linear-in-means peer-effects model. 
    \begin{table}[H]
    \centering
    \caption{Empirical Application: Linear regression on Centrality}\label{tab:empirical_centrality}
        \begin{tabular}{cccccc}
        \toprule
            & \multicolumn{2}{l}{Panel A: Degree centrality}        &       & \multicolumn{2}{l}{Panel B: Eigenvector centrality}   \\
        \midrule
            & \multicolumn{1}{c}{Sampled} & \multicolumn{1}{c}{Imputed} &       & \multicolumn{1}{c}{Sampled} & \multicolumn{1}{c}{Imputed} \\
        $\alpha_C$ & 0.157  & 0.146  &       & 0.163  & 0.160 \\
        $\alpha_1$ & 0.612  & 0.723  &       & 0.026 & 0.028 \\
        \bottomrule
        \end{tabular}
        \vspace{0.5cm}
        \begin{minipage}{\textwidth}
            \footnotesize
            \textbf{Note:} This table provides estimates of regressions of households' microfinance take-up decisions on households' network statistics, using data from \citet{banerjee2013diffusion}. We report results for both sampled (incomplete) networks and imputed networks using our proposed method. The estimates using degree centralities are reported in Panel A. and the estimates using eigenvector centralities are reported in Panel B. 
            $\alpha_C$ represents the intercept and $\alpha_1$ is the coefficient before the centrality measures. 
        \end{minipage}
    \end{table}

    Table~\ref{tab:empirical_centrality} reports estimates of household-level regressions in which the dependent variable is a household's microfinance take-up decision and the regressors are household-level network characteristics. We provide regression results for both sampled (incomplete) networks and imputed network using our proposed method. 
    Panel A shows that, when using the imputed network, the household's take-up probability increases by $7.2\%$ percentage points when its normalized average degree increases by $10\%$ (equivalently, when the household has about $20$ more connections within the village). The estimated effect is weaker when using  sampled networks, with the coefficient $\alpha_1$ decreasing from $0.72$ to $0.61$. Panel B presents results using eigenvector centrality. Both specifications yield similar estimates ($0.026$ using the sampled network and $0.028$ using the imputed network), indicating that households with higher eigenvector centrality are more likely to take up microfinance.

    \begin{table}[htbp]
    \centering
    \caption{Empirical Application: Linear-in-means Peer-effects Model}\label{tab:empirical_linear_in_means}
        \begin{tabular}{lcccccccc}
        \toprule
            & & \multicolumn{1}{l}{$\alpha_C$} & & \multicolumn{1}{l}{$\alpha_{\bar{Y}}$} & &\multicolumn{1}{l}{$\alpha_{W}$} & & \multicolumn{1}{l}{$\alpha_{\bar{W}}$} \\
        \midrule
        Sampled & & 0.18 & &  -0.02 & &  0.06 & & -0.00 \\
        Imputed & & 0.16 & & 0.09  & & 0.06 & & -0.03 \\
        \bottomrule
        \end{tabular}
        \vspace{0.5cm}
        \begin{minipage}{\textwidth}
            \footnotesize
            \textbf{Note:} This table provides estimates of linear-in-means peer-effects model (defined as in Example~\ref{example:linear_in_mean_peer_effect}) using data from \citet{banerjee2013diffusion}.  We report results for both sampled (incomplete) networks and imputed networks using our proposed method. Here, $W$ is a dummy variable, indicating whether a household contains injection nodes. The column $\alpha_C$ reports the estimates of intercept, $\alpha_{\bar{Y}}$ reports the estimates of endogenous effects, and   $\alpha_{W}$ and $\alpha_{\bar{W}}$ report the coefficients on  $W_i$ and $(GW)_i$, respectively. 
        \end{minipage}
    \end{table}

    Table~\ref{tab:empirical_linear_in_means} reports estimates from a household-level linear-in-means peer-effects model, where the dependent variable is the household's microfinance take-up decision, and the regressor $W$ is a dummy variable indicating whether the household contains an injection node. We present regression results based on both the sampled (incomplete) network and the imputed network constructed using our proposed method.

    The coefficient $\alpha_{\bar{Y}}$ captures the endogenous peer effect, measuring the influence of neighbors' participation decisions. Using the imputed network, the estimate of $\alpha_{\bar{Y}}$ is $0.09$, implying that a household's take-up probability increases by approximately $0.9$ percentage points when the fraction of its friends who join the program rises by $10$ percentage points (corresponding to roughly two additional friends joining the program, on average). In contrast, the estimated endogenous effect is $-0.02$ negative when using sampled networks. In addition, both regressions give similar estimates for the direct effect, suggesting that the household's take-up probability increases by $6$ percentage points when it contains an injection individual. Finally, for the contextual peer effect $\alpha_{\bar{W}}$, both sampled and imputed networks produce negative estimates, with the magnitude being larger when using the imputed network.

    \section{Conclusion}\label{sec:conclusion}

    Sampled network data are common in empirical research because collecting full network information is costly, but using sampled networks can lead to biased estimates. We propose a flexible imputation method for sampled networks and show that downstream empirical analysis using the imputed network yields consistent parameter estimates. 
    
    Our method separately imputes the part of each missing link that is explained by covariates and the remaining variation not captured by them: the former via projection onto observed covariates, and the latter using local two-way fixed effects. The imputation method avoids parametric assumptions, does not rely on low-rank restrictions, and flexibly incorporates observed  covariates and unobserved heterogeneity. We establish entrywise convergence rates for the imputed matrix and prove consistency of GMM estimators based on the imputed network. We further derive the convergence rate for the corresponding estimator in the linear-in-mean peer-effects model. Simulations show strong performance of our method both in terms of imputation accuracy and in downstream empirical analysis. 

    The robustness of our imputation comes at the cost of making downstream inference challenging. In the paper, we discuss how undersmoothing can help improve the downstream estimator, and other approaches, such as bias correction may be potentially applicable. Developing inference procedures for downstream estimators  is a potential topic for future research.

    \bibliographystyle{aea}
    \bibliography{ref}
    
\clearpage

\appendix

\numberwithin{proposition}{section}
\numberwithin{lemma}{section}
\numberwithin{theorem}{section}
\numberwithin{corollary}{section}
\numberwithin{example}{section}

\numberwithin{assumption}{section}

    \section{Implementation, Additional Simulation, and Informativeness Condition}\label{appendix:extran}

\subsection{Complete Algorithm}\label{appendix:extra_algorithm}
  
    In the main text, we present a simplified version of our procedure, without sample splitting or cross-validation, in Algorithm~\ref{alg:simple} to facilitate understanding of the main idea behind our approach. In this section, we introduce Algorithm~\ref{alg:cv} and \ref{alg:sp} for cross-validation and sample-splitting. These two algorithms are  employed in the numerical experiments (see Section~\ref{sec:simulation}), and we recommend their use in practice. 

    Algorithm~\ref{alg:cv}  incorporates \emph{leave-one-out} cross-validation for selecting the tuning parameter $h$ but does not use sample splitting.  This algorithm is recommended for empirical applications because, similar to other nonparametric procedures, our method relies critically on the choice of the tuning parameter. However, since Algorithm~\ref{alg:cv} does not employ sample splitting, developing a theoretical analysis based on it is challenging.

    Algorithm~\ref{alg:sp} incorporates both cross-validation and sample splitting and serves as the basis of our theoretical analysis. We randomly split sampled individuals $\mc{S}$ into two subsets, $\mc{S}_1$ and $\mc{S}_2$, of equal sizes. We construct the pseudo-distance using only the network information from $\mc{S}1$. Specifically, for any $i \in \mc{S}_2 \cup \mc{S}^c$ and $i' \in \mc{S}_2$, 
    \begin{align}\label{eq:sp:d}
        \hat{d}_{ii'} = \max_{ k\in \{1, \ldots, N\} \backslash \{i, i'\} } \left|\frac{1}{|\mc{S}_1 |} \sum_{\ell \in \mc{S}_1 } A_{k\ell}(A_{i\ell } - A_{i'\ell})\right|. 
    \end{align}
    We then implement two-way fixed-effects imputation using only the network information from $\mc{S}_2$ (excluding links with individuals in $\mc{S}_1$). That is, for any $i, j \in \mc{S}^c$,
    \begin{equation}\label{eq:sp:Aij}
    \begin{gathered}
        (\hat{\bs{a}}, \hat{\bs{b}}) \in \argmin_{\bs{a}, \bs{b}\in \mb{R}^{|\mc{S}_2| + 1}} \sum_{ \substack{i'\in \mc{S}_2 \cup\{i\}   , j'\in \mc{S}_2 \cup\{j\}, \\
        (i', j')\neq (i, j)}}   \left(A_{i'j'} - \hat{\Pi}_{ i'j'} -  a_{i'} - b_{j'} \right)^2 K_{h}\left(\hat{d}_{ii'}\right)K_{h}\left(\hat{d}_{jj'}\right), \\
        \hat{A}_{ij} = \hat{a}_i + \hat{b}_j + \hat{\Pi}_{ij}. 
    \end{gathered}
    \end{equation} 
    
    However, since sample splitting reduces the effective sample size, and because network sizes are relatively small in most economic applications, Algorithm~\ref{alg:sp}'s finite-sample performance can be worse than Algorithm~\ref{alg:cv} in finite samples (see Section~\ref{sec:simulation}). The details of Algorithm~\ref{alg:cv} and Algorithm~\ref{alg:sp} are provided below.

    \begin{algorithm}[H]
        \caption{Imputation with Cross Validation}\label{alg:cv}
        \begin{algorithmic}
            \REQUIRE Kernel $K(\cdot)$, a list of bandwidths $\{h_q\}_{q=1, \ldots, Q}$, incomplete network $A^{\mr{obs}}$, covariate matrix $X$. 
            \ENSURE Imputed network $\hat{A}$.
                \STATE \textbf{Step 1: Extract information explained by $X$}  
                \STATE Run dyadic nonparametric regression to obtain $\hat{\Pi}(\cdot, \cdot)$. 
                \STATE \textbf{Step 2: Calculate similarity}  
                \STATE For each $i \in 1, \ldots, N$ and $i'\in \mc{S}$, calculate the pseudo distance using observed adjacency matrix: 
                \begin{align*}
                    \hat{d}_{ii'} = \max_{k\in \{1, \ldots, N\}\backslash \{i, i'\}} \left|\frac{1}{n} \sum_{\ell = 1}^{n} A_{k\ell}(A_{i\ell } - A_{i'\ell})\right|. 
                \end{align*} 
                \STATE \textbf{Step 3: Leave-one-out cross validation}   
                \STATE  For each $h_q \in \{h_1, \ldots, h_Q\}$, and for each $i\in \mc{S}$ and $j \in \mc{S}^c$, estimate the local (weighted) two-way fixed-effects model: 
                \begin{align*}
                    (\hat{\bs{a}}(h_q), \hat{\bs{b}}(h_q)) \in \argmin_{\bs{a}, \bs{b} \in \mb{R}^{n}} \sum_{ \substack{ i'\in \mc{S},  j'\in \mc{S} \cup\{j\}, \\ (i', j')\neq (i, j) }}  \left(A_{i'j'} - \hat{\Pi}_{ i'j'} - a_{i'} - b_{j'} \right)^2 K_{h_q}\left(\hat{d}_{ii'}\right)K_{h_q}\left(\hat{d}_{jj'}\right)
                \end{align*}
                to obtain $\bar{A}_{ij}(h_q) = \hat{a}_i(h_q) + \hat{b}_j(h_q) + \hat{\Pi}_{ ij}$. 
                Then the optimal $h^*$ solves 
                \begin{align*}
                    h^* = \argmin_{h_q \in \{h_1, \ldots, h_Q\}} \sum_{i\in \mc{S}, j \in \mc{S}^c}\left(A_{ij} - \bar{A}_{ij}(h_q) \right)^2. 
                \end{align*}
                \STATE \textbf{Step 4: Local imputation}   
                \STATE For each $i\in \mc{S}^c$ and $j \in \mc{S}^c$, estimate the local (weighted) two-way fixed-effects model: 
                \begin{align*}
                    (\hat{\bs{a}}, \hat{\bs{b}}) \in \argmin_{\bs{a}, \bs{b} \in \mb{R}^{n+1}} \sum_{ \substack{i'\in \mc{S}\cup\{i\}   , j'\in \mc{S}\cup\{j\}, \\
                    (i', j')\neq (i, j)}}  \left(A_{i'j'} - \hat{\Pi}_{ i'j'} - a_{i'} - b_{j'} \right)^2 K_{h^*}\left(\hat{d}_{ii'}\right)K_{h^*}\left(\hat{d}_{jj'}\right)
                \end{align*}
                to obtain $\tilde{A}_{ij} = \hat{a}_i + \hat{b}_j + \hat{\Pi}_{ ij}$. By combining the imputed entries with the observed entries of the adjacency matrix, we obtain the (preliminary) imputed adjacency matrix $\tilde{A}$. Truncate $\tilde{A}_{ij}$ to get 
                \begin{align*}
                    \hat{A}_{ij} = \left\{
                        \begin{array}{ll}
                           \tilde{A}_{ij}, & \text{if } \tilde{A}_{ij}\in [0, 1]\\
                            1 & \text{if } \tilde{A}_{ij} > 1 \\
                            0 & \text{if } \tilde{A}_{ij} < 0
                        \end{array}
                    \right., 
                \end{align*}
                and zero the diagonal of $\hat{A}$ such that $\hat{A}_{ii} = 0$ for all $i$. 
        \end{algorithmic}
    \end{algorithm}

    \begin{algorithm}[H]
        \caption{Imputation with Cross Validation and Sample Splitting}\label{alg:sp}
        \begin{algorithmic}
            \REQUIRE Kernel $K(\cdot)$, a list of bandwidths $\{h_q\}_{q=1, \ldots, Q}$, incomplete network $A^{\mr{obs}}$, covariate matrix $X$. Randomly split $\mc{S}$ into two parts,  $\mc{S}_1, \mc{S}_2$, with equal sizes.
            \ENSURE Imputed network $\hat{A}$.
                \STATE \textbf{Step 1: Extract information explained by $X$}  
                \STATE Run dyadic nonparametric regression to obtain $\hat{\Pi}(\cdot, \cdot)$. 
                \STATE \textbf{Step 2: Calculate similarity}  
                \STATE For each $i \in \mc{S}_2 \cup \mc{S}^c$,  and $i'\in \mc{S}_2$, calculate the pseudo distance using observed adjacency matrix: 
                \begin{align*}
                    \hat{d}_{ii'} = \max_{ k\in \{1, \ldots, N\} \backslash \{i, i'\} } \left|\frac{1}{|\mc{S}_1 |} \sum_{\ell \in \mc{S}_1 } A_{k\ell}(A_{i\ell } - A_{i'\ell})\right|.  
                \end{align*}
                \STATE \textbf{Step 3: Leave-one-out cross validation}   
                \STATE For each $h_q \in \{h_1, \ldots, h_Q\}$, and for each $i\in \mc{S}_2$ and $j \in \mc{S}^c$, run the local (weighted) two-way fixed-effect regression:  
                \begin{align*}
                    (\hat{\bs{a}}(h_q), \hat{\bs{b}}(h_q)) \in \argmin_{\bs{a}, \bs{b} \in \mb{R}^{|\mc{S}_2 | }} \sum_{  \substack{i'\in \mc{S}_2,  j'\in \mc{S}_2 \cup\{j\} , \\ (i',j')\neq (i, j) } }  \left(A_{i'j'} - \hat{\Pi}_{ i'j'} - a_{i'} - b_{j'} \right)^2 K_{h_q}\left(\hat{d}_{ii'}\right)K_{h_q}\left(\hat{d}_{jj'}\right)
                \end{align*}
                to obtain $\bar{A}_{ij}(h_q) = \hat{a}_i(h_q) + \hat{b}_j(h_q) + \hat{\Pi}_{X, ij}$.  
                Then the optimal $h^*$ solves 
                \begin{align*}
                    h^* = \argmin_{h_q \in \{h_1, \ldots, h_Q\}} \sum_{i\in \mc{S}_2, j \in \mc{S}^c}\left(A_{ij} - \bar{A}_{ij}(h_q) \right)^2. 
                \end{align*}
                \STATE \textbf{Step 4: Local imputation}   
                \STATE For each $i\in \mc{S}^c $ and $j \in \mc{S}^c$, estimate the local (weighted) two-way fixed-effects regression 
                \begin{align*}
                    (\hat{\bs{a}}, \hat{\bs{b}}) \in \argmin_{\bs{a}, \bs{b} \in \mb{R}^{|\mc{S}_2| + 1}} \sum_{ \substack{i'\in \mc{S}_2 \cup\{i\}   , j'\in \mc{S}_2 \cup\{j\}, \\
                    (i', j')\neq (i, j)}}   \left(A_{i'j'} - \hat{\Pi}_{ i'j'} - a_{i'} - b_{j'} \right)^2 K_{h^*}\left(\hat{d}_{ii'}\right)K_{h^*}\left(\hat{d}_{jj'}\right)
                \end{align*}
                to obtain $\tilde{A}_{ij} = \hat{a}_i + \hat{b}_j + \hat{\Pi}_{ij}$.  By combining the imputed entries with the observed entries of the adjacency matrix, we obtain the (preliminary) imputed adjacency matrix $\tilde{A}$. Truncate $\tilde{A}_{ij}$ to get 
                \begin{align*}
                    \hat{A}_{ij} = \left\{
                        \begin{array}{ll}
                           \tilde{A}_{ij}, & \text{if } \tilde{A}_{ij}\in [0, 1]\\
                            1 & \text{if } \tilde{A}_{ij} > 1 \\
                            0 & \text{if } \tilde{A}_{ij} < 0
                        \end{array}
                    \right., 
                \end{align*}
                and zero the diagonal of $\hat{A}$ such that $\hat{A}_{ii} = 0$ for all $i$. 
        \end{algorithmic}
    \end{algorithm}

\subsection{Dyadic nonparametric regression}\label{appendix:extra_dyadic}

    The extraction of the component of link formation that can be explained by covariates $X$ builds on \citet{graham2021minimax}, who study the asymptotic properties of dyadic nonparametric regression and show that the Nadaraya-Watson estimator achieves the minimax rate. We employ local linear regression in numerical simulations and estimate the model using samples $\{(i, j) \mid i\in \mc{S}_1, \text{or } j\in \mc{S}_1\}$. Specifically, for any $x_1, x_2 \in \mc{X} $, $(\hat{\beta}_0, \hat{\beta}_1)$ solve 
    \begin{align*}
        \argmin_{\substack{ \beta_0\in \mb{R},\\ \beta_1\in \mb{R}^{d_X}} } \sum_{ \substack{ i\in \mc{S}_1, \\ \text{or } j\in \mc{S}_1}} \left(A_{ij} - \beta_0 - \sum_{d = 1}^{d_X} \beta_{1d}\left(\omega(X_{id}, X_{jd}) - \omega(x_{1d}, x_{2d})\right)\right)^2 \prod_{d = 1}^{d_X}K\left(\frac{\omega(X_{id}, X_{jd}) - \omega(x_{1d}, x_{2d})}{h}\right), 
    \end{align*}
    where (i) $K(\cdot)$ is a kernel, (ii) $h>0$ is the bandwidth, and (iii) $\omega(\cdot, \cdot)$ is a user-specified distance function. The nonparametric estimator is given by 
    \begin{align*}
        \hat{\Pi}(x_1, x_2) = \hat{\beta}_0 + \sum_{d=1}^{d_X} \hat{\beta}_{1d} \omega(x_{1d}, x_{2d}). 
    \end{align*}
    
    Our implementation differs from their work in two important respects. First, instead of using the Nadaraya-Watson estimator, we employ local linear regression. Local linear regression is more commonly used in the local smoothing literature and  delivers a second-order approximation error. While in principle one could use higher-order local polynomial regression to obtain higher-order approximation errors, this is not desirable in our setting for two reasons: (i) sample sizes in most applied settings are typically small; and (ii) both the theoretical analysis (see the discussion after Assumption~\ref{assumption:dyadic_nonparametric} and Theorem~\ref{thm:bias_variance}) and the numerical simulations suggest that the estimation error from the dyadic nonparametric regression step is asymptotically negligible relative to that of the local two-way fixed-effects regression.  As a result, reducing the approximation error of the dyadic nonparametric regression step only marginally improves overall performance of the imputation. We therefore restrict attention to local linear regression in our implementation.

    Second, we use homophily effects $\omega(X_i, X_j)$ (defined as in Example~\ref{example:graham}) as the regressor rather than $(X_i, X_j)$. This choice captures homophily in observables in a parsimonious manner and is computationally more efficient. Researchers may prefer different formulations depending on the application.

\subsection{Additional simulations}\label{appendix:extra_simulation}

    In this section, we present (i) simulation results for settings in which the smoothness assumption (Assumption~\ref{assumption:regularity}\ref{item:regularity_assumption_smoothness}) fails, and (ii) the complementary simulation results for the linear-in-means peer-effects model in Section~\ref{sec:simulation}. 

    \bigskip

    The network in the simulation is generated as 
    \begin{equation}\label{eq:MC_formation_absolute}
    \begin{gathered}
        A_{ij} = \bs{1}(\omega(X_i, X_j)'\beta + g(\xi_i, \xi_j) - U_{ij} \geq 0),  \\
        \omega(X_i, X_j)' = \left(|X_{i1} - X_{j1}|, |X_{i2} - X_{j2}|\right), \quad g(\xi_i, \xi_j) = \xi_{i1} +  \xi_{j1} - \frac{1}{8} ( \xi_{i2}  -  \xi_{j2} )^2. 
    \end{gathered}   
    \end{equation}
    The dimensions of the observable and latent characteristics are $d_{X} = d_{\xi} = 2$, and $\{(X_i, \xi_i)\mid i=1,\ldots, N\}$ is i.i.d. across individuals. Covariates and latent factors are generated according to
    \begin{align*}
        X_{id} = \frac{1}{2}(\xi_{i1} + \xi_{i2}) + \epsilon_{X, id}, \quad \forall d=1,2, 
    \end{align*}
    and 
    \begin{align*}
        \xi_i \sim \mc{N}\left( 0, 
            \begin{pmatrix}
                1 & 0 \\
                0 & 1
            \end{pmatrix}\right), 
            \quad
        \epsilon_{X, i1}, \epsilon_{X, i2} \sim \mr{Uniform}([-1, 1]), 
        \quad
        \epsilon_{X, i1}\perp \epsilon_{X, i2}. 
    \end{align*}
    The idiosyncratic shocks $\{U_{ij}\}_{1 \leq i < j \leq N }$ are independently drawn from a standard logistic distribution.  The coefficient vector $\beta$ captures homophily effects on observables and takes values $\beta\in \left\{(-0.5, -0.5), (-2, -2)\right\}$. 

    The network size is $N = 200$, and in the simulations we consider sampling rates $\varphi := n/N \in \{0.2, 0.3, 0.4, 0.5, 0.6, 0.8\}$ to compare the performance of the proposed and alternative methods under different sampling scenarios.   
    We conduct $S = 1,000$ Monte Carlo replications in the numerical experiment and report RMSE defined as in~\eqref{eq:MC_RMSE}. The simulation results are reported in Table~\ref{tab:simulation_imputation_absolute}, with RMSEs presented in units of $0.01$ for ease of reading.

    \begin{table}[h]
    \centering
    \caption{Simulation: Imputation (absolute-distance homophily effects)}\label{tab:simulation_imputation_absolute}%
        \begin{tabular}{cccccccc}
        \toprule
            & \multicolumn{1}{c}{X} & \multicolumn{1}{c}{LR} & \multicolumn{1}{c}{LPCA} & \multicolumn{1}{c}{LTWFE} & \multicolumn{1}{c}{X-LPCA} & \multicolumn{1}{c}{X-LTWFE} & \multicolumn{1}{c}{X-LTWFE-SP} \\
            &   $(\times 0.01)$    &  $(\times 0.01)$     &   $(\times 0.01)$    &   $(\times 0.01)$    &  $(\times 0.01)$     &   $(\times 0.01)$    & $(\times 0.01)$ \\
        \midrule
        $\beta = (-.5, -.5)$ &       &       &       &       &       &       &  \\
        $\varphi = 0.2$ & 22.5    & 14.0  & 14.1  & 13.1  & 15.1  & 10.7  & 13.3 \\
        $\varphi = 0.3$ & 22.4  & 12.2  & 12.3   & 11.6  & 12.0  & 9.2   & 11.3 \\
        $\varphi = 0.4$ & 22.3  & 11.1  & 11.3  & 10.7   & 10.8  & 8.4   & 10.2 \\
        $\varphi = 0.5$ & 22.3  & 10.4  & 10.5  & 10.2  & 9.4   & 7.8   & 9.4 \\
        $\varphi = 0.6$ & 22.2  & 9.8  & 9.9  & 9.8  & 8.7   & 7.4   & 8.9 \\
        $\varphi = 0.8$ & 22.0  & 8.7  & 8.7   & 9.3    & 7.8   & 6.8   & 8.1 \\ 
        $\beta = (-2, -2)$ &       &       &       &       &       &       &  \\
        $\varphi = 0.2$ & 11.9  & 13.2  & 13.4  & 12.7  & 11.1  & 9.2  & 10.0 \\
        $\varphi = 0.3$ & 11.9  & 12.1  & 12.2  & 11.7    & 10.1  & 8.4   & 9.1 \\
        $\varphi = 0.4$ & 11.8  & 11.3  & 11.4  & 11.1  & 9.4  & 7.9   & 8.5 \\
        $\varphi = 0.5$ & 11.8  & 10.6 & 10.7  & 10.7  & 8.8   & 7.5   & 8.1 \\
        $\varphi = 0.6$ & 11.8  & 9.9  & 10.0  & 10.3 & 8.3   & 7.3   & 7.9 \\
        $\varphi = 0.8$ & 11.7  & 8.7   & 8.7   & 9.7    & 7.5   & 6.9   & 7.5 \\
        \bottomrule
        \end{tabular}
        \vspace{0.5cm}
        \begin{minipage}{\textwidth}
            \footnotesize
            \textbf{Note:} We conduct Monte Carlo simulations based on $1,000$ repetitions of the network formation model~\eqref{eq:MC_formation_absolute}. The network size is $N = 200$. We set $\beta\in \left\{(-0.5, -0.5), (-2, -2)\right\}$ and sampling rates $\varphi  \in \{0.2, 0.3, 0.4, 0.5, 0.6, 0.8\}$.  We report RMSE as in~\eqref{eq:MC_RMSE} for the following imputation methods: (1)  covariate-only nonparametric imputation that imputes missing outcomes using $X$ alone (\textbf{X}), (2) 
            the low-rank imputation methods of \citet{bai2021matrix} and \citet{li2023link} (\textbf{LR}), (3)  the local PCA method that ignores covariates (\textbf{LPCA}), (4) the local PCA method using covariates (\textbf{X-LPCA}), (5)  the local two-way fixed-effects regression that ignores covariates (\textbf{LTWFE}), (6) the local two-way fixed-effects regression using covariates (\textbf{X-LTWFE}), and (7) X-LTWFE method with sample-splitting (\textbf{X-LTWFE-SP}). It is important to note that LR, LPCA, LTWFE, X-LPCA, and X-LTWFE are implemented without sample splitting. RMSEs are presented in units of $0.01$ for ease of reading.   
        \end{minipage}
    \end{table}

    The simulations in Table~\ref{tab:simulation_imputation_absolute} display patterns similar to those in Table~\ref{tab:simulation_imputation_basic}, even though the smoothness assumption (Assumption~\ref{assumption:regularity}\ref{item:regularity_assumption_smoothness}) fails. The proposed method outperforms the alternative methods and remains more robust across different levels of network sparsity and sampling rates.

    Table~\ref{tab:simulation_linear_in_means_s1} and \ref{tab:simulation_linear_in_means_s2} report the simulation results for $\alpha_{C}, \alpha_{W,2}, \alpha_{\bar{W},1}, \alpha_{\bar{W},2}$. In most cases, the estimator based on our proposed method (X-LTWFE) performs well, especially when the sampling rate is small. Although LTWFE and X-LPCA may outperform our proposed estimator in some cases when the samplping rate is large ($\varphi \geq 40
    \%$), their performance deteriorates substantially as the sampling rate decreases. In addition, the results in Table~\ref{tab:simulation_linear_in_means_s1} and \ref{tab:simulation_linear_in_means_s2} align with the implications of Theorem~\ref{thm:linear_in_mean}: the bias-to-std ratio of the estimators does not vanish to $0$, which leads to challenges for downstream inference.
    \begin{table}[H]
    \centering
    \caption{Simulation: Linear-in-means peer-effects model (Supplementary results I)}\label{tab:simulation_linear_in_means_s1}
        \begin{tabular}{ccccccc}
            \toprule
            & CD    & X     & LTWFE & X-LPCA & X-LTWFE & X-LTWFE-SP \\
        &       $(\times 0.01)$  &    $(\times 0.01)$   &    $(\times 0.01)$   &   $(\times 0.01)$    &    $(\times 0.01)$   &  $(\times 0.01)$ \\
        \midrule
        \multicolumn{7}{l}{\textbf{Panel A: Estimates $\alpha_{C}$}}\\
        $\varphi = 0.2$ &       &       &       &       &       &  \\
        BIAS  & -1.02 & -2.45 & -22.81 & 20.32 & -13.24 & -10.64 \\
        STD    & 14.39 & 24.71 & 23.96 & 23.69 & 17.81 & 22.13 \\
        $\varphi = 0.3$ &       &       &       &       &       &  \\
        BIAS  & -1.02 & 1.87  & -22.64 & 13.21 & -15.8 & -16.09 \\
        STD    & 14.39 & 20.46 & 18.12 & 18.11 & 15.60  & 17.69 \\
        $\varphi = 0.4$ &       &       &       &       &       &  \\
        BIAS  & -1.02 & 7.24  & -17.02 & 8.38  & -14.66 & -16.43 \\
        STD    & 14.39 & 17.16 & 16.61 & 16.12 & 14.81 & 15.85 \\
        $\varphi = 0.5$ &       &       &       &       &       &  \\
        BIAS  & -1.02 & 10.64 & -11.16 & 5.80   & -12.11 & -14.16 \\
        STD    & 14.39 & 15.40  & 16.21 & 15.56 & 14.56 & 15.11 \\
            &       &       &       &       &       &  \\
        \multicolumn{7}{l}{\textbf{Panel A: Estimates $\alpha_{W,2}$}}\\
        $\varphi = 0.2$ &       &       &       &       &       &  \\
        BIAS  & 0.10   & 13.48 & 43.41 & 19.51 & 19.01 & 20.39 \\
        STD    & 1.57  & 1.96  & 2.24  & 1.87  & 1.77  & 1.92 \\
        $\varphi = 0.3$ &       &       &       &       &       &  \\
        BIAS  & 0.10   & 10.8  & 30.29 & 15.41 & 14.23 & 14.85 \\
        STD    & 1.57  & 1.87  & 2.03  & 1.81  & 1.73  & 1.77 \\
        $\varphi = 0.4$ &       &       &       &       &       &  \\
        BIAS  & 0.10   & 8.24  & 20.3  & 11.4  & 10.23 & 10.47 \\
        STD    & 1.57  & 1.79  & 1.90   & 1.75  & 1.69  & 1.70 \\
        $\varphi = 0.5$ &       &       &       &       &       &  \\
        BIAS  & 0.10   & 5.91  & 12.78 & 7.74  & 6.88  & 6.97 \\
        STD    & 1.57  & 1.73  & 1.77  & 1.69  & 1.66  & 1.65 \\
        \bottomrule
        \end{tabular}\vspace{0.5cm}
        \begin{minipage}{\textwidth}
            \footnotesize
            \textbf{Note:} We conduct Monte Carlo simulations based on $1,000$ repetitions of the network formation model~\eqref{eq:MC_formation}. The network size is $N = 200$. The number of networks is $M = 40$.  We set $\beta = (-0.5, -0.5)$ and sampling rates $\varphi  \in \{0.2, 0.3, 0.4, 0.5\}$.  We report bias and standard deviation of $\hat{\alpha}_{C}$  and $\hat{\alpha}_{W, 2}$ for the following imputation methods: 
                (1) estimates based on complete network data (\textbf{CD}) (2)  covariate-only nonparametric imputation that imputes missing outcomes using $X$ alone (\textbf{X}), (3) the local two-way fixed-effects regression that ignores covariates (\textbf{LTWFE}), (4)  the local PCA method using covariates (\textbf{X-LPCA}), (5)  the local two-way fixed-effects regression using covariates (\textbf{X-LTWFE}), and (6) X-LTWFE method with sample-splitting (\textbf{X-LTWFE-SP}). It is important to note that LTWFE, X-LPCA, and X-LTWFE are implemented without sample splitting. Bias and standard deviation are presented in units of $0.01$ for ease of reading. 
        \end{minipage}
    \end{table}

    \begin{table}[H]
    \centering
   \caption{Simulation: Linear-in-means peer-effects model (Supplementary results II)}\label{tab:simulation_linear_in_means_s2}
        \begin{tabular}{ccccccc}
            \toprule
            & CD    & X     & LTWFE & X-LPCA & X-LTWFE & X-LTWFE-SP \\
        &       $(\times 0.01)$  &    $(\times 0.01)$   &    $(\times 0.01)$   &   $(\times 0.01)$    &    $(\times 0.01)$   &  $(\times 0.01)$ \\
        \midrule
        \multicolumn{7}{l}{\textbf{Panel A: Estimates $\alpha_{\bar{W}, 1}$}}\\
        $\varphi = 0.2$ &       &       &       &       &       &  \\
        BIAS  & -1.54 & -101.43 & -66.82 & -84.64 & -47.86 & -47.89 \\
        STD   & 14.29 & 22.62 & 24.97 & 20.31 & 17.02 & 20.28 \\
        $\varphi = 0.3$ &       &       &       &       &       &  \\
        BIAS  & -1.54 & -92.36 & -34.83 & -58.39 & -35.21 & -36.9 \\
        STD   & 14.29 & 18.70  & 18.08 & 16.21 & 15.23 & 16.69 \\
        $\varphi = 0.4$ &       &       &       &       &       &  \\
        BIAS  & -1.54 & -77.54 & -16.13 & -37.15 & -25.24 & -27.06 \\
        STD   & 14.29 & 15.44 & 16.26 & 14.89 & 14.67 & 15.39 \\
        $\varphi = 0.5$ &       &       &       &       &       &  \\
        BIAS  & -1.54 & -50.38 & -5.88 & -21.92 & -17.05 & -18.76 \\
        STD   & 14.29 & 13.79 & 15.59 & 14.66 & 14.47 & 14.86 \\
            &       &       &       &       &       &  \\
        \multicolumn{7}{l}{\textbf{Panel A: Estimates $\alpha_{\bar{W}, 2}$}}\\
        $\varphi = 0.2$ &       &       &       &       &       &  \\
        BIAS  & -1.69 & -19.57 & -50.39 & 13.96 & -13.15 & 3.66 \\
        STD   & 16.83 & 35.25 & 33.62 & 34.3  & 24.16 & 29.19 \\
        $\varphi = 0.3$ &       &       &       &       &       &  \\
        BIAS  & -1.69 & -12.86 & -32.09 & 3.75  & -16.02 & -5.44 \\
        STD   & 16.83 & 28.67 & 23.96 & 25.42 & 20.40 & 22.46 \\
        $\varphi = 0.4$ &       &       &       &       &       &  \\
        BIAS  & -1.69 & -5.04 & -18.38 & -0.28 & -14.64 & -8.44 \\
        STD   & 16.83 & 23.61 & 20.78 & 21.52 & 18.75 & 19.68 \\
        $\varphi = 0.5$ &       &       &       &       &       &  \\
        BIAS  & -1.69 & 0.78  & -8.47 & -0.07 & -11.43 & -8.04 \\
        STD   & 16.83 & 20.63 & 19.78 & 19.95 & 18.09 & 18.54 \\
        \bottomrule
        \end{tabular}\vspace{0.5cm}
        \begin{minipage}{\textwidth}
            \footnotesize
            \textbf{Note:} We conduct Monte Carlo simulations based on $1,000$ repetitions of the network formation model~\eqref{eq:MC_formation}. The network size is $N = 200$. The number of networks is $M = 40$.  We set $\beta = (-0.5, -0.5)$ and sampling rates $\varphi  \in \{0.2, 0.3, 0.4, 0.5\}$.  We report bias and standard deviation of $\hat{\alpha}_{\bar{W}, 1}$  and $\hat{\alpha}_{\bar{W}, 2}$ for the following imputation methods: 
                (1) estimates based on complete network data (\textbf{CD}) (2)  covariate-only nonparametric imputation that imputes missing outcomes using $X$ alone (\textbf{X}), (3) the local two-way fixed-effects regression that ignores covariates (\textbf{LTWFE}), (4)  the local PCA method using covariates (\textbf{X-LPCA}), (5)  the local two-way fixed-effects regression using covariates (\textbf{X-LTWFE}), and (6) X-LTWFE method with sample-splitting (\textbf{X-LTWFE-SP}). It is important to note that LTWFE, X-LPCA, and X-LTWFE are implemented without sample splitting. Bias and standard deviation are presented in units of $0.01$ for ease of reading. 
        \end{minipage}
    \end{table}%

\subsection{Discussion on Informativeness Condition}\label{appendix:extra_informativeness}

    As discussed in the main text, Assumption~\ref{assumption:informativeness} plays a key role in establishing entrywise bounds for the imputation error. In this subsection, we (i) discuss the relationship between the population pseudo-distance $d(\zeta_i, \zeta_{i'})$ and the squared $L_2$ distance $L_2^2(\zeta_i, \zeta_{i'})$, (ii) provide a set of primitive sufficient conditions for this assumption and verify that it holds under commonly used network formation models, and (iii) present examples in which the condition may fail. While this assumption is particularly convenient for theoretical analysis, it should not be interpreted as necessary for good imputation performance. Even when the informativeness condition fails, our simulation results in Appendix~\ref{appendix:extra_simulation} show that our proposed method  still performs well in practice and outperforms alternative approaches.

\subsubsection{Discussion on $d(\zeta_i, \zeta_{i'})$ and  $L_2^2(\zeta_i, \zeta_{i'})$}
    
    As discussed in the main text, although  $L^2_2(\zeta_i, \zeta_{i'})$ provides a natural proxy for latent similarity, it is not directly estimable. To see this, consider the sample analogue based on observed network links:
    \begin{align*}
        \frac{1}{n}\sum_{\ell = 1}^{n} (A_{i\ell} - A_{i'\ell})^2  = \frac{1}{n} \sum_{\ell = 1}^{n} (f_{i\ell} - f_{i'\ell})^2 + \frac{2}{n}\sum_{\ell = 1}^{n} (f_{i\ell} - f_{i'\ell})  (\epsilon_{i\ell} - \epsilon_{i'\ell}) + \frac{1}{n}\sum_{\ell = 1}^{n} (\epsilon_{i\ell} - \epsilon_{i'\ell})^2 + O\left(1/n\right). 
    \end{align*}
    Here, we have (i)  $\frac{1}{n} \sum_{\ell = 1}^{n} (f_{i\ell} - f_{i'\ell})^2 \cp L_2^2(\zeta_i, \zeta_{i'}) $ by law of large number, (ii) $\frac{2}{n}\sum_{\ell = 1}^{n} (f_{i\ell} - f_{i'\ell})  (\epsilon_{i\ell} - \epsilon_{i'\ell})\cp 0$ by law of large number and the exogeneity of $\{\epsilon_{ij}\}_{1\leq i < j\leq N}$ (see Assumption\ref{assumption:regularity}\ref{item:regularity_assumption_independence}), (iii) $\frac{1}{n}\sum_{\ell = 1}^{n} (\epsilon_{i\ell} - \epsilon_{i'\ell})^2 \cp \mb{E}(\epsilon_{i\ell}^2\mid \zeta_i) + \mb{E}(\epsilon^2_{i'\ell}\mid \zeta_{i'})$ by law of large number, and (iv) the fourth term arises because self-loops are ruled out. Therefore, 
    \begin{align*}
        \frac{1}{n}\sum_{\ell = 1}^{n} (A_{i\ell} - A_{i'\ell})^2 \cp L_2^2(\zeta_i, \zeta_{i'}) + \mb{E}(\epsilon_{i\ell}^2\mid \zeta_i) + \mb{E}(\epsilon^2_{i'\ell}\mid \zeta_{i'}),
    \end{align*}
    and the additional variance terms, $\mb{E}(\epsilon_{i\ell}^2\mid \zeta_i)$ and $\mb{E}(\epsilon^2_{i'\ell}\mid \zeta_{i'})$,  do not vanish asymptotically and therefore contaminate the estimation. In general, these terms depend on latent characteristics and cannot be removed. Only under homoskedasticity can one recover $L_2^2(\zeta_i, \zeta_{i'})$ up to an additive constant. 

    In contrast, $d(\cdot, \cdot)$ admits a consistent estimator based on the observed network and provides an upper bound for $L_2^2(\zeta_i, \zeta_{i'})$. By definition,  for any $\zeta_1, \zeta_2 \in \mr{supp}(\zeta)$, 
    \begin{align*}
        d(\zeta_i, \zeta_{i'}) \geq \left| \int f(\zeta_{1} , \tilde{\zeta}) (f(\zeta_{i} , \tilde{\zeta}) - f(\zeta_{i'}, \tilde{\zeta})) \mr{d}\mb{P}(\tilde{\zeta})\right|, \\
        d(\zeta_i, \zeta_{i'}) \geq \left|\int  f(\zeta_{2} , \tilde{\zeta}) (f(\zeta_{i} , \tilde{\zeta}) - f(\zeta_{i'}, \tilde{\zeta})) \mr{d}\mb{P}(\tilde{\zeta})\right|. 
    \end{align*}
    Taking the average of the two inequalities and applying the triangle inequality yields 
    \begin{align*}
        d(\zeta_i, \zeta_{i'}) \geq & \frac{1}{2} \left(\left| \int f(\zeta_{1} , \tilde{\zeta}) (f(\zeta_{i} , \tilde{\zeta}) - f(\zeta_{i'}, \tilde{\zeta})) \mr{d}\mb{P}(\tilde{\zeta})\right| + \left| \int f(\zeta_{2} , \tilde{\zeta}) (f(\zeta_{i} , \tilde{\zeta}) - f(\zeta_{i'}, \tilde{\zeta})) \mr{d}\mb{P}(\tilde{\zeta})\right|\right) \\
        \geq & \frac{1}{2} \left| \int (f(\zeta_{1} , \tilde{\zeta}) - f(\zeta_{2} , \tilde{\zeta}) ) (f(\zeta_{i} , \tilde{\zeta}) - f(\zeta_{i'}, \tilde{\zeta})) \mr{d}\mb{P}(\tilde{\zeta})\right| , \quad \forall \zeta_1, \zeta_2 \in \mr{supp}(\zeta). 
    \end{align*}
    Taking the supremum over $\zeta_1, \zeta_2 \in \mr{supp}(\zeta)$, we obtain
    \begin{align}\label{eq:verification_transofmation}
        d(\zeta_i, \zeta_{i'}) \geq \sup_{\zeta_1, \zeta_2 \in \mr{supp}(\zeta)} \frac{1}{2} \left| \int (f(\zeta_{1} , \tilde{\zeta}) - f(\zeta_{2} , \tilde{\zeta}) ) (f(\zeta_{i} , \tilde{\zeta}) - f(\zeta_{i'}, \tilde{\zeta})) \mr{d}\mb{P}(\tilde{\zeta})\right|. 
    \end{align} 
    The inequality~\eqref{eq:verification_transofmation} is useful in verifying informativeness condition. A useful implication of this inequality is that 
    \begin{align}\label{eq:verification_transofmation_quadratic}
        d(\zeta_i, \zeta_{i'}) \geq  \frac{1}{2} \left| \int (f(\zeta_{i} , \tilde{\zeta}) - f(\zeta_{i'} , \tilde{\zeta}) )^2 \mr{d}\mb{P}(\tilde{\zeta})\right| = \frac{1}{2} L_2^2(\zeta_i, \zeta_{i'}),   
    \end{align}
    which shows that controlling $d(\zeta_i, \zeta_{i'})$ is sufficient for controlling $L_2^2(\zeta_i, \zeta_{i'})$. 

    The following Lemma shows that, when $\zeta$ has finite support, as in the stochastic block model in Example~\ref{example:SBM}, the informativeness condition is automatically satisfied. 
    \begin{lemma}\label{lemma:SBM_informativeness}
        Let $\zeta_i \in \{\bar{\zeta}_1, \ldots, \bar{\zeta}_G\}$ be i.i.d. random variables. Then Assumption~\ref{assumption:informativeness} holds, except in the degenerate case where there exist $g \neq g'$ such that $f(\bar{\zeta}_g, \zeta) = f(\bar{\zeta}_{g'}, \zeta)$ for all $\zeta \in \{\bar{\zeta}_1, \ldots, \bar{\zeta}_G\}$.
    \end{lemma}
    This lemma shows that when $\zeta$ takes values in a finite set, the informativeness condition is satisfied as long as different $\zeta$ have different linking patterns. Even when two different values of $\zeta$ generate identical link formation probabilities, they can be merged into a single type without affecting the observed linking structure. Therefore, the stochastic block model always satisfies the informativeness condition.

\subsubsection{Sufficient conditions for Assumption~\ref{assumption:informativeness}} 

    While the informativeness condition is automatically satisfied when $\zeta$ has finite support, verifying this condition becomes more challenging when $\zeta$ is multi-dimensional and continuously distributed. 
    In the following proposition, we provide a set of sufficient conditions under which the informativeness condition holds. These conditions require that the graphon $f$ is sufficiently smooth and that each coordinate of $\zeta$ has a non-degenerate marginal effect on link formation.

    \begin{proposition}\label{proposition:verification}
        Under Assumption~\ref{assumption:regularity}, suppose further that:
        \begin{enumerate}[label=(\roman*)]
            \item $\mr{supp}(\zeta)$ is convex and $\mr{affine}(\mr{supp}(\zeta)) = \mb{R}^{d_{\zeta}}$;
            \item for any $\delta >0$ and any $\zeta_{i}, \zeta_{i'}\in  \mr{supp}(\zeta)$ such that $ \| \zeta_{i} - \zeta_{i'} \| \geq \delta$, 
            \begin{align}\label{eq:verification_1}
                \int \left(f(\zeta_{i} , \tilde{\zeta}) - f(\zeta_{i'}, \tilde{\zeta})\right)^2 \mr{d}\mb{P}(\tilde{\zeta}) >0; 
            \end{align} 
            \item there exists a constant $\underline{\lambda}>0$ such that 
            \begin{align}\label{eq:verification_2}
                \inf_{\zeta \in \mr{supp}(\zeta)}   \lambda_{\min} \left(\int   \nabla f(\zeta, \tilde{\zeta}) \nabla' f(\zeta, \tilde{\zeta}) \mr{d}\mb{P}(\tilde{\zeta})\right)  \geq  \underline{\lambda}, 
            \end{align}
            where $\lambda_{\min}(\cdot)$ denotes the smallest singular value of a matrix. 
        \end{enumerate}
        Then Assumption~\ref{assumption:informativeness} holds. 
    \end{proposition}
    
    This proposition states that, under the compact support condition and the second-order smoothness condition imposed in Assumption~\ref{assumption:regularity}, conditions (i)-(iii) imply that the pseudo-distance is informative about latent proximity. Here, condition (i) is a mild regularity for $\zeta$'s support. Condition (ii) is a global identification condition. It states that distinct latent factors, $\zeta_{i}$ and $\zeta_{i'}$, cannot generate identical linking probabilities almost everywhere, which ensures that the $L_2^2$ distance between $f(\zeta_i,\cdot)$ and $f(\zeta_{i'},\cdot)$ is strictly positive whenever $\zeta_i \neq \zeta_{i'}$. This condition is essential, as any two latent factors that generate identical linking probabilities are observationally equivalent and cannot be distinguished using network data. At the same time, it is a mild requirement in practice and is satisfied in most commonly used network formation processes. Condition (iii) can be regarded as a local non-degeneracy condition, which requires that the sensitivity matrix,  $\int \nabla f(\zeta, \tilde{\zeta}) \nabla' f(\zeta, \tilde{\zeta})  \mr{d}\mb{P}(\tilde{\zeta})$,  is uniformly positive definite. Intuitively, this means that variation in every direction of $\zeta$ leads to non-degenerated variation in link formation. 

    \bigskip

    We now use Proposition~\ref{proposition:verification} to verify the informativeness condition for a class of commonly used network formation models.
     \begin{lemma}\label{lemma:verification_sigle_index_quadratic}
        Under Assumption~\ref{assumption:regularity}, suppose that: (i) $\mr{supp}(\zeta)$ is convex, compact, and $\mr{affine}(\mr{supp}(\zeta)) = \mb{R}^{d_{\zeta}}$; (ii) $f(\zeta_i, \zeta_{i'})$ takes the single-index form, i.e., there exists a strictly monotonic transformation $g: \mb{R}\mapsto \mb{R}_{+}$ with $g(\cdot) >0$ such that $f(\zeta_i, \zeta_{i'}) = g\left(\zeta_{i1} + \zeta_{i'1} + \sum_{d=2}^{d_{\zeta}} \beta_d \left(\zeta_{id} - \zeta_{i'd}\right)^2\right)$;
        (iii) $\beta_d \neq 0$ for each $d = 2, \ldots, d_{\zeta}$; and (iv) the covariance matrix of $\left(\zeta_2, \ldots, \zeta_{d_{\zeta}}\right)$ 
        is positive definite. Then, Assumption~\ref{assumption:informativeness} holds.
    \end{lemma}
    
    Lemma~\ref{lemma:verification_sigle_index_quadratic} verifies the informativeness condition for a common class of single-index network formation models. Condition (i) is mild and requires that the support of the latent factors is compact and has full affine dimension, which rules out redundant latent dimensions. Condition (ii) specifies that the network formation process follows a single-index model with a linear index.. We also require that the transformation function $g(\cdot)$ is strictly monotone, as in most standard discrete choice models. Condition (iii) states that if some $\beta_d$ were zero, then the $d$-th coordinate would not affect link formation and could not be recovered from network data. This condition is not restrictive, as such coordinates can be omitted without loss of generality. Condition (iv) is standard and requires sufficient variation in each latent coordinate. This condition is easily satisfied when each dimension is independent and non-degenerate.

    The following corollary follows directly from Lemma~\ref{lemma:verification_sigle_index_quadratic} and shows that, in Example~\ref{example:graham}, the informativeness condition is satisfied when the idiosyncratic shock $U_{ij}$ follows a standard logistic or normal distribution, and the index includes both degree heterogeneity and homophily effects with quadratic distance. 

    \begin{corollary}\label{corollary:logit_informativeness}
        Let $\zeta$ be a $d_{\zeta}$-dimensional random vector, and suppose that the components of $\zeta = (\zeta_{1}, \ldots, \zeta_{d_{\zeta}})'$ are mutually independent with $\mr{supp}(\zeta_{d}) = [a_d, b_d]$ and  $\mr{Var}(\zeta_{d}) > 0$ for each $d$. Then Assumption~\ref{assumption:informativeness} holds when 
        \begin{align*}
            f(\zeta_i, \zeta_{i'})= \frac{\exp\left(\zeta_{i1} + \zeta_{i'1} + \sum_{d=2}^{d_{\zeta}} \beta_{d} (\zeta_{id} - \zeta_{i'd})^2 \right)}{1 + \exp\left(\zeta_{i1} + \zeta_{i'1} + \sum_{d=2}^{d_{\zeta}} \beta_{d} (\zeta_{id} - \zeta_{i'd})^2 \right)}, \quad \beta_2, \ldots, \beta_{d_{\zeta}} \neq 0,
        \end{align*}
        or
        \begin{align*}
            f(\zeta_i, \zeta_{i'}) = \Phi\left(\zeta_{i1} + \zeta_{i'1} + \sum_{d=2}^{d_{\zeta}} \beta_{d} (\zeta_{id} - \zeta_{i'd})^2 \right), \quad \beta_2, \ldots, \beta_{d_{\zeta}} \neq 0,
        \end{align*} 
        where $\Phi(\cdot)$ denotes the cumulative distribution function of the standard normal distribution.
    \end{corollary}

\subsubsection{Other examples} 

    We have discussed how to verify the informativeness condition when the graphon $f(\cdot)$ is twice differentiable. We now turn to the case in which $f$ is not sufficiently smooth. This case is empirically relevant in many applications. For example, in Example~\ref{example:graham}, when homophily effect is modeled using absolute distance rather than quadratic distance, the resulting graphon is no longer differentiable. Verifying the informativeness condition under non-smooth graphon is generally more challenging. We begin by establishing the following lemma, which shows that when $\zeta$ is one-dimensional, the informativeness condition can still be verified even under absolute distance.

    \begin{lemma}\label{lemma:verification_absolute_logit}
        Let $\zeta$ be a random variable  with $\zeta \sim \mr{Uniform}[0, 1]$. Then Assumption~\ref{assumption:informativeness} holds when 
        \begin{align*}
            f(\zeta_i, \zeta_{i'})= \frac{\exp\left( -| \zeta_{i} - \zeta_{i'}| \right)}{1 + \exp\left(-| \zeta_{i} - \zeta_{i'}|  \right)}. 
        \end{align*}
    \end{lemma}
    However, when $\zeta$ is multidimensional, the informativeness condition generally fails when the homophily effect is modeled using absolute distance. The following lemma provides a counterexample.
    \begin{lemma}[Counterexample]\label{lemma:verification_impossible}
        Let $\zeta  $ be a $d_{\zeta}$-dimensional random vector with $d_{\zeta} \geq 2$, and suppose that the components of $\zeta = (\zeta_{1}, \ldots, \zeta_{d_{\zeta}})'$ are mutually independent with $\zeta_{d} \sim \mr{Uniform}[0, 1]$ for each $d$. Then Assumption~\ref{assumption:informativeness} fails when 
        \begin{align*}
            f(\zeta_i, \zeta_{i'})= \frac{\exp\left( -\sum_{d=1}^{d_{\zeta}} | \zeta_{id} - \zeta_{i'd}| \right)}{1 + \exp\left(-\sum_{d=1}^{d_{\zeta}} | \zeta_{id} - \zeta_{i'd}|  \right)}. 
        \end{align*}
    \end{lemma}
    In fact, using a similar argument, one can further show that, in Example~\ref{example:graham}, the informativeness condition fails when the index includes additive degree heterogeneity and homophily effects based on absolute distance. This is also the case considered in the simulations. 

    While the informativeness assumption is convenient for theoretical analysis, it should not be interpreted as necessary for good imputation performance. For example, in related settings, \citet{zhang2017estimating} establish $2,\infty$-norm error bounds for graphon estimation without relying on an informativeness-type condition\footnote{\citet{zhang2017estimating}  study graphon estimation under complete networks, whereas our setting focuses on imputing missing links under incomplete network observations.}. 
    Our simulation results in Appendix~\ref{appendix:extra_simulation} show that, even when the informativeness condition fails, our proposed method still performs well in practice and outperforms alternative approaches.

\clearpage

    \section{Consistency of Imputation}\label{appendix:imputation}

    \begin{prooflmm}{lemma:consistency_hat_d}
        In the proof, we aim to establish 
        \begin{align*}
            \mb{P}\left(\max_{  i \in \mc{S}^c,  i' \in \mc{S}_2 } \left| \hat{d}_{ii'} - d(\zeta_i, \zeta_{i'}) \right| \geq \gamma_1 \frac{\log N }{N^{1/d_{\zeta}}} + \gamma_2 \frac{\log N }{\sqrt{n}}   \right) \leq 1- n^{-\frac{1}{2}}. 
        \end{align*}
        In fact, we can show that for any $\delta>0$, there exist corresponding constants $\gamma_1, \gamma_2>0$ such that  
        \begin{align*}
            \mb{P}\left(\max_{  i \in \mc{S}^c,  i' \in \mc{S}_2 } \left| \hat{d}_{ii'} - d(\zeta_i, \zeta_{i'}) \right| \geq \gamma_1 \frac{\log N }{N^{1/d_{\zeta}}} + \gamma_2 \frac{\log N }{\sqrt{n}}   \right) \leq 1- \delta n^{-\frac{1}{2}}.
        \end{align*}
        In the following proof, we use $\delta$ to denote a generic positive constant that may vary from line to line and absorbs multiplicative or additive constants, with the associated constants $\gamma_1$ and $\gamma_2$ adjusted accordingly. 

        Recall that for any $i\in \mc{S}^c, i' \in \mc{S}_2$, 
        \begin{align*}
            \hat{d}_{ii'} = \max_{ k\in \{1, \ldots, N\}\backslash\{i, i'\} } \left|\frac{1}{|\mc{S}_1| }\sum_{\ell \in \mc{S}_1} A_{k\ell}(A_{i\ell} - A_{i'\ell})\right|. 
        \end{align*}
        The population counterpart of $\hat{d}_{ii'}$ is  
        \begin{align*}
            d(\zeta_i, \zeta_{i'}) := \sup_{\zeta \in \mr{supp}(\zeta) } \left| \int f(\zeta, \tilde{\zeta})  (f(\zeta_i, \tilde{\zeta}) - f(\zeta_{i'}, \tilde{\zeta})  ) \mr{d}\mb{P}(\tilde{\zeta})\right|. 
        \end{align*}
        We define  
        \begin{align*}
            \tilde{d}_{ii'} := \max_{k\in \{1, \ldots, N\}\backslash\{i, i'\} } \left|\frac{1}{|\mc{S}_1 |} \sum_{\ell \in \mc{S}_1\backslash \{k\} } f(\zeta_k, \zeta_\ell)  (f(\zeta_i, \zeta_\ell) - f(\zeta_{i'}, \zeta_\ell)  ) \right|. 
        \end{align*}
        Then, by the triangle inequality, it follows that 
        \begin{align*}
            \max_{i \in \mc{S}^c, i'\in \mc{S}_2 } \left| \hat{d}_{ii'} - d(\zeta_i, \zeta_{i'})\right| \leq \max_{i \in \mc{S}^c, i'\in \mc{S}_2 } \left| \hat{d}_{ii'} - \tilde{d}_{ii'}\right| +  \max_{i \in \mc{S}^c, i'\in \mc{S}_2 } \left| \tilde{d}_{ii'} - d(\zeta_i, \zeta_{i'})\right|. 
        \end{align*}
        \paragraph{Bound on $\max_{i \in \mc{S}^c, i'\in \mc{S}_2 } \left| \hat{d}_{ii'} - \tilde{d}_{ii'}\right|$}       For any $i, i'$ and any $k\in \{1, \ldots, N\}\backslash\{i, i'\}$, by the reverse triangle inequality and the triangle inequality, it follows that, 
        \begin{align*}
            & \left|  \left|\frac{1}{|\mc{S}_1 |} \sum_{\ell \in \mc{S}_1  } A_{k\ell}  (A_{i \ell}  - A_{i' \ell}   ) \right|  -    \left|\frac{1}{|\mc{S}_1|} \sum_{\ell \in \mc{S}_1\backslash \{k\} }  f(\zeta_k, \zeta_\ell)  (f(\zeta_i, \zeta_\ell) - f(\zeta_{i'}, \zeta_\ell)  ) \right|\right| \\
            \leq &    \left|\frac{1}{|\mc{S}_1|} \sum_{\ell \in \mc{S}_1 \backslash \{ k\} }  \epsilon_{k \ell}  (f(\zeta_i, \zeta_\ell) - f(\zeta_{i'}, \zeta_\ell)  ) \right|  +  \left|\frac{1}{|\mc{S}_1 |} \sum_{\ell \in \mc{S}_1 \backslash \{ k\}}  f(\zeta_k, \zeta_\ell)  (\epsilon_{i \ell} - \epsilon_{i' \ell}  ) \right| +   \left|\frac{1}{|\mc{S}_1 |} \sum_{\ell \in \mc{S}_1 \backslash \{ k\}}  \epsilon_{k \ell}  (\epsilon_{i \ell} - \epsilon_{i'\ell}  )\right|. 
        \end{align*}
          
        \begin{align*}
            \max_{i \in \mc{S}^c, i'\in \mc{S}_2 } \left| \hat{d}_{ii'} - \tilde{d}_{ii'}\right| 
            \leq &  \underbrace{\max_{i \in \mc{S}^c, i'\in \mc{S}_2 } \max_{k\in \{1, \ldots, N\}\backslash\{i, i'\} }  \left|\frac{1}{|\mc{S}_1|} \sum_{\ell \in \mc{S}_1 \backslash \{ k\} }  \epsilon_{k \ell}  (f(\zeta_i, \zeta_\ell) - f(\zeta_{i'}, \zeta_\ell)  ) \right|}_{B_1}  \\
            & + \underbrace{\max_{i \in \mc{S}^c, i'\in \mc{S}_2 } \max_{k\in \{1, \ldots, N\}\backslash\{i, i'\} } \left|\frac{1}{|\mc{S}_1 |} \sum_{\ell \in \mc{S}_1 \backslash \{ k\}}  f(\zeta_k, \zeta_\ell)  (\epsilon_{i \ell} - \epsilon_{i' \ell}  ) \right|}_{B_2}  \\
            & +   \underbrace{\max_{i \in \mc{S}^c, i'\in \mc{S}_2 } \max_{k\in \{1, \ldots, N\}\backslash\{i, i'\}} \left|\frac{1}{|\mc{S}_1 |} \sum_{\ell \in \mc{S}_1 \backslash \{ k\}}  \epsilon_{k \ell}  (\epsilon_{i \ell} - \epsilon_{i'\ell}  )\right|}_{B_3}. 
        \end{align*}
        Since the support of $\zeta$ is compact (Assumption~\ref{assumption:regularity}\ref{item:regularity_assumption_compact_support}) and the graphon $f$ is smooth (Assumption~\ref{assumption:regularity}\ref{item:regularity_assumption_smoothness}), $f$ is bounded. Thus, there exists a constant $\rho_f$ such that $\|f\|_{\infty} \leq \rho_f$. 
        Since the random variables $\{\zeta_i\}_{i=1}^N$ and $\{\epsilon_{ij}\}_{1\leq i<j\leq N}$ are independent (Assumption~\ref{assumption:regularity}\ref{item:regularity_assumption_independence}), by Bernstein's inequality (Lemma~\ref{lemma:Bernstein}), we obtain that, for any $t>0$,
        \begin{align*}
            \mb{P}\left(B_1 \geq t \mid \{\zeta_i\}_{i=1}^{N}\right) \leq & \mb{P}\left( \bigcup_{i \in \mc{S}^c, i'\in \mc{S}_2 }\bigcup_{k\in \{1, \ldots, N\}\backslash\{i, i'\}} \left\{\left|\frac{1}{|\mc{S}_1|} \sum_{\ell \in \mc{S}_1 \backslash \{ k\} }  \epsilon_{k \ell}  (f(\zeta_i, \zeta_\ell) - f(\zeta_{i'}, \zeta_\ell)  ) \right| \geq t \right\} \mid \{\zeta_i\}_{i=1}^{N}  \right) \\
            \leq & \sum_{i \in \mc{S}^c, i'\in \mc{S}_2 } \sum_{ k\in \{1, \ldots, N\}\backslash\{i, i'\} } \mb{P}\left( \left|\frac{1}{|\mc{S}_1|} \sum_{\ell \in \mc{S}_1 \backslash \{ k\} }  \epsilon_{k \ell}  (f(\zeta_i, \zeta_\ell) - f(\zeta_{i'}, \zeta_\ell)  ) \right| \geq t  \mid \{\zeta_i\}_{i=1}^{N} \right) \\
            \leq & N^3 \sup_{\substack{ i \in \mc{S}^c, i'\in \mc{S}_2, \\ k\in \{1, \ldots, N\}\backslash\{i, i'\} }}\mb{P}\left( \left|\frac{1}{|\mc{S}_1|} \sum_{\ell \in \mc{S}_1 \backslash \{ k\} }  \epsilon_{k \ell}  (f(\zeta_i, \zeta_\ell)  - f(\zeta_{i'}, \zeta_\ell)  ) \right| \geq t  \mid \{\zeta_i\}_{i=1}^{N} \right) \\
            \leqtext{(i)}  & 2 N^3 \exp\left(-\frac{n t^2}{16\sigma^2 \rho_f^2 + 8 \sigma \rho_f t }\right),  
        \end{align*}
        where we employ $|\mc{S}_1| = |\mc{S}_2| = n/2$ to obtain (ii). Since the inequality above holds for any realization of $\{\zeta_i\}_{i=1}^{N}$, we can drop the conditioning and obtain $\mb{P}\left(B_1 \geq t \right) \leq 2 N^3 \exp\left(-\frac{n t^2}{16\sigma^2 \rho_f^2 + 8 \sigma \rho_f t }\right) $.  In addition, since the random variables $\{\zeta_i\}_{i=1}^N$ and $\{\epsilon_{ij}\}_{1\leq i<j\leq N}$ are independent (Assumption~\ref{assumption:regularity}\ref{item:regularity_assumption_independence}), by Bernstein's inequality (Lemma~\ref{lemma:Bernstein}), we obtain that, for any $t>0$,
        \begin{align*}
            \mb{P}\left(B_2 \geq t  \mid \{\zeta_i\}_{i=1}^{N} \right) \leq & \mb{P}\left( \bigcup_{i \in \mc{S}^c, i'\in \mc{S}_2 }\bigcup_{k\in \{1, \ldots, N\}\backslash\{i, i'\}} \left\{\left|\frac{1}{|\mc{S}_1 |} \sum_{\ell \in \mc{S}_1 \backslash \{ k\}}  f(\zeta_k, \zeta_\ell)  (\epsilon_{i \ell} - \epsilon_{i' \ell}  ) \right|\geq t \right\}  \mid \{\zeta_i\}_{i=1}^{N} \right) \\
            \leq & \sum_{i \in \mc{S}^c, i'\in \mc{S}_2 } \sum_{k\in \{1, \ldots, N\}\backslash\{i, i'\} } \mb{P}\left( \left|\frac{1}{|\mc{S}_1 |} \sum_{\ell \in \mc{S}_1 \backslash \{ k\}}  f(\zeta_k, \zeta_\ell)  (\epsilon_{i \ell} - \epsilon_{i' \ell}  ) \right|\geq t     \mid \{\zeta_i\}_{i=1}^{N} \right) \\
            \leq & N^3 \sup_{\substack{ i \in \mc{S}^c, i'\in \mc{S}_2, \\ k\in \{1, \ldots, N\}\backslash\{i, i'\} }} \mb{P}\left(\left|\frac{1}{|\mc{S}_1 |} \sum_{\ell \in \mc{S}_1 \backslash \{ k\}}  f(\zeta_k, \zeta_\ell)  (\epsilon_{i \ell} - \epsilon_{i' \ell}  ) \right|\geq t   \mid \{\zeta_i\}_{i=1}^{N} \right) \\
            \leq & 2 N^3 \exp\left(-\frac{n t^2}{16\sigma^2 \rho_f^2 + 8 \sigma \rho_f t }\right).   
        \end{align*}
        Similarly, we drop the conditioning and obtain $\mb{P}\left(B_2 \geq t \right) \leq 2 N^3 \exp\left(-\frac{n t^2}{16\sigma^2 \rho_f^2 + 8 \sigma \rho_f t }\right)$,  because the inequality above holds for any realization of $\{\zeta_i\}_{i=1}^{N}$. 
        Based on the similar argument, we have that, for any $t>0$, 
        \begin{align*}
            \mb{P}\left(B_3 \geq t \right) \leq & \mb{P}\left( \bigcup_{i \in \mc{S}^c, i'\in \mc{S}_2 }\bigcup_{k\in \{1, \ldots, N\}\backslash\{i, i'\}} \left\{\left|\frac{1}{|\mc{S}_1 |} \sum_{\ell \in \mc{S}_1 \backslash \{ k\}}  \epsilon_{k\ell}  (\epsilon_{i \ell} - \epsilon_{i' \ell}  ) \right|\geq t \right\}  \right) \\
            \leq & \sum_{i \in \mc{S}^c, i'\in \mc{S}_2 } \sum_{k\in \{1, \ldots, N\}\backslash\{i, i'\}} \mb{P}\left( \left|\frac{1}{|\mc{S}_1 |} \sum_{\ell \in \mc{S}_1 \backslash \{ k\}}   \epsilon_{k\ell}  (\epsilon_{i \ell} - \epsilon_{i' \ell}  ) \right|\geq t    \right) \\
            \leq & N^3 \sup_{\substack{ i \in \mc{S}^c, i'\in \mc{S}_2, \\ k\in \{1, \ldots, N\}\backslash\{i, i'\} }} \mb{P}\left(\left|\frac{1}{|\mc{S}_1 |} \sum_{\ell \in \mc{S}_1 \backslash \{ k\}}   \epsilon_{k\ell}  (\epsilon_{i \ell} - \epsilon_{i' \ell}  ) \right|\geq t   \right) \\
            \leq & 2 N^3 \exp\left(-\frac{n t^2}{16\sigma^4 + 8 \sigma^2 t }\right).    
        \end{align*}
        Combining the probability bounds for $B_1$, $B_2$,  and $B_3$, we conclude that there exists a sufficiently large constant $C_1>0$ such that, when $t = C_1 \frac{\log N }{\sqrt{n}}$,   
        \begin{align*}
            \mb{P}\left(\max_{i \in \mc{S}^c, i'\in \mc{S}_2 }  \left| \hat{d}_{ii'} - \tilde{d}_{ii'}\right|  \geq  C_1 \frac{ \log N }{\sqrt{n}}  \right) \leq  \delta n^{-1/2}. 
        \end{align*}

        \paragraph{Bound on $\max_{i \in \mc{S}^c, i'\in \mc{S}_2 } \left| \tilde{d}_{ii'} - d(\zeta_i, \zeta_{i'})\right|$} For any $i\in \mc{S}^c$ and $ i' \in \mc{S}_2$,  
        \begin{align*}
            \left| \tilde{d}_{ii'} - d(\zeta_i, \zeta_{i'})\right|  \leq &    \max_{k\in \{1, \ldots, N\}\backslash\{i, i'\} } \left|\frac{1}{|\mc{S}_1 |} \sum_{\ell \in \mc{S}_1 \backslash \{k\}}  f(\zeta_k, \zeta_\ell)  (f(\zeta_i, \zeta_\ell) - f(\zeta_{i'}, \zeta_\ell)  ) -  \int  f(\zeta_k, \tilde{\zeta})  (f(\zeta_i, \tilde{\zeta}) - f(\zeta_{i'}, \zeta_\ell)  ) \mr{d}\mb{P}(\tilde{\zeta})\right|  \\
            & +   \sup_{\zeta \in \mr{supp}(\zeta) } \left| \int  f(\zeta, \tilde{\zeta})  (f(\zeta_i, \tilde{\zeta}) - f(\zeta_{i'}, \tilde{\zeta})  ) \mr{d}\mb{P}(\tilde{\zeta})\right|   \\
            & - \max_{k\in \{1, \ldots, N\}\backslash\{i, i'\} }   \left| \int  f(\zeta_k, \tilde{\zeta})  (f(\zeta_i, \tilde{\zeta}) - f(\zeta_{i'}, \tilde{\zeta})  ) \mr{d}\mb{P}(\tilde{\zeta})\right|. 
        \end{align*}
        It follows that $ \max_{i \in \mc{S}^c, i'\in \mc{S}_2 }\left| \tilde{d}_{ii'} - d(\zeta_i, \zeta_{i'})\right| $ is bounded by  
        \begin{align*}
            & \underbrace{\max_{i \in \mc{S}^c, i'\in \mc{S}_2 } \max_{k\in \{1, \ldots, N\}\backslash\{i, i'\} } \left|\frac{1}{|\mc{S}_1 |} \sum_{\ell\in \mc{S}_1}  f(\zeta_k, \tilde{\zeta})  (f(\zeta_i, \tilde{\zeta}) - f(\zeta_{i'}, \tilde{\zeta})  ) -  \int f(\zeta_k, \tilde{\zeta})  (f(\zeta_i, \tilde{\zeta}) - f(\zeta_{i'}, \tilde{\zeta})  ) \mr{d}\mb{P}(\tilde{\zeta})\right|}_{Q_1}  + \frac{4\rho_f^2}{n} \\
            & + \underbrace{\max_{i \in \mc{S}^c, i'\in \mc{S}_2 } \left(\sup_{\zeta \in \mr{supp}(\zeta)}  \left| \int f(\zeta, \tilde{\zeta})  (f(\zeta_i, \tilde{\zeta}) - f(\zeta_{i'}, \tilde{\zeta})  ) \mr{d}\mb{P}(\tilde{\zeta})\right|   - \max_{k\in \{1, \ldots, N\}\backslash\{i, i'\}} \left| \int  f(\zeta_k, \tilde{\zeta})  (f(\zeta_i, \tilde{\zeta}) - f(\zeta_{i'}, \tilde{\zeta})  ) \mr{d}\mb{P}(\tilde{\zeta})\right| \right)}_{Q_2}. 
        \end{align*}
        The term $\frac{4\rho_f^2}{n}$ arises from the fact that $A_{ii} = 0$ instead of $f_{ii} + \epsilon_{ii}$. 
        Note that for any $t >0$, $\mb{P}\left(Q_1 \geq t \right)$ can be rewritten and bounded by 
        \begin{align*}
             & \mb{P}\left( \bigcup_{ \substack{i \in \mc{S}^c, \\ i'\in \mc{S}_2 } }\bigcup_{k\in \{1, \ldots, N\}\backslash\{i, i'\}} \left\{\left|\frac{1}{|\mc{S}_1 |}   \sum_{\ell\in \mc{S}_1}  f(\zeta_k, \zeta_\ell)  (f(\zeta_i, \zeta_\ell) - f(\zeta_{i'}, \zeta_\ell)  ) - \mb{E}(f(\zeta_k, \zeta_\ell)  (f(\zeta_i, \zeta_\ell) - f(\zeta_{i'}, \zeta_\ell)  ) \mid \zeta_i, \zeta_{i'}, \zeta_{k})\right|\geq t \right\}  \right) \\
             \leq & \sum_{ \substack{i \in \mc{S}^c, \\ i'\in \mc{S}_2 } }\sum_{k\in \{1, \ldots, N\}\backslash\{i, i'\}} \mb{P}\left(  \left|\frac{1}{|\mc{S}_1 |}   \sum_{\ell\in \mc{S}_1}  f(\zeta_k, \zeta_\ell)  (f(\zeta_i, \zeta_\ell) - f(\zeta_{i'}, \zeta_\ell)  ) - \mb{E}(f(\zeta_k, \zeta_\ell)  (f(\zeta_i, \zeta_\ell) - f(\zeta_{i'}, \zeta_\ell)  ) \mid \zeta_i, \zeta_{i'}, \zeta_{k})\right|\geq t   \right) \\
             \leq & N^3 \sup_{\substack{ i \in \mc{S}^c, i'\in \mc{S}_2, \\ k\in \{1, \ldots, N\}\backslash\{i, i'\} }} \mb{P}\left(  \left|\frac{1}{|\mc{S}_1 |}   \sum_{\ell\in \mc{S}_1}  f(\zeta_k, \zeta_\ell)  (f(\zeta_i, \zeta_\ell) - f(\zeta_{i'}, \zeta_\ell)  ) - \mb{E}(f(\zeta_k, \zeta_\ell)  (f(\zeta_i, \zeta_\ell) - f(\zeta_{i'}, \zeta_\ell)  ) \mid \zeta_i, \zeta_{i'}, \zeta_{k})\right|\geq t   \right) \\
             \leqtext{(i)} & 2 N^3 \exp\left(-\frac{n t^2}{16\rho_f^4 + 8  \rho^2_f t }\right).
        \end{align*}
        Here, since the random variables $\{\zeta_i\}_{i=1}^N$  are independent (Assumption~\ref{assumption:regularity}\ref{item:regularity_assumption_independence}), we employ Bernstein's inequality (Lemma~\ref{lemma:Bernstein}) to obtain inequality (i).  
        Then, there exists a sufficiently large constant $C_2>0$ such that, when $t = C_2 \frac{\log N }{\sqrt{n}}$, we have  
        \begin{align*}
            \mb{P}\left(Q_1 \geq C_2 \frac{ \log N }{\sqrt{n}}  \right)  \leq  \delta n^{-1/2}. 
        \end{align*}
        To establish a bound for $Q_2$, let $\zeta^*$ be a maximizer of $d(\zeta_i, \zeta_{i'})$ such that  
        \begin{align*}
            d(\zeta_i, \zeta_{i'}) = \left| \int  f(\zeta^*, \tilde{\zeta})  (f(\zeta_i, \tilde{\zeta}) - f(\zeta_{i'}, \tilde{\zeta})  ) \mr{d}\mb{P}(\tilde{\zeta})\right|. 
        \end{align*}
        Then, we have  
        \begin{align*}
            Q_2 & \leq \max_{i \in \mc{S}^c, i'\in \mc{S}_2 } \left( \left| \int f(\zeta^*, \tilde{\zeta})  (f(\zeta_i, \tilde{\zeta}) - f(\zeta_{i'}, \tilde{\zeta})  ) \mr{d}\mb{P}(\tilde{\zeta})\right| - \min_{k\in \{1, \ldots, N\}\backslash\{i, i'\}}\left| \int f(\zeta_k, \tilde{\zeta})  (f(\zeta_i, \tilde{\zeta}) - f(\zeta_{i'}, \tilde{\zeta})  ) \mr{d}\mb{P}(\tilde{\zeta})\right| \right) \\
            & \leq \max_{i \in \mc{S}^c, i'\in \mc{S}_2 } \min_{k\in \{1, \ldots, N\}\backslash\{i, i'\}}  \left| \int (f(\zeta^*, \tilde{\zeta}) -  f(\zeta_k, \tilde{\zeta}))(f(\zeta_i, \tilde{\zeta}) -  f(\zeta_{i'}, \tilde{\zeta}))\mr{d}\mb{P}(\tilde{\zeta}) \right|  \\
             & \leqtext{(i)}  \max_{i \in \mc{S}^c, i'\in \mc{S}_2 } \min_{k\in \{1, \ldots, N\}\backslash\{i, i'\}} 2 \rho_f L \sum_{d=1}^{d_{\zeta}} | \zeta^*_{d} - \tilde{\zeta}_{d}|\\
            & \leq  \max_{i \in \mc{S}^c, i'\in \mc{S}_2 } \min_{k\in \{1, \ldots, N\}\backslash\{i, i'\}} 2 \rho_f L d_{\zeta}^{1/2} \|\zeta^* - \zeta_k\| \\
            & \leq 2 \rho_f L d_{\zeta}^{1/2}  \max_{i \in \mc{S}^c, i'\in \mc{S}_2 } \sup_{\zeta \in \mr{supp}(\zeta)} \min_{k\in \{1, \ldots, N\}\backslash\{i, i'\}} \|\zeta - \zeta_k\|. 
        \end{align*}
        Here, inequality (i) follows from Assumption~\ref{assumption:regularity}\ref{item:regularity_assumption_smoothness}, which states that $f$ is  Lipschitz with constant $L>0$. 
        By Assumption~\ref{assumption:regularity}\ref{item:regularity_assumption_compact_support}, there exists a $t$-covering of the support of $\zeta$ consisting of $\mc{J} \leq 4^{d_{\zeta}}d_{\zeta}^{d_{\zeta}/2}t^{-d_{\zeta}}$ open balls with radius $t$, denoted by $\{\bar{\zeta}_j\}_{j=1}^{\mc{J}}$. Therefore,  
        \begin{align*}
            \sup_{\zeta\in \mr{supp}(\zeta) }  \min_{k\in \{1, \ldots, N\}\backslash\{i, i'\} }   \|\zeta - \zeta_k\| \leq \max_{j=1,\ldots, \mc{J}}  \min_{k\in \{1, \ldots, N\}\backslash\{i, i'\} }  \|\bar{\zeta}_j - \zeta_k\|  + t. 
        \end{align*}
        It follows that, for any $t \leq (2\underline{c})^{1/d_{\zeta}}$,  
        \begin{align*}
            \mb{P}\left( \max_{i \in \mc{S}^c, i'\in \mc{S}_2 }  \sup_{\zeta \in \mr{supp}(\zeta) } \min_{ k\in \{1, \ldots, N\}\backslash\{i, i'\}  } \|\zeta - \zeta_k\|  \geq 2 t  \right) & \leq \mb{P}\left( \bigcup_{i \in \mc{S}^c, i'\in \mc{S}_2 }  \left\{\sup_{\zeta \in \mr{supp}(\zeta) } \min_{ k\in \{1, \ldots, N\}\backslash\{i, i'\}  } \|\zeta - \zeta_k\|  \geq 2 t \right\} \right)\\
            & \leq \sum_{i \in \mc{S}^c, i'\in \mc{S}_2 } \mb{P}\left(   \sup_{\zeta \in \mr{supp}(\zeta) } \min_{ k\in \{1, \ldots, N\}\backslash\{i, i'\} } \|\zeta - \zeta_k\|  \geq 2 t  \right)\\
            & \leq N^2 \mb{P}\left(   \sup_{\zeta \in \mr{supp}(\zeta) } \min_{ k\in \{1, \ldots, N\}\backslash\{i, i'\}  } \|\zeta - \zeta_k\|  \geq 2 t  \right)\\
             & \leq N^2 \mb{P}\left(\max_{j=1,\ldots, \mc{J}} \min_{ k\in \{1, \ldots, N\}\backslash\{i, i'\}} \|\bar{\zeta}_j - \zeta_k\|  \geq t  \right) \\
            & \leq N^2 \mc{J}\mb{P}\left( \min_{ k\in \{1, \ldots, N\}\backslash\{i, i'\} } \|\bar{\zeta}_j - \zeta_k\|  \geq t  \right) \\
            & \eqtext{(i)} N^2 \mc{J} \left(\mb{P}\left(  \|\bar{\zeta}_j - \zeta\| \geq t\right)\right)^{N-2} \\
            & \leqtext{(ii)}  4 N^2 \mc{J} \left(1-\underline{c}t^{d_{\zeta}}\right)^N \\
            & \leq 4^{d_{\zeta} + 1}d_{\zeta}^{d_{\zeta}/2}  N^2 t^{-d_{\zeta}} \left(1-\underline{c}t^{d_{\zeta}}\right)^N, 
        \end{align*}
        where equality (i) follows from the independence of $\{\zeta_i\mid i=1, \ldots, N\}$ (Assumption~\ref{assumption:regularity}\ref{item:regularity_assumption_independence}), and inequality $(ii)$ follows from Assumption~\ref{assumption:regularity}\ref{item:regularity_assumption_compact_support}. 
        Therefore, there exists a constant $C_3>0$ such that, when $t = C_3 \frac{ \log N}{N^{1/d_{\zeta}}}$, 
        \begin{align*}
            \mb{P}\left(Q_2  \geq C_3 \frac{ \log N}{N^{1/d_{\zeta}}} \right)  \leq \delta  n^{-1/2}. 
        \end{align*}
        Combining the above probability bounds and letting $\gamma_1 = C_3$ and $\gamma_2  = C_1 + C_2$, we obtain   
        \begin{align*}
            \mb{P}\left(\max_{i \in \mc{S}^c, i'\in \mc{S}_2 } \left| \hat{d}_{ii'} - d(\zeta_i, \zeta_{i'}) \right| \geq \gamma_1  \frac{ \log N}{N^{1/d_{\zeta}}} + \gamma_2 \frac{\log N}{\sqrt{n}}      + \frac{4\rho_f^2}{n}\right) \leq \delta n^{-1/2}. 
        \end{align*}
        Since $\delta$ is a generic positive constant, we set $\delta = 1$ to complete the proof of Lemma~\ref{lemma:consistency_hat_d}. 
        
        Finally, since $\frac{4\rho_f^2}{n}$ is asymptotically negligible relative to $\gamma_2 \frac{\log N}{\sqrt{n}} $, it is sufficient to focus on the event 
        \begin{equation}\label{def:consistency_hat_d}
            \mc{A}:= \left\{\max_{i \in \mc{S}^c, i'\in \mc{S}_2 } \left| \hat{d}_{ii'} - d(\zeta_i, \zeta_{i'}) \right| \geq \gamma_1  \frac{ \log N}{N^{1/d_{\zeta}}} + \gamma_2 \frac{\log N}{\sqrt{n}} \right\} = \left\{\max_{i \in \mc{S}^c, i'\in \mc{S}_2 } \left| \hat{d}_{ii'} - d(\zeta_i, \zeta_{i'}) \right| \geq \delta_{N, n}    \right\} . 
        \end{equation} 
    \end{prooflmm}

    \begin{lemma}\label{lemma:kernel_hat_d}
        Conditional on the event $\mc{A}$ defined in \eqref{def:consistency_hat_d},  for any  $h>0$ such that $\delta_{N, n}/h\rightarrow 0$, and for all 
        $i \in \mc{S}^c,  i' \in \mc{S}_2$, if $\hat{d}_{ii'} \leq h$, then there exists a constant $\eta > 0$ such that
        \begin{align*}
            \|\zeta_i -\zeta_{i'} \| \leq \eta h. 
        \end{align*}
    \end{lemma}
    \begin{prooflmm}{lemma:kernel_hat_d}
        Note that for all $i \in \mc{S}^c$ and $ i'\in \mc{S}_2$, since $\delta_{N, n}/h\rightarrow 0$, there exists a constant $C_1>1$ such that 
        \begin{align*}
            \left\{\hat{d}_{ii'} \leq h  \right\} \subseteq \left\{d(\zeta_i, \zeta_{i'})  \leq h + \delta_{N, n} \right\} \subseteq \left\{d(\zeta_i, \zeta_{i'})  \leq C_1 h \right\}. 
        \end{align*}
        Also, by Assumption~\ref{assumption:informativeness}, we have 
        \begin{align*}
            \left\{d(\zeta_i, \zeta_{i'})  \leq C_1 h \right\} \subseteq \left\{\|\zeta_i - \zeta_{i'}\|  \leq \lambda^{-1} C_1 h \right\}. 
        \end{align*}
        Let $\eta = \lambda^{-1} C_1$. This completes the proof. 
    \end{prooflmm}

    \begin{proofthm}{thm:bias_variance} 
        In the proof, we establish that~\eqref{eq:thm:bias_variance:bvbound} holds on the event $\mc{A}$ (defined in~\eqref{def:consistency_hat_d}, with $\mb{P}(\mc{A}^c)\leq n^{-1/2}$ by Lemma~\ref{lemma:consistency_hat_d}). The probability bound~\eqref{eq:thm:bias_variance:pbound} then follows directly from the same proof strategy. Since,  under Assumption~\ref{assumption:dyadic_nonparametric}, the bias and variance of the dyadic nonparametric estimator $\hat{\Pi}(\cdot, \cdot)$ achieve faster convergence rates than those characterized in Theorem~\ref{thm:bias_variance}, we ignore the first step for analytical convenience and focus on 
        \begin{equation}\label{eq:sp:Aij_only}
        \begin{gathered}
            (\hat{\bs{a}}, \hat{\bs{b}}) \in \argmin_{\bs{a}, \bs{b} \in \mb{R}^{|\mc{S}_2| + 1}} \sum_{ \substack{i'\in \mc{S}_2 \cup\{i\}   , j'\in \mc{S}_2 \cup\{j\}, \\
            (i', j')\neq (i, j)}}   \left(A_{i'j'} - a_{i'} - b_{j'} \right)^2 K_{h}\left(\hat{d}_{ii'}\right)K_{h}\left(\hat{d}_{jj'}\right), \\
            \hat{A}_{ij} = \hat{a}_i + \hat{b}_j.  
        \end{gathered}
        \end{equation} 
        In the following proof, we work with this formulation. It is easy to verify that the estimators in~\eqref{eq:sp:Aij} and~\eqref{eq:sp:Aij_only} share the same asymptotic properties. 

        Fix $i, j \in \mc{S}^c$. For any $i'\in \mc{S}_2 \cup\{i\}$ and  $j'\in \mc{S}_2 \cup\{j\}$, for notational simplicity, define  
        \begin{gather*}
            K_{h, i'}^{(i)} := K\left(d(\zeta_i, \zeta_{i'})/h\right),\quad K_{h, j'}^{(j)} := K\left(d(\zeta_j, \zeta_{j'})/h\right), \quad K_{h, i'j'}^{(ij)} := K_{h, i'}^{(i)} K_{h, j'}^{(j)}, \\
            \hat{K}^{(i)}_{h, i'} := K\left(\hat{d}_{ii'}/h\right),\quad \hat{K}_{h, j'}^{(j)} := K\left(\hat{d}_{jj'}/h\right), \quad \hat{K}_{h, i'j'}^{(ij)} := \hat{K}_{h, i'}^{(i)} \hat{K}_{h, j'}^{(j)}. 
        \end{gather*}
        By Lemma~\ref{lemma:TWFE_solution}, the two-way fixed-effects estimator~\eqref{eq:sp:Aij_only} admits the following decomposition: 
        \begin{equation}\label{eq:TWFE_solution}
        \begin{aligned}
            \hat{A}_{ij} = & \frac{\sum_{i'\in \mc{S}_2 }\hat{K}_{h, i'}^{(i)} A_{i'j} }{\sum_{i'\in \mc{S}_2 }\hat{K}_{h, i'}^{(i)} } +  \frac{\sum_{j'\in \mc{S}_2  }\hat{K}_{h, j'}^{(j)} A_{ij'} }{\sum_{j'\in \mc{S}_2 }\hat{K}_{h, j'}^{(j)} } - \frac{\sum_{i', j'\in \mc{S}_2   }\hat{K}^{(ij)}_{h, i'j'} A_{i'j'} }{ \sum_{i', j'\in \mc{S}_2   }\hat{K}^{(ij)}_{h, i'j'} } \\
            = &   \underbrace{\frac{\sum_{i'\in \mc{S}_2  }\hat{K}_{h, i'}^{(i)} f_{i'j} }{\sum_{i'\in \mc{S}_2  }\hat{K}_{h, i'}^{(i)} } +  \frac{\sum_{j'\in \mc{S}_2  }\hat{K}_{h, j'}^{(j)} f_{ij'} }{\sum_{j'\in \mc{S}_2  }\hat{K}_{h, j'}^{(j)} } - \frac{\sum_{i', j'\in \mc{S}_2   }\hat{K}_{h, i'j'}^{(ij)} f_{i'j'} }{ \sum_{i' , j'\in \mc{S}_2   }\hat{K}_{h, i'j'}^{(ij)} } }_{Q^{(ij)}_1} \\
            & + \underbrace{ \frac{\sum_{i'\in \mc{S}_2  }\hat{K}_{h, i'}^{(i)} \epsilon_{i'j} }{\sum_{i'\in \mc{S}_2  }\hat{K}_{h, i'}^{(i)} } +  \frac{\sum_{j'\in \mc{S}_2  }\hat{K}_{h, j'}^{(j)} \epsilon_{ij'} }{\sum_{j'\in \mc{S}_2  }\hat{K}_{h, j'}^{(j)} } - \frac{\sum_{i', j'\in \mc{S}_2   }\hat{K}_{h, i'j'}^{(ij)} \epsilon_{i'j'} }{ \sum_{i', j'\in \mc{S}_2  }\hat{K}_{h, i'j'}^{(ij)} } }_{Q^{(ij)}_2} - 
            \underbrace{\frac{\sum_{i'\in \mc{S}_2  }\hat{K}^{(ij)}_{h, i'i'} f_{i'i'} }{ \sum_{i' , j'\in \mc{S}_2   }\hat{K}^{(ij)}_{h, i'j'} } }_{Q^{(ij)}_3}. 
        \end{aligned}
        \end{equation}
        Here, $Q_3^{(ij)}$ arises from the fact that the diagonal elements of $A$ are zero.  In addition, by Assumption~\ref{assumption:regularity}\ref{item:regularity_assumption_smoothness}, the Taylor expansion of $f_{i'j'}$ around $(\zeta_i, \zeta_j)$ is given by 
        \begin{align*}
            f_{i'j'} = &  f(\zeta_i, \zeta_j) + \nabla'_{\zeta_i } f(\zeta_i, \zeta_j) (\zeta_{i'} - \zeta_i) + \nabla'_{\zeta_j } f(\zeta_i, \zeta_j) (\zeta_{j'} - \zeta_j) \\
            & +  (\zeta_{i'}' - \zeta_i', \zeta_{j'}' - \zeta_j')\nabla^{2} f(\zeta_i + \eta_{1, i'j'}'(\zeta_{i'} - \zeta_i), \zeta_j + \eta_{2, i'j'}'(\zeta_{j'}- \zeta_j) ) (\zeta_{i'}' - \zeta_i', \zeta_{j'}' - \zeta_j')', 
        \end{align*}
        where $\eta_{1, i'j'}, \eta_{2, i'j'} \in [0, 1]^{d_{\zeta}}$. 
        We apply the Taylor expansion to each term in $Q_1$ to obtain
        \begin{align*}
            \frac{\sum_{i'\in \mc{S}_2  }\hat{K}_{h, i'}^{(i)} f_{i'j} }{\sum_{i'\in \mc{S}_2  }\hat{K}_{h, i'}^{(i)} } = & f(\zeta_i, \zeta_j) + \frac{\sum_{i'\in \mc{S}_2  }\hat{K}_{h, i'}^{(i)} \nabla'_{\zeta_i } f(\zeta_i, \zeta_j) (\zeta_{i'} - \zeta_i)  }{\sum_{i'\in \mc{S}_2  }\hat{K}_{h, i'}^{(i)} } + \hat{R}^{(ij)}_1, \\
            \frac{\sum_{j'\in \mc{S}_2  }\hat{K}_{h, j'}^{(j)} f_{ij'} }{\sum_{j'\in \mc{S}_2  }\hat{K}_{h, j'}^{(j)} } = & f(\zeta_i, \zeta_j) + \frac{\sum_{j'\in \mc{S}_2  }\hat{K}_{h, j'}^{(j)} \nabla'_{\zeta_j } f(\zeta_i, \zeta_j) (\zeta_{j'} - \zeta_j)  }{\sum_{j'\in \mc{S}_2  }\hat{K}_{h, j'}^{(j)} } + \hat{R}^{(ij)}_2,  
        \end{align*}
        and 
        \begin{align*}
            \frac{\sum_{i'\in \mc{S}_2  , j'\in \mc{S}_2   }\hat{K}_{h, i'j'}^{(ij)} f_{i'j'} }{ \sum_{i'\in \mc{S}_2  , j'\in \mc{S}_2   }\hat{K}_{h, i'j'}^{(ij)} } = f(\zeta_i, \zeta_j) + \frac{\sum_{i'\in \mc{S}_2  }\hat{K}_{h, i'}^{(i)} \nabla'_{\zeta_i } f(\zeta_i, \zeta_j) (\zeta_{i'} - \zeta_i)  }{\sum_{i'\in \mc{S}_2  }\hat{K}_{h, i'}^{(i)} } + \frac{\sum_{j'\in \mc{S}_2  }\hat{K}_{h, j'}^{(j)} \nabla'_{\zeta_j } f(\zeta_i, \zeta_j) (\zeta_{j'} - \zeta_j)  }{\sum_{j'\in \mc{S}_2  }\hat{K}_{h, j'}^{(j)} }  + \hat{R}^{(ij)}_3. 
        \end{align*}
        Here, $\hat{R}^{(ij)}_1, \hat{R}^{(ij)}_2, \hat{R}^{(ij)}_3$ are second-order remainder terms: 
        \begin{equation}\label{eq:definition_R_hat}
        \begin{gathered}
            \hat{R}^{(ij)}_1 = \frac{\sum_{i'\in \mc{S}_2 }\hat{K}_{h, i'}^{(i)} (\zeta_i - \zeta_{i'})'H^{(ij)}_{i'}(\zeta_i - \zeta_{i'}) }{ \sum_{i'\in \mc{S}_2 }\hat{K}^{(i)}_{h, i'} }, \quad \hat{R}^{(ij)}_2 = \frac{\sum_{j'\in \mc{S}_2  }\hat{K}_{h, j'}^{(j)} (\zeta_j - \zeta_{j'})'H_{j'}^{(ij)}(\zeta_j - \zeta_{j'}) }{ \sum_{j'\in \mc{S}_2 }\hat{K}_{h, j'}^{(j)} }, \\
            \hat{R}^{(ij)}_3 = \frac{\sum_{i', j'\in \mc{S}_2 }\hat{K}_{h, i'j'}^{(ij)}(\zeta'_{i'} - \zeta'_i, \zeta'_{j'} - \zeta'_j) H_{i'j'}^{(ij)}(\zeta'_{i'} - \zeta'_i, \zeta'_{j'} - \zeta'_j)' }{\sum_{i', j'\in \mc{S}_2 }\hat{K}^{(ij)}_{h, i'j'} }, 
        \end{gathered}
        \end{equation}
        where  $H_{i'}^{(ij)}\in \mb{R}^{d_{\zeta}\times d_{\zeta}}$, $ H_{j'}^{(ij)}\in \mb{R}^{d_{\zeta}\times d_{\zeta}}$, and $ H_{i'j'}^{(ij)} \in \mb{R}^{2d_{\zeta}\times 2d_{\zeta}}$ are Hessians defined as
        \begin{equation}\label{eq:definition_H}
        \begin{gathered}
            H_{i'}^{(ij)} = \nabla^{2}_{\zeta_i, \zeta_i } f(\zeta_i + \eta_{1, i'j}'(\zeta_{i'} - \zeta_i), \zeta_j ), \quad H_{j'}^{(ij)} = \nabla^{2}_{\zeta_j, \zeta_j } f(\zeta_i, \zeta_j + \eta_{2, ij'}'(\zeta_{j'} - \zeta_j) ), \\
           H_{i'j'}^{(ij)} =  \nabla^{2} f(\zeta_i + \eta_{1, i'j'}'(\zeta_{i'} - \zeta_i), \zeta_j + \eta_{2, i'j'}'(\zeta_{j'}- \zeta_j) ). 
        \end{gathered}
        \end{equation}
        Then it is straightforward to verify that 
        \begin{align*}
            \hat{A}_{ij} = \underbrace{f_{ij} + \hat{R}^{(ij)}_1 + \hat{R}^{(ij)}_2 - \hat{R}^{(ij)}_3}_{Q^{(ij)}_1} + Q^{(ij)}_2 - Q^{(ij)}_3. 
        \end{align*}

        \paragraph{Bias} 
        Note that 
        \begin{align*}
            \mb{E}\left(\hat{A}_{ij} - f_{ij} \mid \{\zeta_i\}_{i=1}^{N} \right) = \mb{E}(\hat{R}^{(ij)}_1 + \hat{R}^{(ij)}_2 + \hat{R}^{(ij)}_3 \mid \{\zeta_i\}_{i=1}^{N} )  + \mb{E}(Q^{(ij)}_2  \mid \{\zeta_i\}_{i=1}^{N} ) - \mb{E}(Q^{(ij)}_3  \mid \{\zeta_i\}_{i=1}^{N} ) . 
        \end{align*}
        Let $\rho_{f''} := \sup_{\zeta_1, \zeta_2\in \mr{supp}(\zeta)}\lambda_{\max}(\nabla^2_{\zeta_1, \zeta_2}f(\zeta_1, \zeta_2))$, where $\lambda_{\max}(\cdot)$ denotes the largest singular value of a matrix. Under the boundedness condition on the second derivative of the graphon (Assumption~\ref{assumption:regularity}\ref{item:regularity_assumption_smoothness}), $\rho_{f''} <\infty$. In addition, by Lemma~\ref{lemma:kernel_hat_d} and the compact support of the kernel (Assumption~\ref{assumption:kernel}), on the event $\mc{A}$, which occurs with probability at least $1 - \delta n^{-1/2}$ (see Lemma~\ref{lemma:consistency_hat_d})\footnote{
            Although the Lemma~\ref{lemma:consistency_hat_d} states the bound $\mb{P}(\mc{A}) \geq 1 - n^{-1/2}$, it follows easily from the proof of Lemma~\ref{lemma:kernel_hat_d} that for any $\delta > 0$, there exist corresponding constants $\gamma_1$ and $\gamma_2$ such that $\mb{P}(\mc{A}) \geq 1 - \delta n^{-1/2}$. 
        }, 
        \begin{align*}
            |\hat{R}^{(ij)}_1| =  \left| \frac{\sum_{i'\in \mc{S}_2  }\bs{1}(\|\zeta_i - \zeta_{i'}\| \leq \eta h ) \hat{K}^{(i)}_{h, i'} (\zeta_i - \zeta_{i'})'H^{(ij)}_{i'}(\zeta_i - \zeta_{i'}) }{ \sum_{i'\in \mc{S}_2 }\hat{K}^{(i)}_{h, i'} }\right| \leq \rho_{f''} \eta^2 h^2.
        \end{align*}
        Here and in the following proof, we use $\delta$ to denote a generic positive constant that may vary from line to line and absorb multiplicative or additive constants.
        
        By the same argument, we obtain that, with probability at least $1 - \delta n^{-1/2}$, $\hat{R}^{(ij)}_2 \leq \rho_{f''} \eta^2h^2$ and $\hat{R}^{(ij)}_3 \leq 4\rho_{f''} \eta^2h^2$. Thus, with probability at least $1-\delta n^{-1/2}$,    
        \begin{align*}
            \left|\mb{E}(\hat{R}^{(ij)}_1 + \hat{R}^{(ij)}_2 + \hat{R}^{(ij)}_3 \mid \{\zeta_i\}_{i=1}^{N} ) \right|  \leq 6\rho_{f''} \eta^2h^2. 
        \end{align*}
       Since, on the event $\mc{A}$, the inequality above holds for any $\zeta_i, \zeta_j \in \{\zeta_i\}_{i\in \mc{S}^c}$, it follows that, with probability at least $1-\delta n^{-1/2}$, 
        \begin{align}\label{eq:thm:bias_variance:bound_Q1}
            \max_{i, j\in \mc{S}^c }\left|\mb{E}(\hat{R}^{(ij)}_1 + \hat{R}^{(ij)}_2 + \hat{R}^{(ij)}_3 \mid \{\zeta_i\}_{i=1}^{N} ) \right|  \leq 6\rho_{f''} \eta^2h^2. 
        \end{align}
        In addition, Assumption~\ref{assumption:regularity}\ref{item:regularity_assumption_independence} and sample-splitting (see Algorithm~\ref{alg:sp}) guarantee that $\left\{\epsilon_{i'j'} \mid i',j'\in \mc{S}_2\right\}$ are independent of $\{\hat{K}^{(i)}_{i'}, \hat{K}^{(j)}_{j'} \mid i',j'\in \mc{S}_2\}$. Then, we have, for any $\zeta_i, \zeta_j \in \{\zeta_i\}_{i\in \mc{S}^c}$, 
        \begin{align}\label{eq:thm:bias_variance:bound_Q2}
            \max_{i, j\in \mc{S}^c } \mb{E}(Q_2^{(ij)} \mid \{\zeta_i\}_{i=1}^{N}) = 0. 
        \end{align}
        To bound $\mb{E}(Q_3^{(ij)} \mid \{\zeta_i\}_{i=1}^{N}) $, let $\rho_K$ denote $\|K(\cdot)\|_{\max}$.  Then, with probability at least $1-\delta n^{-1/2}$, 
        \begin{equation}\label{eq:thm:bias_variance:Q3}
            \begin{aligned}
            \max_{i, j\in \mc{S}^c }\left| \frac{\sum_{i'\in \mc{S}_2  }\hat{K}^{(ij)}_{h, i'i'} f_{i'i'} }{ \sum_{i', j'\in \mc{S}_2   } \hat{K}^{(ij)}_{h, i'j'} } \right| \leqtext{(i)} 
            & \frac{1}{\underline{D}_2 n^2 h^{2d_{\zeta}}} \max_{i, j\in \mc{S}^c } \left|\sum_{i'\in \mc{S}_2  }\hat{K}^{(ij)}_{h, i'i'} f_{i'i'} \right|  \\
            \leq & \frac{\rho_K \rho_f}{\underline{D}_2 n^2 h^{2d_{\zeta}}} \max_{i, j\in \mc{S}^c } \left|\sum_{i'\in \mc{S}_2  }\hat{K}^{(i)}_{h, i'}  + \sum_{j'\in \mc{S}_2  }\hat{K}^{(j)}_{h, j'}  \right| \\
            \leqtext{(ii)} & \frac{2 \rho_K \rho_f}{\underline{D}_2 n^2 h^{2d_{\zeta}}} \bar{C}_2 n h^{d_{\zeta}} \\
            = & \frac{2 \bar{C}_2  \rho_K \rho_f}{\underline{D}_2 n h^{d_{\zeta}}}. 
        \end{aligned}
        \end{equation} 
        Here, (i) holds with probability at least $1-\delta n^{-1/2}$ because of~\eqref{eq:lemma_numerator_2_K_hat} in Lemma~\ref{eq:lemma_numerator_2_K}, and (ii) holds with probability at least $1-\delta n^{-1/2}$ because of~\eqref{eq:lemma_numerator_1_K_hat} in Lemma~\ref{eq:lemma_numerator_1_K}. It follows that, with probability at least $1-\delta n^{-1/2}$, 
        \begin{equation}\label{eq:thm:bias_variance:bound_Q3}
             \max_{i, j\in \mc{S}^c } \left|\mb{E}\left(Q_3^{(ij)} \mid \{\zeta_i\}_{i=1}^{N}\right) \right| \leq \frac{2 \bar{C}_2  \rho_K \rho_f}{\underline{D}_2 n h^{d_{\zeta}}}. 
        \end{equation}
        Therefore, combining~\eqref{eq:thm:bias_variance:bound_Q1}, \eqref{eq:thm:bias_variance:bound_Q2}, and \eqref{eq:thm:bias_variance:bound_Q3}, we conclude that, with probability at least $1-\delta n^{-1/2}$, 
        \begin{align*}
            \max_{i, j\in \mc{S}^c} \left| \mb{E}(\hat{A}_{ij} - f_{ij} \mid \{\zeta_i\}_{i=1}^{N})\right|   = O\left(h^2 + \frac{1}{nh^{d_{\zeta}}}\right). 
        \end{align*}

        \paragraph{Variance} For notational simplicity, define the population counterparts of $(\hat{R}^{(ij)}_1, \hat{R}^{(ij)}_2, \hat{R}^{(ij)}_3 )$ (evaluated at the population pseudo distance $d(\zeta_i, \zeta_{i'})$) as follows: 
        \begin{equation}\label{eq:definition_R}
        \begin{gathered}
            R_1^{(ij)} = \frac{\sum_{i'\in \mc{S}_2 }K^{(i)}_{h, i'} (\zeta_i - \zeta_{i'})'H^{(ij)}_{i'}(\zeta_i - \zeta_{i'}) }{ \sum_{i'\in \mc{S}_2   }K^{(i)}_{h, i'} }, \quad R_2^{(ij)} = \frac{\sum_{j'\in \mc{S}_2   }K^{(j)}_{h, j'} (\zeta_j - \zeta_{j'})'H^{(ij)}_{j'}(\zeta_j - \zeta_{j'}) }{ \sum_{j'\in \mc{S}_2   }K^{(j)}_{h, j'} }, \\
            R_3^{(ij)} = \frac{\sum_{i', j'\in \mc{S}_2   }K^{(ij)}_{h, i'j'}(\zeta_{i'}' - \zeta_i', \zeta_{j'}' - \zeta'_j) H^{(ij)}_{i'j'}(\zeta_{i'}' - \zeta_i', \zeta_{j'}' - \zeta'_j)' }{\sum_{i' , j'\in \mc{S}_2   }K^{(ij)}_{h, i'j'} }. 
        \end{gathered}
        \end{equation}
        Then, 
        \begin{align*}
            \hat{A}_{ij} = f_{ij} + \sum_{k=1, 2, 3}R_{k}^{(ij)} + \sum_{k=1, 2, 3}(\hat{R}^{(ij)}_{k} - R^{(ij)}_{k})  + Q^{(ij)}_2 - Q^{(ij)}_3, 
        \end{align*}
        and, by Cauchy-Schwarz inequality, the variance of the imputation estimator is bounded as
        \begin{equation}\label{eq:thm_bias_variance_var_decomposition}
        \begin{aligned}
            \mr{Var}\left(\hat{A}_{ij} \mid  \{\zeta_i\}_{i=1}^{N} \right)  \leq  3 \mr{Var}\left(\sum_{k=1, 2, 3}(\hat{R}^{(ij)}_{k} - R^{(ij)}_{k}) \mid  \{\zeta_i\}_{i=1}^{N} \right) +  3 \mr{Var}\left(Q^{(ij)}_2 \mid  \{\zeta_i\}_{i=1}^{N} \right)  + 3 \mr{Var}\left(Q^{(ij)}_3 \mid  \{\zeta_i\}_{i=1}^{N} \right).  
        \end{aligned}  
        \end{equation}
        \paragraph{Step 1: Bound on $\max_{i, j\in \mc{S}^c} \mr{Var}\left(\sum_{k=1, 2, 3}(\hat{R}^{(ij)}_{k} - R^{(ij)}_{k}) \mid  \{\zeta_i\}_{i=1}^{N} \right) $} 
        Note that for any positive constant $\delta$, there exist 
        constants $\underline{C}_1, \underline{C}_2, M >0$ such that, 
        with probability at least $1- \delta n^{-1/2} $, 
        \begin{equation}\label{eq:thm_bias_variance_R_R_hat}
        \begin{aligned}
            & \max_{i, j\in \mc{S}^c }\mr{Var}\left(\hat{R}^{(ij)}_1 - R^{(ij)}_1 \mid  \{\zeta_i\}_{i=1}^{N} \right)\\
             = & \max_{i, j\in \mc{S}^c } \mr{Var}\left(\frac{\sum_{i'\in \mc{S}_2 }\hat{K}^{(i)}_{h, i'} (\zeta_i - \zeta_{i'})'H^{(ij)}_{i'}(\zeta_i - \zeta_{i'}) }{ \sum_{i'\in \mc{S}_2  }\hat{K}^{(i)}_{h, i'} }  -  \frac{\sum_{i'\in \mc{S}_2   }K^{(i)}_{h, i'} (\zeta_i - \zeta_{i'})'H^{(ij)}_{i'}(\zeta_i - \zeta_{i'}) }{ \sum_{i'\in \mc{S}_2  }K^{(i)}_{h, i'} } \mid \{\zeta_i\}_{i=1}^{N}\right) \\
            \leqtext{(i)} &  \frac{1}{\min\{\underline{C}_1, \underline{C}_2\} n^2 h^{2d_{\zeta}}} \max_{i, j\in \mc{S}^c } \mb{E}\left(\left(\sum_{i'\in \mc{S}_2  } \bs{1}(d(\zeta_i, \zeta_{i'})\leq h \vee \hat{d}_{ii'}\leq h)(\zeta_i - \zeta_{i'})'H^{(ij)}_{i'}(\zeta_i - \zeta_{i'}) (\hat{K}^{(i)}_{h, i'} - K^{(i)}_{h, i'}) \right)^2 \mid \{\zeta_i\}_{i=1}^{N} \right) \\
            \leqtext{(ii)} & \frac{\bar{K}^2 (\delta_{N, n}/h)^2 }{\min\{\underline{C}_1, \underline{C}_2\} n^2 h^{2d_{\zeta}}} \max_{i, j\in \mc{S}^c } \mb{E}\left(\left(\sum_{i'\in \mc{S}_2 } \bs{1}(d(\zeta_i, \zeta_{i'})\leq h  \vee \hat{d}_{ii'}\leq h)(\zeta_i - \zeta_{i'})'H^{(ij)}_{i'}(\zeta_i - \zeta_{i'})  \right)^2 \mid  \{\zeta_i\}_{i=1}^{N} \right) \\
            \leqtext{(iii)} &  \frac{\bar{K}^2 (\delta_{N, n}/h)^2 \rho_{f''}^2 h^4 \max\{\eta^4, \lambda^{-4}\} }{\min\{\underline{C}_1, \underline{C}_2\} n^2 h^{2d_{\zeta}}} \max_{i, j\in \mc{S}^c } \mb{E}\left(\left(\sum_{i'\in \mc{S}_2  } \bs{1}(d(\zeta_i, \zeta_{i'})\leq h  \vee \hat{d}_{ii'}\leq h) \right)^2 \mid  \{\zeta_i\}_{i=1}^{N}  \right)\\
            \leq &  \frac{2 \bar{K}^2 (\delta_{N, n}/h)^2 \rho_{f''}^2 h^4 \max\{\eta^4, \lambda^{-4}\} }{\min\{\underline{C}_1, \underline{C}_2\} n^2 h^{2d_{\zeta}}} \\
            & \max_{i, j\in \mc{S}^c } \left( \mb{E}\left(\left(\sum_{i'\in \mc{S}_2  } \bs{1}(d(\zeta_i, \zeta_{i'})\leq h  )  \right)^2 \mid \{\zeta_i\}_{i=1}^{N} \right) + \mb{E}\left(\left(\sum_{i'\in \mc{S}_2  } \bs{1}( \hat{d}_{ii'}\leq h) \right)^2\mid \{\zeta_i\}_{i=1}^{N} \right)\right) \\
            \leqtext{(iv)} & \frac{2 \bar{K}^2 (\delta_{N, n}/h)^2 \rho_{f''}^2 h^4 \max\{\eta^4, \lambda^{-4}\} }{\min\{\underline{C}_1, \underline{C}_2\} n^2 h^{2d_{\zeta}}} M n^2 h^{2d_{\zeta}} \\
            = & O(\delta_{N, n}^2 h^2 ) . 
        \end{aligned}    
        \end{equation}
        Here, inequality (i) is from \eqref{eq:lemma_numerator_1_K} and \eqref{eq:lemma_numerator_1_K_hat} in Lemma~\ref{lemma:numerator_1}, inequality (ii) holds because $K(\cdot)$ is Lipschitz (Assumption~\ref{assumption:kernel}) and Lemma~\ref{lemma:consistency_hat_d}, inequality (iii) follows from compact support of kernel function (Assumption~\ref{assumption:kernel}) and Lemma~\ref{lemma:kernel_hat_d}. Inequality (iv) holds with probability at least $1-\delta n^{-1/2}$ (or equivalently, on the event $\mc{A}$ as in~\eqref{def:consistency_hat_d}),  following from \eqref{eq:lemma_numerator_1_K} and \eqref{eq:lemma_numerator_1_K_hat} in Lemma~\ref{lemma:numerator_1}, by replacing $K(\cdot)$ with the uniform kernel on $[-1, 1]$. 
        By the same argument, we can also show that, with probability at least $1-\delta n^{-1/2}$,  $\max_{i, j\in \mc{S}^c }\mr{Var}\left(\hat{R}^{(ij)}_2 - R^{(ij)}_2 \mid  \{\zeta_i\}_{i=1}^{N} \right) =  O(\delta_{N, n}^2h^2)$ and $\max_{i, j\in \mc{S}^c }\mr{Var}\left(\hat{R}^{(ij)}_3 - R_3^{(ij)} \mid  \{\zeta_i\}_{i=1}^{N} \right) =  O(\delta_{N, n}^2h^2)$.  Therefore, with probability at least $1-\delta n^{-1/2}$, 
        \begin{align}\label{eq:thm:bias_variance:v_Q1}
            \max_{i, j \in \mc{S}^c } \mr{Var}\left(\sum_{k=1, 2, 3}(\hat{R}^{(ij)}_{k} - R^{(ij)}_{k}) \mid  \{\zeta_i\}_{i=1}^{N} \right) = O(\delta_{N, n}^2h^2). 
        \end{align}

        \paragraph{Step 2: Bound on $\max_{i, j\in \mc{S}^c} \mr{Var}(Q_2^{(ij)} \mid \{\zeta_i\}_{i = 1}^{N}) $} Note that for $i, j \in \mc{S}^c$,  (i) $\{\epsilon_{i'j' }\mid i'\in \mc{S}_2, j'\in \mc{S}_2\}$ is independent of $\left\{\hat{d}_{ii'} \mid i = 1, \ldots, N, i'\in \mc{S}_2\right\}$ conditional on $\{\zeta_i\}_{i = 1}^{N}$ (Assumption~\ref{assumption:regularity}\ref{item:regularity_assumption_independence}); (ii) $\hat{K}^{(ij)}_{h, i'j'}$ is uniformly bounded (Assumption~\ref{assumption:kernel}); (iii) conditional $\{\zeta_i\}_{i = 1}^{N}$, $\{\epsilon_{i'j' }\mid i', j'\in \mc{S}_2\}$ is a collection of independent sub-Gaussian random variables with $\sigma = 1$; (iv) fixing $\zeta_i,\zeta_j\in \mc{S}^c$, we have $\mr{Var}(\hat{K}^{(i)}_{h, i'} \epsilon_{i'j}\mid \{\zeta_i\}_{i\in \mc{S}^c} ) = O(h^{d_{\zeta}})$, $\mr{Var}(\hat{K}^{(j)}_{h, j'} \epsilon_{ij'}\mid \{\zeta_i\}_{i\in \mc{S}^c} ) = O(h^{d_{\zeta}})$, and $\mr{Var}(\hat{K}^{(ij)}_{h, i'j'} \epsilon_{i'j'}\mid \{\zeta_i\}_{i\in \mc{S}^c} ) = O(h^{2d_{\zeta}})$.  Then, by Bernstein's inequality (Lemma~\ref{lemma:Bernstein}), there exists a constant $\bar{M}$ such that for any $t>0$,  
        \begin{align*}
            \mb{P}\left(\left|\sum_{i'\in \mc{S}_2  }\hat{K}^{(i)}_{h, i'} \epsilon_{i'j}\right|   \geq t \mid \{\zeta_i\}_{i =1}^{N}\right) & \leq 2\exp\left(\frac{-t^2}{\bar{M}^2 n h^{d_{\zeta} } + \frac{1}{3}\bar{M}t}\right), \\
            \mb{P}\left(\left|\sum_{j'\in \mc{S}_2 }\hat{K}_{h, j'}^{(j)} \epsilon_{ij'} \right|  \geq t  \mid \{\zeta_i\}_{i =1}^{N} \right) & \leq 2\exp\left(\frac{-t^2}{\bar{M}^2 n h^{d_{\zeta} } + \frac{1}{3}\bar{M}t}\right), \\
            \mb{P}\left(\left|\sum_{i', j'\in \mc{S}_2  }\hat{K}^{(ij)}_{h, i'j'} \epsilon_{i'j'}\right|  \geq t \mid \{\zeta_i\}_{i =1}^{N} \right) & \leq 2 \exp\left(\frac{-t^2}{\bar{M}^2 n^2 h^{2d_{\zeta} } + \frac{1}{3}\bar{M}t}\right), 
        \end{align*}
        and 
        \begin{align*}
            \mb{P}\left(\max_{i \in \mc{S}^c} \left|\sum_{i'\in \mc{S}_2  }\hat{K}^{(i)}_{h, i'} \epsilon_{i'j}\right|   \geq t \mid  \{\zeta_i\}_{i =1}^{N} \right) & \leq 2 N \exp\left(\frac{-t^2}{\bar{M}^2 n h^{d_{\zeta} } + \frac{1}{3}\bar{M}t}\right),  \\
            \mb{P}\left(\max_{i \in \mc{S}^c} \left|\sum_{j'\in \mc{S}_2 }\hat{K}_{h, j'}^{(j)} \epsilon_{ij'} \right|  \geq t  \mid \{\zeta_i\}_{i =1}^{N} \right) & \leq 2 N \exp\left(\frac{-t^2}{\bar{M}^2 n h^{d_{\zeta} } + \frac{1}{3}\bar{M}t}\right),  \\
            \mb{P}\left(\max_{i, j  \in \mc{S}^c} \left|\sum_{i', j'\in \mc{S}_2  }\hat{K}^{(ij)}_{h, i'j'} \epsilon_{i'j'}\right|  \geq t \mid \{\zeta_i\}_{i =1}^{N} \right) & \leq 2 N^2 \exp\left(\frac{-t^2}{\bar{M}^2 n^2 h^{2d_{\zeta} } + \frac{1}{3}\bar{M}t}\right).
        \end{align*}
        Then, there exist constants $M_1, M_2>0$ such that, when $t = M_1 \sqrt{nh^{d_{\zeta}}\log N }, M_2 nh^{d_{\zeta}}\sqrt{\log N  } $, we have 
        \begin{equation}\label{eq:thm_bias_variance_bound_Bernstein}
        \begin{aligned}
            \mb{P}\left(\max_{i \in \mc{S}^c} \left|\sum_{i'\in \mc{S}_2 }\hat{K}^{(i)}_{h, i'} \epsilon_{i'j}\right|   \geq M_1 \sqrt{nh^{d_{\zeta}}\log N } \mid \{\zeta_i\}_{i =1}^{N} \right) & \leq \delta n^{-1/2}, \\
            \mb{P}\left(\max_{i \in \mc{S}^c}\left|\sum_{j'\in \mc{S}_2 }\hat{K}^{(j)}_{h, j'} \epsilon_{ij'} \right| \geq M_1 \sqrt{nh^{d_{\zeta}}\log N } \mid  \{\zeta_i\}_{i =1}^{N} \right) & \leq \delta  n^{-1/2}, \\
            \mb{P}\left(\max_{i, j \in \mc{S}^c}\left|\sum_{i', j'\in \mc{S}_2 }\hat{K}^{(ij)}_{h, i'j'} \epsilon_{i'j'} \right|  \geq M_2 nh^{d_{\zeta}}\sqrt{\log N}   \mid \{\zeta_i\}_{i =1}^{N} \right) & \leq \delta  n^{-1/2}. 
        \end{aligned}    
        \end{equation}
        
        Therefore, with probability at least $1-\delta n^{-1/2}$, we have 
        \begin{equation}\label{eq:thm_bias_variance_var_product_1}
        \begin{aligned}
            \max_{i, j \in \mc{S}^c} \mr{Var}\left(\frac{\sum_{i'\in \mc{S}_2 }\hat{K}^{(i)}_{h, i'} \epsilon_{i'j}}{\sum_{i'\in \mc{S}_2  }\hat{K}^{(i)}_{h, i'} } \mid  \{\zeta_i\}_{i=1}^{N} \right) \leqtext{(i)} & \frac{1}{\underline{C}_2^2 n^2 h^{2d_{\zeta}}}\mb{E}\left(\left(\sum_{i'\in \mc{S}_2  }\hat{K}^{(i)}_{h, i'} \epsilon_{i'j}  \right)^2 \mid  \{\zeta_i\}_{i=1}^{N}  \right) \\
            \leqtext{(ii)} & \frac{M_1^2 nh^{d_{\zeta}}\log N}{\underline{C}_2^2 n^2 h^{2d_{\zeta}}}\\
            = & \frac{M_1^2 }{\underline{C}_2^2 }\frac{\log N }{n h^{d_{\zeta}}}. 
        \end{aligned}
        \end{equation}
        Here, (i) holds with probability at least $1-\delta n^{-1/2}$ because of \eqref{eq:lemma_numerator_1_K_hat} in Lemma~\ref{lemma:numerator_1}, and (ii), which also holds with probability at least $1-\delta n^{-1/2}$,  follows from \eqref{eq:thm_bias_variance_bound_Bernstein}. 
        Using the same argument, we also show that $\max_{i, j \in \mc{S}^c} \mr{Var}\left(\frac{\sum_{j'\in \mc{S}_2 }\hat{K}^{(j)}_{h, j'} \epsilon_{ij'}}{\sum_{j'\in \mc{S}_2  }\hat{K}^{(j)}_{h, j'} } \mid  \{\zeta_i\}_{i=1}^{N} \right) \leq \frac{M_1^2 }{\underline{C}_2^2 }\frac{\log N}{n h^{d_{\zeta}}} $ with probability at least $1-\delta n^{-1/2}$. 

        In addition, with probability at least $1-\delta n^{-1/2}$, we have 
        \begin{equation}\label{eq:thm_bias_variance_var_product_2}
        \begin{aligned}
            \max_{i, j \in \mc{S}^c} \mr{Var}\left(\frac{\sum_{i', j'\in \mc{S}_2 \ }\hat{K}^{(ij)}_{h, i'j'} \epsilon_{i'j'}}{\sum_{i', j'\in \mc{S}_2 }\hat{K}^{(ij)}_{h, i'j'} } \mid \{\zeta_i\}_{i=1}^{N}\right) \leqtext{(i)} & \frac{1}{\underline{D}_2^2 n^4 h^{4d_{\zeta}}}\mb{E}\left(\left(\sum_{i', j'\in \mc{S}_2 }\hat{K}^{(ij)}_{h, i'j'} \epsilon_{i'j'}  \right)^2\mid \{\zeta_i\}_{i=1}^{N}\right)  \\
            \leqtext{(ii)} & \frac{M_2^2 n^2h^{2d_{\zeta}}\log N}{\underline{D}_2^2 n^4 h^{4d_{\zeta}}}\\
            = & \frac{M_2^2 }{\underline{D}_2^2 }\frac{\log N}{n^2 h^{2d_{\zeta}}}. 
        \end{aligned}
        \end{equation}
        Here, (i) holds with probability at least $1-\delta n^{-1/2}$ because of \eqref{eq:lemma_numerator_2_K_hat} in Lemma~\ref{lemma:numerator_2}, and (ii), which holds with probability $1-\delta n^{-1/2}$,  follows from \eqref{eq:thm_bias_variance_bound_Bernstein}. Therefore, with probability at least $1-\delta n^{-1/2}$, 
        \begin{align}\label{eq:thm:bias_variance:v_Q2}
            \max_{i, j\in \mc{S}^c} \mr{Var}(Q^{(ij)}_2 \mid \{\zeta_i\}_{i = 1}^{N})  = O\left(\frac{\log N}{nh^{d_{\zeta}}} + \frac{\log N}{n^2h^{2d_{\zeta}}}\right) = O\left(\frac{\log N}{nh^{d_{\zeta}}} \right). 
        \end{align}

        \paragraph{Step 3: Bound on $\max_{i, j\in \mc{S}^c}\mr{Var}(Q^{(ij)}_3 \mid \{\zeta_i\}_{i = 1}^{N}) $} Note that, with probability at least $1-\delta n^{-1/2}$,   
        \begin{align}\label{eq:thm:bias_variance:v_Q3}
            \max_{i, j\in \mc{S}^c}\mr{Var}(Q^{(ij)}_3 \mid \{\zeta_i\}_{i = 1}^{N})  \leq \max_{i, j\in \mc{S}^c}\mb{E}\left(\left(Q_3^{(ij)}\right)^2 \mid \{\zeta_i\}_{i = 1}^{N}\right)\leqtext{(i)} \frac{4\bar{C}^2_2 \rho^2_K \rho^2_f}{\underline{D}^2_2 n^2h^{2d_{\zeta}}}, 
        \end{align}
        where (i) follows from~\eqref{eq:thm:bias_variance:Q3}. 
        
        \bigskip
        
        Therefore, combining~\eqref{eq:thm_bias_variance_var_decomposition}, \eqref{eq:thm:bias_variance:v_Q1}, \eqref{eq:thm:bias_variance:v_Q2}, and \eqref{eq:thm:bias_variance:v_Q3}, we obtain that, with probability at least $1-\delta n^{-1/2}$,   
        \begin{align*}
            \max_{i, j\in \mc{S}^c} \mr{Var}\left(\hat{A}_{ij} - f_{ij}\mid \{\zeta_i\}_{i=1}^{N}\right) = O\left( \delta_{N, n}^2 h^2 +  \frac{\log N}{nh^{d_{\zeta}}} + \frac{1}{n^2h^{2d_{\zeta}}}\right) = O\left( \delta_{N, n}^2 h^2 +  \frac{\log N}{nh^{d_{\zeta}}} \right). 
        \end{align*}
        This completes the proof. 
    \end{proofthm}

    \begin{lemma}\label{lemma:TWFE_solution}
        The solution to~\eqref{eq:sp:Aij_only} is given by~\eqref{eq:TWFE_solution}.
    \end{lemma}
    \begin{prooflmm}{lemma:TWFE_solution}
        Without loss of generality, we impose the normalization $\sum_{j' \in \mc{S}_2} \hat{K}^{(j)}_{h, j'} \hat{b}_{j'} = 0$. For any $i, j\in \mc{S}^c$, the first-order conditions with respect to $a_i$ and $b_j$ are given by 
        \begin{gather*}
            \sum_{j'\in \mc{S}_2} \hat{K}^{(j)}_{h, j'} A_{ij'} - \sum_{j'\in \mc{S}_2} \hat{K}^{(j)}_{h, j'} \hat{a}_i - \underbrace{\sum_{j'\in \mc{S}_2} \hat{K}^{(j)}_{h, j'} \hat{b}_{j'}}_{=0} = 0,  \\
            \sum_{i'\in \mc{S}_2} \hat{K}^{(i)}_{h, i'} A_{i' j} - \sum_{i'\in \mc{S}_2} \hat{K}^{(i)}_{h, i'} \hat{b}_j - \sum_{i'\in \mc{S}_2} \hat{K}^{(i)}_{h, i'} \hat{a}_{i'} = 0.  
        \end{gather*}
        Therefore, we have 
        \begin{align*}
            \hat{a}_i = \frac{\sum_{j'\in \mc{S}_2} \hat{K}^{(j)}_{h, j'} A_{ij'}}{\sum_{j'\in \mc{S}_2} \hat{K}^{(j)}_{h, j'} }, \quad \hat{b}_j = \frac{\sum_{i'\in \mc{S}_2} \hat{K}^{(i)}_{h, i'} A_{i' j} - \sum_{i'\in \mc{S}_2} \hat{K}^{(i)}_{h, i'} \hat{a}_{i'}  }{\sum_{i'\in \mc{S}_2} \hat{K}^{(i)}_{h, i'}  }. 
        \end{align*}
        For any $\tilde{i} \in \mc{S}_2$, the first-order condition with respect to $a_{\tilde{i}}$ is given by
        \begin{align*}
            \sum_{j'\in \mc{S}_2} \hat{K}^{(j)}_{h, j'} A_{\tilde{i}j'} - \sum_{j'\in \mc{S}_2} \hat{K}^{(j)}_{h, j'} \hat{a}_{\tilde{i}} - \underbrace{\sum_{j'\in \mc{S}_2} \hat{K}^{(j)}_{h, j'} \hat{b}_{j'}}_{=0} + \hat{K}^{(j)}_{h, j}A_{\tilde{i}j} - \hat{K}^{(j)}_{h, j} \hat{a}_{\tilde{i}}- \hat{K}^{(j)}_{h, j}\hat{b}_j = 0. 
        \end{align*} 
        Thus, 
        \begin{align*}
            \hat{a}_{\tilde{i}} = \frac{\sum_{j'\in \mc{S}_2} \hat{K}^{(j)}_{h, j'} A_{\tilde{i}j'} + \hat{K}^{(j)}_{h, j}A_{\tilde{i}j} - \hat{K}^{(j)}_{h, j}\hat{b}_j}{\sum_{j'\in \mc{S}_2} \hat{K}^{(j)}_{h, j'}  + \hat{K}^{(j)}_{h, j}}. 
        \end{align*}
        Then, 
        \begin{align*}
            \sum_{i'\in \mc{S}_2} \hat{K}^{(i)}_{h, i'}  \hat{b}_j & = \sum_{i'\in \mc{S}_2} \hat{K}^{(i)}_{h, i'} A_{i' j} - \sum_{i'\in \mc{S}_2} \hat{K}^{(i)}_{h, i'} \frac{\sum_{j'\in \mc{S}_2} \hat{K}^{(j)}_{h, j'} A_{i'j'} + \hat{K}^{(j)}_{h, j}A_{i'j} - \hat{K}^{(j)}_{h, j}\hat{b}_j}{\sum_{j'\in \mc{S}_2} \hat{K}^{(j)}_{h, j'}  + \hat{K}^{(j)}_{h, j}}  \\
            \Rightarrow \hat{b}_j & = \frac{\sum_{i'\in \mc{S}_2} \hat{K}^{(i)}_{h, i'} A_{i' j} }{\sum_{i'\in \mc{S}_2} \hat{K}^{(i)}_{h, i'}} - \frac{\sum_{i'\in \mc{S}_2}\sum_{j'\in \mc{S}_2}\hat{K}^{(i)}_{h, i'}\hat{K}^{(j)}_{h, j'}A_{i'j'}}{\sum_{i'\in \mc{S}_2}\sum_{j'\in \mc{S}_2}\hat{K}^{(i)}_{h, i'}\hat{K}^{(j)}_{h, j'}}. 
        \end{align*}
        Let $\hat{A}_{ij} = \hat{a}_i + \hat{b}_j$. This completes the proof. 
    \end{prooflmm}

    \begin{lemma}\label{lemma:numerator_1}
        Under the conditions in Theorem~\ref{thm:bias_variance}, for any constant $\delta >0$,  there exist constants $0 < \underline{C}_1 <\bar{C}_1  < \infty$ and $0 < \underline{C}_2 <\bar{C}_2  < \infty$ such that 
        \begin{equation}\label{eq:lemma_numerator_1_K}
        \begin{aligned}
            \mb{P} \left(\underline{C}_1  nh^{d_{\zeta}} \leq  \min_{i\in \mc{S}^c } \sum_{i'\in \mc{S}_2 }K^{(i)}_{h, i'}   \leq  \max_{i\in \mc{S}^c } \sum_{i'\in \mc{S}_2 }K^{(i)}_{h, i'}  \leq \bar{C}_1 nh^{d_{\zeta}}  \right) \geq 1 - \delta n^{-1/2},
        \end{aligned}
        \end{equation}
        and 
        \begin{equation}\label{eq:lemma_numerator_1_K_hat}
        \begin{aligned}
            \mb{P} \left( \underline{C}_2  nh^{d_{\zeta}} \leq \min_{i\in \mc{S}^c} \sum_{i'\in \mc{S}_2 }\hat{K}^{(i)}_{h, i'}  \leq \max_{i\in \mc{S}^c} \sum_{i'\in \mc{S}_2 }\hat{K}_{h, i'}^{(i)} \leq  \bar{C}_2 nh^{d_{\zeta}}\right) \geq 1 - \delta n^{-1/2}.  
        \end{aligned}
        \end{equation}
        In addition, there exists a constant $\bar{C}_3>0$ such that
        \begin{equation}\label{eq:lemma_numerator_1_product}
        \begin{aligned}
            \mb{P}\bigg(& \max_{i, j \in \mc{S}^c}\left| \sum_{i'\in \mc{S}_2 } K^{(i)}_{h, i'}(\zeta_{i} - \zeta_{i'})' H^{(ij)}_{i'} (\zeta_{i} - \zeta_{i'})- \sum_{i'\in \mc{S}_2 } \mb{E}(K^{(i)}_{h, i'}(\zeta_{i} - \zeta_{i'})' H^{(ij)}_{i'} (\zeta_{i} - \zeta_{i'})\mid \{\zeta_i\}_{i\in \mc{S}^c} )\right|   \\ 
            & \geq \bar{C}_3 \sqrt{nh^{d_{\zeta}+4}\log N }   \mid \{\zeta_i\}_{i\in \mc{S}^c} \bigg) 
            \leq  \delta n^{-1/2}. 
        \end{aligned}
        \end{equation}
    \end{lemma}
    \begin{prooflmm}{lemma:numerator_1}
        Fixing $\{\zeta_i\}_{i\in \mc{S}^c}$, by Assumption~\ref{assumption:regularity}\ref{item:regularity_assumption_compact_support} and Assumption~\ref{assumption:kernel}, it is straightforward to verify that for any $i\in \mc{S}^c$, $\mr{Var}(K^{(i)}_{h, i'} \mid \{\zeta_i\}_{i\in \mc{S}^c}) = O(h^{d_{\zeta}})$. 
        In addition, by the boundedness of $K^{(i)}_{h, i'}$ (Assumption~\ref{assumption:kernel}),   we apply Bernstein's inequality for independent random variables  (Lemma~\ref{lemma:Bernstein}) to demonstrate that there exists a constant $M>0$ such that for any $t>0$, 
        \begin{equation*}
             \mb{P}\left(\left|\sum_{i'\in \mc{S}_2 }\left(K^{(i)}_{h, i'} - \mb{E}(K^{(i)}_{h, i'}  \mid \{\zeta_i\}_{i\in \mc{S}^c} ) \right) \right|   \geq t    \mid \{\zeta_i\}_{i\in \mc{S}^c} \right) \leq 2\exp\left(\frac{-t^2}{M^2 n h^{d_{\zeta}} + \frac{1}{3}Mt}\right). 
        \end{equation*}
        Therefore, 
        \begin{align*}
             \mb{P}\left(\max_{i \in \mc{S}^c }\left|\sum_{i'\in \mc{S}_2 }\left(K^{(i)}_{h, i'} - \mb{E}(K^{(i)}_{h, i'}  \mid \{\zeta_i\}_{i\in \mc{S}^c}) \right) \right|   \geq t   \mid \{\zeta_i\}_{i\in \mc{S}^c}  \right) \leq 2N \exp\left(\frac{-t^2}{M^2 n h^{d_{\zeta}} + \frac{1}{3}Mt}\right). 
        \end{align*}
        It follows that we can choose a sufficiently large constant $C_1>0$  such that when $t = C_1\sqrt{nh^{d_{\zeta}}\log N }$, 
        \begin{align}\label{eq:lemma_numerator_1_eq_1}
            \mb{P}\left(\max_{i \in \mc{S}^c } \left|\sum_{i'\in \mc{S}_2 }\left(K^{(i)}_{h, i'} - \mb{E}(K^{(i)}_{h, i'}  \mid \{\zeta_i\}_{i\in \mc{S}^c}) \right) \right|  \geq C_1 \sqrt{nh^{d_{\zeta}}\log N } \right) \leq \delta n^{-1/2}. 
        \end{align} 
        In addition, 
        \begin{equation}\label{eq:lemma_numerator_1_eq_2}
        \begin{aligned}
            \max_{i\in \mc{S}^c } \mb{E}(K^{(i)}_{h, i'} \mid \{\zeta_i\}_{i\in \mc{S}^c} ) = \max_{i\in \mc{S}^c } \mb{E}(K^{(i)}_{h, i'} \mid \zeta_i)  & =  \max_{i\in \mc{S}^c } \int  K\left(\frac{d(\zeta_i, \zeta_{i'})}{h}\right) \mr{d}\mb{P} (\zeta_{i'}) \\
            & = \max_{i\in \mc{S}^c } \int_{ \{ d(\zeta_i, \zeta_{i'}) \leq h\}}  K\left(\frac{d(\zeta_i, \zeta_{i'})}{h}\right) \mr{d}\mb{P} (\zeta_{i'}) \\
            & \geq \max_{i\in \mc{S}^c } \int_{  \left\{d(\zeta_i, \zeta_{i'}) \leq \frac{K(0)}{2 \bar{K}} h \right\} }  K\left(\frac{d(\zeta_i, \zeta_{i'})}{h}\right) \mr{d}\mb{P} (\zeta_{i'}) \\
            & \geqtext{(i)} C_2\max_{i\in \mc{S}^c }   \int_{  \left\{d(\zeta_i, \zeta_{i'}) \leq \frac{K(0)}{2 \bar{K}} h \right\} }   \mr{d}\mb{P} (\zeta_{i'})  \\ 
            &  \geqtext{(ii)} C_2\max_{i\in \mc{S}^c }  \mb{P}\left(\|\zeta_i - \zeta_{i'}\|\leq \lambda^{-1} \frac{{K(0)}}{2 \bar{K}} h  \mid \zeta_i \right)  \\ 
            & \geqtext{(iii)} C_2 \underline{c}\left(\lambda^{-1} \frac{K(0)}{2 \bar{K}} h\right)^{d_{\zeta}}. 
        \end{aligned}   
        \end{equation} 
        Inequality (i) follows from the Lipschitz continuity of kernel $K(\cdot)$ and $K(0)>0$ (see Assumption~\ref{assumption:kernel}), inequality (ii) follows from the informativeness condition (Assumption~\ref{assumption:informativeness}), and inequality (iii) holds because of Assumption~\ref{assumption:regularity}\ref{item:regularity_assumption_compact_support}. In addition, by the boundedness of kernel $K(\cdot)$, there exists a constant $C_3>0$ such that 
        \begin{equation}\label{eq:lemma_numerator_1_eq_3}
            \max_{i\in \mc{S}^c }\mb{E}(K^{(i)}_{h, i'} \mid \{\zeta_i\}_{i\in \mc{S}^c}) \leq C_3 \bar{c} (\lambda^{-1} h)^{d_{\zeta}}. 
        \end{equation}
        Combining \eqref{eq:lemma_numerator_1_eq_1}, \eqref{eq:lemma_numerator_1_eq_2}, and \eqref{eq:lemma_numerator_1_eq_3}, we have 
        \begin{align*}
            \mb{P}\bigg( & C_2 \underline{c}n\left(\lambda^{-1} \frac{K(0)}{2 \bar{K}} h\right)^{d_{\zeta}} - C_1 \sqrt{nh^{d_{\zeta}}\log N  } \leq \min_{i\in \mc{S}^c }\sum_{i'\in \mc{S}_2 } K^{(i)}_{h, i'} \leq \max_{i\in \mc{S}^c }\sum_{i'\in \mc{S}_2 } K^{(i)}_{h, i'} \leq  \\
            &  C_3 \bar{c} n(\lambda^{-1} h)^{d_{\zeta}} + C_1 \sqrt{nh^{d_{\zeta}}\log N  }\mid \{\zeta_i\}_{i\in \mc{S}^c } \bigg) \geq 1 - \delta n^{-1/2}
        \end{align*}
        Therefore, there exist constants $0 < \underline{C}_1 <\bar{C}_1  < \infty$ such that\footnote{
            $\sqrt{nh^{d_{\zeta}}\log N}$ is asymptotically negligible compared to $nh^{d_{\zeta}}$ by conditions in Theorem~\ref{thm:bias_variance}. 
        }
        \begin{align*}
            \mb{P} \left(\underline{C}_1  nh^{d_{\zeta}} \leq  \min_{i\in \mc{S}^c } \sum_{i'\in \mc{S}_2 }K^{(i)}_{h, i'}   \leq  \max_{i\in \mc{S}^c } \sum_{i'\in \mc{S}_2 }K^{(i)}_{h, i'}  \leq \bar{C}_1 nh^{d_{\zeta}} \mid \{\zeta_i\}_{i\in \mc{S}^c } \right) \geq 1 - \delta n^{-1/2}
        \end{align*}
        Since the inequality above holds for any realization of $\{\zeta_i\}_{i\in \mc{S}^c }$, we can drop the conditioning and obtain \eqref{eq:lemma_numerator_1_K}. 
        
        Now we turn to the derivation of~\eqref{eq:lemma_numerator_1_K_hat}. On the event $\mc{A}$, or equivalently, which occurs with probability at least $1 - \delta n^{-1/2}$, we have
        \begin{equation*}
        \begin{aligned}
            \max_{i\in \mc{S}^c} \left|\sum_{i'\in \mc{S}_2 }(\hat{K}^{(i)}_{h, i'} - K^{(i)}_{h, i'})\right|  \leqtext{(i)} & \max_{i\in \mc{S}^c} \sum_{i'\in \mc{S}_2 }  \bs{1}(\hat{d}_{ii'}\leq h \vee d(\zeta_i, \zeta_{i'})\leq h) \bar{K} |\hat{d}_{ii'} - d(\zeta_i, \zeta_{i'})|/h \\
            \leqtext{(ii)} & \max_{i\in \mc{S}^c} \sum_{i'\in \mc{S}_2 }  \bs{1}(\hat{d}_{ii'}\leq h \vee d(\zeta_i, \zeta_{i'})\leq h ) \bar{K} \delta_{N,n}/h, 
        \end{aligned}
        \end{equation*} 
        where inequality (i) follows from Assumption~\ref{assumption:kernel}, and inequality (ii) comes from the definition of $\mc{A}$ in~\eqref{def:consistency_hat_d}. Since we require $\delta_{N, n} /h\rightarrow 0$ in Theorem~\ref{thm:bias_variance}, we  employ the same argument used to prove~\eqref{eq:lemma_numerator_1_eq_2} and~\eqref{eq:lemma_numerator_1_eq_2} to show that (the proof is analogous and is therefore omitted) there exist constants $0< C_4 < C_5 <\infty$ such that\footnote{
            Here we drop the conditioning because it is straightforward to verify that the analysis does not depend on the specific realization of $\{\zeta_i\}_{i \in \mc{S}^c}$. 
        }
        \begin{equation}\label{eq:lemma_numerator_1_eq_4}
        \begin{aligned}
            & \mb{P}\left( C_4  n h^{d_{\zeta}}   \leq \min_{i\in \mc{S}^c} \sum_{i'\in \mc{S}_2 }  \bs{1}(\hat{d}_{ii'}\leq h \vee d(\zeta_i, \zeta_{i'})\leq h ) \leq  \max_{i\in \mc{S}^c} \sum_{i'\in \mc{S}_2 }  \bs{1}(\hat{d}_{ii'}\leq h \vee d(\zeta_i, \zeta_{i'})\leq h ) \leq  C_5  n h^{d_{\zeta}} \right) \geq 1 - \delta n^{-1/2} \\
            \Rightarrow & \mb{P}\left(   \max_{i\in \mc{S}^c} \left|\sum_{i'\in \mc{S}_2 }(\hat{K}^{(i)}_{h, i'} - K^{(i)}_{h, i'})\right| \leq C_5  n h^{d_{\zeta}-1} \delta_{N, n} \right) \geq 1 -  \delta n^{-1/2}. 
        \end{aligned}   
        \end{equation} 

        Combining \eqref{eq:lemma_numerator_1_K} and \eqref{eq:lemma_numerator_1_eq_4}, there exist constants  $0 < \underline{C}_2 <\bar{C}_2  < \infty$ such that
        \begin{align*}
            \mb{P} \left( \underline{C}_2  nh^{d_{\zeta}} \leq \min_{i\in \mc{S}^c} \sum_{i'\in \mc{S}_2 }\hat{K}^{(i)}_{h, i'}  \leq \max_{i\in \mc{S}^c} \sum_{i'\in \mc{S}_2 }\hat{K}_{h, i'}^{(i)} \leq  \bar{C}_2 nh^{d_{\zeta}}\right) \geq 1 - \delta n^{-1/2}. 
        \end{align*}
        Therefore, we obtain \eqref{eq:lemma_numerator_1_K_hat}. 

        \bigskip

        To prove \eqref{eq:lemma_numerator_1_product}, fix $\{\zeta_i\}_{i\in \mc{S}^c}$. By Assumption~\ref{assumption:regularity}\ref{item:regularity_assumption_compact_support}, \ref{item:regularity_assumption_smoothness}, and  Assumption~\ref{assumption:kernel}, it is straightforward to verify that $\mr{Var}(K^{(i)}_{h, i'}(\zeta_{i} - \zeta_{i'})' H^{(ij)}_{i'} (\zeta_{i} - \zeta_{i'})v\mid \{\zeta_i\}_{i\in \mc{S}^c}) = O(h^{d_{\zeta } + 4})$. 
        In addition, by the uniform boundedness of $K(\cdot)$, $\zeta_{i}$,  and $H^{(ij)}_{i'}$ (see Assumption~\ref{assumption:regularity}\ref{item:regularity_assumption_compact_support}, \ref{item:regularity_assumption_smoothness}, and  Assumption~\ref{assumption:kernel}), we apply  Bernstein inequality for independent random variables (Lemma~\ref{lemma:Bernstein}) to demonstrate that there exists a constant $M>0$ such that for any $t>0$,  
        \begin{align*}
            & \mb{P}\left(\left| \sum_{i'\in \mc{S}_2 } K^{(i)}_{h, i'}(\zeta_{i} - \zeta_{i'})' H^{(ij)}_{i'} (\zeta_{i} - \zeta_{i'})- \sum_{i'\in \mc{S}_2 } \mb{E}(K^{(i)}_{h, i'}(\zeta_{i} - \zeta_{i'})' H^{(ij)}_{i'} (\zeta_{i} - \zeta_{i'})\mid \{\zeta_i\}_{i\in \mc{S}^c} )\right|   \geq t  \mid \{\zeta_i\}_{i\in \mc{S}^c} \right) \\
            \leq & 2\exp\left(\frac{-t^2}{M^2 n h^{d_{\zeta} + 4} + \frac{1}{3}Mt}\right). 
        \end{align*}
        Therefore, 
        \begin{align*}
            & \mb{P}\left(\max_{i\in \mc{S}^c}\left| \sum_{i'\in \mc{S}_2 } K^{(i)}_{h, i'}(\zeta_{i} - \zeta_{i'})' H^{(ij)}_{i'} (\zeta_{i} - \zeta_{i'})- \sum_{i'\in \mc{S}_2 } \mb{E}(K^{(i)}_{h, i'}(\zeta_{i} - \zeta_{i'})' H^{(ij)}_{i'} (\zeta_{i} - \zeta_{i'})\mid \{\zeta_i\}_{i\in \mc{S}^c} )\right|   \geq t  \mid \{\zeta_i\}_{i\in \mc{S}^c} \right) \\
            \leq & 2 N \exp\left(\frac{-t^2}{M^2 n h^{d_{\zeta} + 4} + \frac{1}{3}Mt}\right). 
        \end{align*}
        It follows that there exists a sufficiently large constant $\bar{C}_3 >0$  such that when $t = \bar{C}_3\sqrt{nh^{d_{\zeta}+4}\log N }$, 
        \begin{align*}
            \mb{P}\bigg(& \max_{i\in \mc{S}^c}\left| \sum_{i'\in \mc{S}_2 } K^{(i)}_{h, i'}(\zeta_{i} - \zeta_{i'})' H^{(ij)}_{i'} (\zeta_{i} - \zeta_{i'})- \sum_{i'\in \mc{S}_2 } \mb{E}(K^{(i)}_{h, i'}(\zeta_{i} - \zeta_{i'})' H^{(ij)}_{i'} (\zeta_{i} - \zeta_{i'})\mid \{\zeta_i\}_{i\in \mc{S}^c} )\right|   \\ 
            & \geq \bar{C}_3 \sqrt{nh^{d_{\zeta}+4}\log N }   \mid \{\zeta_i\}_{i\in \mc{S}^c} \bigg) 
            \leq  \delta n^{-1/2}. 
        \end{align*} 
        This establishes \eqref{eq:lemma_numerator_1_product}. 
    \end{prooflmm}

    \begin{lemma}\label{lemma:numerator_2}
        Under the conditions in Theorem~\ref{thm:bias_variance}, for any constant $\delta>0$,  there exist constants $0 < \underline{D}_1 <\bar{D}_1  < \infty$ and $0 < \underline{D}_2 <\bar{D}_2  < \infty$ such that 
         \begin{equation}\label{eq:lemma_numerator_2_K}
            \mb{P}\bigg(  \underline{D}_1  n^2h^{2d_{\zeta}} \leq \min_{i, j\in \mc{S}^c}\sum_{i',  j'\in \mc{S}_2} K^{(ij)}_{h, i'j'} \leq \max_{i, j\in \mc{S}^c}\sum_{i',  j'\in \mc{S}_2} K^{(ij)}_{h, i'j'}  \geq \bar{D}_1 n^2h^{2d_{\zeta}} \bigg) 
            \geq  1 - \delta n^{-1/2}, 
        \end{equation}
        and 
        \begin{equation}\label{eq:lemma_numerator_2_K_hat}
            \mb{P}\bigg(  \underline{D}_2  n^2h^{2d_{\zeta}} \leq \min_{i, j\in \mc{S}^c}\sum_{i',  j'\in \mc{S}_2} \hat{K}^{(ij)}_{h, i'j'} \leq \max_{i, j\in \mc{S}^c}\sum_{i',  j'\in \mc{S}_2} \hat{K}^{(ij)}_{h, i'j'}  \leq \bar{D}_2 n^2h^{2d_{\zeta}} \bigg) 
            \geq  1 - \delta n^{-1/2}.  
        \end{equation}
        In addition, there exists a constant $\bar{D}_3>0$ such that 
        \begin{equation} 
        \begin{aligned}\label{eq:lemma_numerator_2_product}
            \mb{P}\bigg( & \max_{i, j\in \mc{S}^c }\bigg| \sum_{i', j'\in \mc{S}_2} \big(K^{(ij)}_{h, i'j'}(\zeta'_{i} - \zeta_{i'}', \zeta_{j}' - \zeta_{j'}') H^{(ij)}_{i'j'} (\zeta'_{i} - \zeta_{i'}', \zeta_{j}' - \zeta_{j'}')' \\
            & - \mb{E}\big(K^{(ij)}_{h, i'j'}(\zeta'_{i} - \zeta_{i'}', \zeta_{j}' - \zeta_{j'}') H^{(ij)}_{i'j'} (\zeta'_{i} - \zeta_{i'}', \zeta_{j}' - \zeta_{j'}')'  \mid \{\zeta_i\}_{i\in \mc{S}^c})  \big) \bigg|  \\
            & \leq \bar{D}_3 \sqrt{n^3h^{3d_{\zeta}+4}\log N }   \mid \{\zeta_i\}_{i\in \mc{S}^c }\bigg) \geq  1-\delta n^{-1/2}. 
        \end{aligned}
        \end{equation}
    \end{lemma}
    \begin{prooflmm}{lemma:numerator_2}
        Fix $\zeta_i, \zeta_j \in \{\zeta_i\}_{i\in \mc{S}^c}$. Because  $\mb{E}\left(\mb{E}\left(K^{(ij)}_{h, i'j'} \mid \zeta_i, \zeta_j, \zeta_{i'} \right)^2  \mid \zeta_i, \zeta_j\right) = O(h^{3d_{\zeta}})$ and $K^{(ij)}_{h, i'j'}$ is uniformly bounded,  we apply Bernstein's inequality for U-statistics (see \citet[Theorem~2]{arcones1995bernstein}) to show that there exists a sufficiently large constant $M > 0$ such that, for any $\zeta_i, \zeta_j \in \{\zeta_i\}_{i \in \mc{S}^c}$ and any $t > 0$,
        \begin{align*}
            \mb{P}\left(\left| \sum_{i', j'\in \mc{S}_2  } K^{(ij)}_{h, i'j'} - \mb{E}\left( \sum_{i', j'\in \mc{S}_2  } K^{(ij)}_{h, i'j'} \mid \{\zeta_i\}_{i\in \mc{S}^c}  \right)\right| \geq t \mid \{\zeta_i\}_{i\in \mc{S}^c} \right)  \leq 4\exp\left(\frac{-t^2}{M^2 n^3 h^{3d_{\zeta}} + M n t} \right),
        \end{align*} 
        and 
        \begin{align*}
           \mb{P}\left( \max_{i, j\in \mc{S}^c}\left| \sum_{i', j'\in \mc{S}_2  } K^{(ij)}_{h, i'j'} - \mb{E}\left( \sum_{i', j'\in \mc{S}_2  } K^{(ij)}_{h, i'j'} \mid \{\zeta_i\}_{i\in \mc{S}^c}  \right)\right| \geq t \mid \{\zeta_i\}_{i\in \mc{S}^c} \right)  \leq 4N^2 \exp\left(\frac{-t^2}{M^2 n^3 h^{3d_{\zeta}} + M n t} \right).
        \end{align*}
        It follows that there exists a constant $D_1 > 0$ such that, when $t = D_1 \sqrt{n^3 h^{3d_{\zeta}} \log N}$, 
        \begin{align}\label{eq:lemma_numerator_2_eq_1}
            \mb{P}\left(\max_{i, j\in \mc{S}^c} \left|\sum_{i', j'\in \mc{S}_2  } K^{(ij)}_{h, i'j'} - \mb{E}\left( \sum_{i', j'\in \mc{S}_2  } K^{(ij)}_{h, i'j'} \mid \{\zeta_i\}_{i\in \mc{S}^c} \right)\right|  \geq D_1 \sqrt{n^3 h^{3d_{\zeta}} \log N }  \mid \{\zeta_i\}_{i\in \mc{S}^c} \right) \leq \delta n^{-1/2}. 
        \end{align}
        In addition,  
        \begin{equation}\label{eq:lemma_numerator_2_eq_2}
        \begin{aligned}
            & \max_{i, j\in \mc{S}^c}  \mb{E}(K^{(ij)}_{h, i'j'} \mid \{\zeta_i\}_{i\in \mc{S}^c} ) \\
            = & \max_{i, j\in \mc{S}^c}  \int  K\left(\frac{d(\zeta_i, \zeta_{i'})}{h}\right) K\left(\frac{d(\zeta_j, \zeta_{j'})}{h}\right) \mr{d}\mb{P} (\zeta_{i'})\mr{d}\mb{P} (\zeta_{j'}) \\
            = & \max_{i, j\in \mc{S}^c}  \int_{ \{d(\zeta_i, \zeta_{i'}) \leq h\}}  K\left(\frac{d(\zeta_i, \zeta_{i'})}{h}\right) \mr{d}\mb{P} (\zeta_{i'}) \int_{\{d(\zeta_j, \zeta_{j'}) \leq h \} }  K\left(\frac{d(\zeta_j, \zeta_{j'})}{h}\right) \mr{d}\mb{P} (\zeta_{j'}) \\
            \geq & \max_{i, j\in \mc{S}^c}  \int_{\left\{d(\zeta_i, \zeta_{i'}) \leq \frac{K(0)}{2\bar{K}}h \right\}}  K\left(\frac{d(\zeta_i, \zeta_{i'})}{h}\right) \mr{d}\mb{P} (\zeta_{i'}) \int_{\left\{d(\zeta_j, \zeta_{j'}) \leq \frac{K(0)}{2\bar{K}}h \right\}}  K\left(\frac{d(\zeta_j, \zeta_{j'})}{h}\right) \mr{d}\mb{P} (\zeta_{j'}) \\
            \geqtext{(i)} &  D_2^2  \max_{i\in \mc{S}^c}  \int_{\left\{d(\zeta_i, \zeta_{i'}) \leq \frac{K(0)}{2\bar{K}}h \right\}} \mr{d}\mb{P} (\zeta_{i'}) \max_{ j\in \mc{S}^c} \int_{\left\{d(\zeta_j, \zeta_{j'}) \leq \frac{K(0)}{2\bar{K}}h \right\}}  \mr{d}\mb{P} (\zeta_{j'}) \\
            \geqtext{(ii)} &  D_2^2  \max_{i\in \mc{S}^c} \mb{P}\left( \|\zeta_i - \zeta_{i'}\|\leq \lambda^{-1}\frac{K(0)}{2\bar{K}}h \mid \zeta_i \right) \max_{j\in \mc{S}^c} \mb{P}\left( \|\zeta_j - \zeta_{j'}\|\leq \lambda^{-1}\frac{K(0)}{2\bar{K}}h  \mid \zeta_j \right) \\
            \geqtext{(iii)} & D^2_2 \underline{c}^2\left(\lambda^{-1} \frac{K(0)}{2\bar{K}}h\right)^{2 d_{\zeta}}. 
        \end{aligned}   
        \end{equation} 
        Inequality (i) follows from the Lipschitz continuity of kernel $K(\cdot)$ and $K(0)>0$ (see Assumption~\ref{assumption:kernel}), inequality (ii) follows from the informativeness condition (Assumption~\ref{assumption:informativeness}), and inequality (iii) holds because of Assumption~\ref{assumption:regularity}\ref{item:regularity_assumption_compact_support}.  In addition, by the boundedness of kernel $K(\cdot)$, there exists a constant $D_3>0$ such that  
        \begin{equation}\label{eq:lemma_numerator_2_eq_3}
            \max_{i, j\in \mc{S}^c} \mb{E}(K^{(ij)}_{h, i'j'} \mid \{\zeta_i\}_{i\in \mc{S}^c}) \leq D_3 \bar{c}^2 (\lambda^{-1} h)^{2d_{\zeta}}. 
        \end{equation}
        Combining \eqref{eq:lemma_numerator_2_eq_1}, \eqref{eq:lemma_numerator_2_eq_2}, and \eqref{eq:lemma_numerator_2_eq_3}, 
        \begin{align*}
            \mb{P}\bigg( & D^2_2 \underline{c}^2n^2\left(\lambda^{-1} \frac{K(0)}{2\bar{K}}h\right)^{2 d_{\zeta}} - D_1 \sqrt{n^3 h^{3d_{\zeta}}\log N } \leq \min_{i, j\in \mc{S}^c}\sum_{i',  j'\in \mc{S}_2} K^{(ij)}_{h, i'j'} \leq \max_{i, j\in \mc{S}^c}\sum_{i',  j'\in \mc{S}_2} K^{(ij)}_{h, i'j'}  \\  
            & \leq D_3 \bar{c}^2 n^2(\lambda^{-1} h)^{2d_{\zeta}} + C_1 \sqrt{n^3h^{3d_{\zeta}} \log N  }\mid \{\zeta_i\}_{i\in \mc{S}^c} \bigg) 
            \geq  1 - \delta n^{-1/2}.
        \end{align*}
        Therefore, there exist constants $0 < \underline{C}_1 <\bar{C}_1  < \infty$ such that\footnote{
            $\sqrt{n^3 h^{3d_{\zeta}}\log N}$ is asymptotically negligible compared to $n^2h^{2d_{\zeta}}$ by conditions in Theorem~\ref{thm:bias_variance}. 
        }
        \begin{align*}
            \mb{P}\bigg( & \underline{D}_1  n^2h^{2d_{\zeta}} \leq \min_{i, j\in \mc{S}^c}\sum_{i',  j'\in \mc{S}_2} K^{(ij)}_{h, i'j'} \leq \max_{i, j\in \mc{S}^c}\sum_{i',  j'\in \mc{S}_2} K^{(ij)}_{h, i'j'}  \leq \bar{D}_1 n^2h^{2d_{\zeta}} \mid \{\zeta_i\}_{i\in \mc{S}^c} \bigg) 
            \geq  1 - \delta n^{-1/2}.
        \end{align*}
        Since the inequality above holds for any realization of $\{\zeta_i\}_{i\in \mc{S}^c }$, we can drop the conditioning and obtain \eqref{eq:lemma_numerator_2_K}. 
        
        \bigskip 

        Now we turn to the derivation of \eqref{eq:lemma_numerator_2_K_hat}. Let $\rho_{K} : = \|K(\cdot)\|_{\infty} < \infty$.  On the event $\mc{A}$, which occurs with probability at least $1 - \delta n^{-1/2}$, we have 
        \begin{equation*}
        \begin{aligned}
            & \max_{i, j\in \mc{S}^c} \left|\sum_{i',j'\in \mc{S}_2 }(\hat{K}^{(ij)}_{h, i'j'} - K^{(ij)}_{h, i'j'})\right|\\
            \leqtext{(i)} & \max_{i, j\in \mc{S}^c} \sum_{i', j'\in \mc{S}_2 }  \bs{1}(\hat{d}_{ii'}\leq h \vee d(\zeta_i, \zeta_{i'})\leq h) \bs{1}(\hat{d}_{jj'}\leq h \vee d(\zeta_j, \zeta_{j'})\leq h)   \bar{K} \rho_{K} \frac{|\hat{d}_{ii'} - d(\zeta_i, \zeta_{i'})| + |\hat{d}_{jj'} - d(\zeta_j, \zeta_{j'})|}{h} \\
            & +  \max_{i, j\in \mc{S}^c} \sum_{i', j'\in \mc{S}_2 }  \bs{1}(\hat{d}_{ii'}\leq h \vee d(\zeta_i, \zeta_{i'})\leq h) \bs{1}(\hat{d}_{jj'}\leq h \vee d(\zeta_j, \zeta_{j'})\leq h)   \bar{K}^2 \frac{|\hat{d}_{ii'} - d(\zeta_i, \zeta_{i'})| |\hat{d}_{jj'} - d(\zeta_j, \zeta_{j'})|}{h^2}  \\
            \leqtext{(ii)} & \max_{i, j\in \mc{S}^c} \sum_{i', j'\in \mc{S}_2 }  \bs{1}(\hat{d}_{ii'}\leq h \vee d(\zeta_i, \zeta_{i'})\leq h) \bs{1}(\hat{d}_{jj'}\leq h \vee d(\zeta_j, \zeta_{j'})\leq h)   \left(\bar{K} \rho_{K} \frac{\delta_{N, n}}{h}  + \bar{K}^2 \frac{\delta_{N, n}^2}{h^2}\right), 
        \end{aligned}
        \end{equation*}
        where inequality (i) follows from Assumption~\ref{assumption:kernel}, and inequality (ii) comes from the definition of $\mc{A}$ in~\eqref{def:consistency_hat_d}. Since we require $\delta_{N, n} /h\rightarrow 0$ in Theorem~\ref{thm:bias_variance}, we  employ the same argument used to prove~\eqref{eq:lemma_numerator_2_eq_2} and~\eqref{eq:lemma_numerator_2_eq_3} to show that (the proof is analogous and is therefore omitted) there exist constants $0<D_4 < D_5 <\infty$  such that\footnote{
            Here we drop the conditioning because it is straightforward to verify that the analysis does not depend on the specific realization of $\{\zeta_i\}_{i \in \mc{S}^c}$. 
        }
        \begin{equation}\label{eq:lemma_numerator_2_eq_4}
        \begin{aligned}
            \mb{P}\bigg( & D_4  n^2 h^{2d_{\zeta}}   \leq \min_{i, j\in \mc{S}^c} \sum_{i', j'\in \mc{S}_2 }  \bs{1}(\hat{d}_{ii'}\leq h \vee d(\zeta_i, \zeta_{i'})\leq h) \bs{1}(\hat{d}_{jj'}\leq h \vee d(\zeta_j, \zeta_{j'})\leq h) \leq \\
            & \max_{i, j\in \mc{S}^c} \sum_{i', j'\in \mc{S}_2 }  \bs{1}(\hat{d}_{ii'}\leq h \vee d(\zeta_i, \zeta_{i'})\leq h) \bs{1}(\hat{d}_{jj'}\leq h \vee d(\zeta_j, \zeta_{j'})\leq h)    \leq  D_5  n^2 h^{2d_{\zeta}}  \bigg)  \leq 1 - \delta n^{-1/2} \\
            \Rightarrow \mb{P}\bigg( &  \max_{i, j\in \mc{S}^c}\left|\sum_{i', j'\in \mc{S}_2 }(\hat{K}^{(ij)}_{h, i'j'} - K^{(ij)}_{h, i'j'})\right|  \leq  D_5  n^{2} h^{2d_{\zeta}-1} \delta_{N, n} \bigg)  \geq 1 - \delta n^{-1/2}.
        \end{aligned}   
        \end{equation} 
        Then, by combining~\eqref{eq:lemma_numerator_2_K} and~\eqref{eq:lemma_numerator_2_eq_4}, there exist constants $0 < \underline{D}_2 <\bar{D}_2  < \infty$ such that 
        \begin{align*}
            \mb{P}\bigg(  \underline{D}_2  n^2h^{2d_{\zeta}} \leq \min_{i, j\in \mc{S}^c}\sum_{i',  j'\in \mc{S}_2} \hat{K}^{(ij)}_{h, i'j'} \leq \max_{i, j\in \mc{S}^c}\sum_{i',  j'\in \mc{S}_2} \hat{K}^{(ij)}_{h, i'j'}  \leq \bar{D}_2 n^2h^{2d_{\zeta}} \bigg) 
            \geq  1 - \delta n^{-1/2}. 
        \end{align*}
        This completes the proof of~\eqref{eq:lemma_numerator_2_K_hat}.

        \bigskip 

        To prove \eqref{eq:lemma_numerator_2_product}, let $g^{(ij)}_{i'j'}$  denote $K^{(ij)}_{h, i'j'}(\zeta_{i}' - \zeta_{i'}', \zeta_{j}' - \zeta_{j'}') H^{(ij)}_{i'j'} (\zeta_{i}' - \zeta_{i'}', \zeta_{j}' - \zeta_{j'}')' $ for notational simplicity.  Fixing any $\zeta_i, \zeta_j\in \{\zeta_i\}_{i\in \mc{S}^c}$, by the boundedness of $K^{(i)}_{h, i'}$ and $H^{(i)}_{i'}$ (Assumption~\ref{assumption:kernel} and Assumption~\ref{assumption:regularity}\ref{item:regularity_assumption_smoothness}),  we have that $g^{(ij)}_{ i'j'}$ is uniformly bounded and $\mb{E}\left(\mb{E}\left(g^{(ij)}_{i'j'} \mid \zeta_i, \zeta_j, \zeta_{i'}\right)^2 \mid \zeta_i, \zeta_j\right) = O(h^{3d_{\zeta} + 4})$. Then, we apply Bernstein's inequality for U-statistics (see \citet[Theorem~2]{arcones1995bernstein}) to show that there exists a sufficiently large constant $M > 0$ such that for any $t > 0$, 
        \begin{align*}
            \mb{P}\left(\left| \sum_{i', j'\in \mc{S}_2 } \left(g^{(ij)}_{i'j'} - \mb{E}\left(g^{(ij)}_{i'j'}\mid \{\zeta_i\}_{i\in \mc{S}^c} \right)\right)   \right|\geq t \mid \{\zeta_i\}_{i\in \mc{S}^c} \right)  \leq 4\exp\left(\frac{-t^2}{M^2 n^3 h^{3d_{\zeta} + 4} + M n h^2 t} \right), 
        \end{align*}   
        and 
        \begin{align*}
            \mb{P}\left(\max_{i, j\in \mc{S}^c }\left| \sum_{i', j'\in \mc{S}_2 } \left(g^{(ij)}_{i'j'} - \mb{E}\left(g^{(ij)}_{i'j'}\mid \{\zeta_i\}_{i\in \mc{S}^c} \right)\right)   \right|\geq t \mid \{\zeta_i\}_{i\in \mc{S}^c} \right)  \leq 4N^2 \exp\left(\frac{-t^2}{M^2 n^3 h^{3d_{\zeta} + 4} + M n h^2 t} \right). 
        \end{align*}   
        It follows that we can choose a sufficiently large constant $\bar{D}_3  > 0$  such that when $t = \bar{D}_3\sqrt{n^3h^{3d_{\zeta}+4}\log N }$,  
        \begin{align*}
            \mb{P}\left(\max_{i, j\in \mc{S}^c }\left| \sum_{i', j'\in \mc{S}_2 } \left(g^{(ij)}_{i'j'} - \mb{E}\left(g^{(ij)}_{i'j'}\mid \{\zeta_i\}_{i\in \mc{S}^c} \right)\right)   \right| \geq  \bar{D}_3\sqrt{n^3h^{3d_{\zeta}+4}\log N } \mid \{\zeta_i\}_{i\in \mc{S}^c} \right) \leq \delta n^{-1/2}.
        \end{align*} 
        This completes the proof of~\eqref{eq:lemma_numerator_2_product}. 
    \end{prooflmm}

\clearpage

    \section{Proofs for Empirical Applications}\label{appendix:empirical}

    \begin{proofthm}{thm:consistency_plug_in}
        Consider the following decomposition, 
        \begin{align*}
            & \sup_{\alpha \in \mr{supp}(\alpha)} \left\| \psi(\hat{A},  Y_m, W_m, \alpha ) - \mb{E}\left(\psi\left(A_m,   W_m, Y_m, \alpha_0 \right)\right) \right\|\\
            \leq & \sup_{\alpha \in \mr{supp}(\alpha)}  \left\| \frac{1}{M}\sum_{m=1}^{M}\psi(\hat{A},  W_m, Y_m, \alpha ) -  \frac{1}{M}\sum_{m=1}^{M}\psi(P_m,   W_m, Y_m, \alpha ) \right\| \\
            & + \sup_{\alpha \in \mr{supp}(\alpha)} \left\| \frac{1}{M}\sum_{m=1}^{M} \psi(P_m,   W_m, Y_m, \alpha ) -  \frac{1}{M}\sum_{m=1}^{M}\psi(A_m,   W_m, Y_m, \alpha ) \right\| \\
            & + \sup_{\alpha \in \mr{supp}(\alpha)}  \left\| \frac{1}{M}\sum_{m=1}^{M}\psi(A_m,   W_m, Y_m, \alpha ) - \mb{E}\left(\psi\left(A_m,   W_m, Y_m, \alpha_0 \right)\right) \right\|. 
        \end{align*}
        For the first term on the RHS, 
        \begin{align*}
            & \sup_{\alpha \in \mr{supp}(\alpha)} \left\| \frac{1}{M}\sum_{m=1}^{M}\psi(\hat{A}_m,   W_m, Y_m, \alpha ) - \frac{1}{M}\sum_{m=1}^{M} \psi(P_m,   W_m, Y_m, \alpha ) \right\| \\
            \leqtext{(i)} & \frac{1}{M}\sum_{m=1}^{M} L(  W_m, Y_m) N_m^{-1} \|\hat{A}_m - P_m\|_{\mr{F}}\\ 
            \leqtext{(ii)} & \frac{1}{M}\sum_{m=1}^{M} L(  W_m, Y_m) \sup_{i, j =1, \ldots, N_m}(\hat{A}_{m, ij} - P_{m, ij}) \\
            \leqtext{(iii)} & \left(\frac{1}{M}\sum_{m=1}^{M} L^2(  W_m, Y_m)\right) \left(\frac{1}{M}\sum_{m=1}^{M} \sup_{i, j =1, \ldots, N_m}(\hat{A}_{m, ij} - P_{m, ij})^2 \right).  
        \end{align*}
        Here, inequality (i) follows from  Assumption~\ref{assumption:consistency_empirics}\ref{item:consistency_empirics_assumption_ULLN}, inequality (ii) and (iii) follow from Cauchy-Schwarz inequality. By the law of large numbers and Assumption~\ref{assumption:consistency_empirics}\ref{item:consistency_empirics_assumption_ULLN}, we have 
        \begin{align*}
            \left(\frac{1}{M}\sum_{m=1}^{M} L(  W_m, Y_m)\right)^2 \cp \mb{E}L^2(W_m, Y_m) <\infty. 
        \end{align*}
        In addition, Theorem~\ref{thm:bias_variance} ensures that $\left(\frac{1}{M}\sum_{m=1}^{M} \sup_{i, j =1, \ldots, N_m}(\hat{A}_{m, ij} - P_{m, ij})^2 \right)\cp 0$. Therefore, 
        \begin{align}\label{eq:consistency_1}
            \sup_{\alpha \in \mr{supp}(\alpha)} \left\| \frac{1}{M}\sum_{m=1}^{M}\psi(\hat{A}_m,   W_m, Y_m, \alpha ) - \frac{1}{M}\sum_{m=1}^{M} \psi(P_m,   W_m, Y_m, \alpha ) \right\| \cp 0. 
        \end{align}
        For the second term, Assumption~\ref{assumption:consistency_empirics}\ref{item:consistency_empirics_assumption_graphon} ensures that 
        \begin{align}\label{eq:consistency_2}
            \sup_{\alpha \in \mr{supp}(\alpha)} \left\| \frac{1}{M}\sum_{m=1}^{M} \psi(P_m,   W_m, Y_m, \alpha ) -  \frac{1}{M}\sum_{m=1}^{M}\psi(A_m,   W_m, Y_m, \alpha ) \right\|  \cp 0. 
        \end{align} 
        For the last term, by Assumption~\ref{assumption:consistency_empirics}\ref{item:consistency_empirics_assumption_compactness_and_integrability}, we have the uniform law of large numbers: 
        \begin{align}\label{eq:consistency_3}
            \sup_{\alpha \in \mr{supp}(\alpha)} \left\| \frac{1}{M}\sum_{m=1}^{M}\psi(A_m,  W_m, Y_m, \alpha ) -  \mb{E}\left(\psi\left(A_m,   W_m, Y_m, \alpha_0 \right)\right)  \right\| \cp 0. 
        \end{align}

        Thus, \eqref{eq:consistency_1}, \eqref{eq:consistency_2}, and \eqref{eq:consistency_3} imply that 
        \begin{align}\label{eq:consistency_4}
            \sup_{\alpha \in \mr{supp}(\alpha)} \left\| \psi(\hat{A},  Y_m, W_m, \alpha ) - \mb{E}\left(\psi\left(A_m,   W_m, Y_m, \alpha_0 \right)\right) \right\| \cp 0. 
        \end{align}
        Combining Assumptions~\ref{assumption:consistency_empirics}\ref{item:consistency_empirics_assumption_asymptotics}--\ref{item:consistency_empirics_assumption_W_n}, and the uniform convergence~\eqref{eq:consistency_4}, one cam employ \citet[Theorem~2.1]{newey1994large} to conclude that $\hat{\alpha} \cp \alpha_0$. 
    \end{proofthm}

    \begin{proofthm}{thm:linear_in_mean}
        For any $m\in \{1, \ldots, M\}$ and $i = 1, \ldots, N_m$, consider the following expansion of $(\hat{G}_mW_m - G_mW_m)_{ i}$ up to second order\footnote{
            It is straightforward to verify that, by~\eqref{eq:lemma_sum_G_ij_bias}, \eqref{eq:lemma_sum_G_ij_variance}, and \eqref{eq:lemma_sum_G_ij_bound}, higher order expansion does not (asymptotically) contribute the bias and variance of $(\hat{G}_mW _m- G_mW_m)_{i}$. 
        }, 
        \begin{align*}
            (\hat{G}_mW_m - G_mW_m)_{i} = & \sum_{ j = 1, \ldots, N_m }(\hat{G}_{m, ij} - G_{m, ij})W_{m, j} \\
            = & \frac{\sum_{j = 1, \ldots, N_m } A_{ij} W_{m, j} + \sum_{j = 1, \ldots, N_m}(\hat{A}_{m, ij} - A_{m, ij}) W_{m, j} }{\sum_{ j = 1, \ldots, N_m }A_{m, ij} + \left(\sum_{j = 1, \ldots, N }\hat{A}_{m, ij}- \sum_{j = 1, \ldots, N_m}A_{m, ij}\right)} - \frac{\sum_{j = 1, \ldots, N_m} A_{m, ij} W_{m, j} }{\sum_{j = 1, \ldots, N_m}A_{m, ij}} \\
            \approx & \frac{\sum_{j = 1, \ldots, N_m}(\hat{A}_{m, ij} - A_{m, ij}) W_{m, j}}{\sum_{j = 1, \ldots, N_m}A_{m, ij}} - \frac{\sum_{j = 1, \ldots, N_m}A_{m, ij}W_{m, j} }{\left(\sum_{j = 1, \ldots, N_m}A_{m, ij}\right)^2 }\sum_{j = 1, \ldots, N_m}(\hat{A}_{m, ij} - A_{m,ij})  \\
            & - \frac{ 1}{\left(\sum_{j = 1, \ldots, N_m}A_{m, ij}\right)^2 }\left(\sum_{j = 1, \ldots, N_m}(\hat{A}_{m, ij} - A_{m, ij})\right)\left(\sum_{j = 1, \ldots, N_m}(\hat{A}_{m, ij} - A_{m, ij})W_{m, j}\right) \\
            & + \frac{\sum_{j = 1, \ldots, N_m}A_{m, ij}W_{m, j} }{\left(\sum_{j = 1, \ldots, N_m}A_{m, ij}\right)^3 }\left(\sum_{j = 1, \ldots, N_m}(\hat{A}_{m, ij} - A_{m, ij})\right)^2. 
        \end{align*}
        Since $\sum_{j\in \mc{S}}A_{m, ij} \asymp N_m$ with probability at least $1-\delta n_m^{-1/2}$, and $W$ is uniformly bounded, $\sum_{j\in \mc{S}}A_{m, ij}W_{m, j} = O(N_m)$ with probability at least $1-\delta n_m^{-1/2}$. Then, by~\eqref{eq:lemma_sum_G_ij_bias}, \eqref{eq:lemma_sum_G_ij_variance}, and \eqref{eq:lemma_sum_G_ij_bound}, we obtain that, with probability at least $1-\delta n_m^{-1/2}$, 
        \begin{align*}
            \mb{E}\left((\hat{G}_mW_m - G_mW_m)_{i} \mid \{\zeta_{m, i}, W_{m, i}\}_{i=1}^{N_m}\right) = O\left(  h^2 + \frac{\log N_m}{n^2_m h^{2d_{\zeta}}} \right), \\
            \mr{Var}\left((\hat{G}_mW_m - G_mW_m)_{m, i} \mid \{\zeta_{m, i}, W_{m, i}\}_{i=1}^{N_m}\right) = O\left(  \delta_{N, n}^2 h^2 + \frac{\log N_m }{n_m^2 h^{2d_{\zeta}}} \right). 
        \end{align*}
        One can also show that, using concentration inequalities in Lemma~\ref{lemma:numerator_1} and Lemma~\ref{lemma:numerator_2}, $(\hat{G}_mW_m - G_mW_m)_{ i} = O\left(h^2 + \frac{\log N_m}{n^2_m h^{2d_{\zeta}}} \right)$ with probability at least $1-\delta n_m^{-1/2}$. 
        Also, it is straightforward to extend the result to $(\hat{G}_mY_m - G_mY_m)_i$, such that
        \begin{align*}
            \mb{E}\left((\hat{G}_mY_m - G_mY_m)_{i} \mid \{\zeta_{m, i}, W_{m, i}\}_{i=1}^{N_m}\right) = O\left(  h^2 + \frac{\log N_m}{n^2_m h^{2d_{\zeta}}} \right),  \\
            \mr{Var}\left((\hat{G}_mY_m - G_mY_m)_{m, i} \mid \{\zeta_{m, i}, W_{m, i}\}_{i=1}^{N_m}\right) = O\left(  \delta_{N, n}^2 h^2 + \frac{\log N_m }{n_m^2 h^{2d_{\zeta}}} \right), 
        \end{align*} 
        and $(\hat{G}_mW_m - G_mW_m)_{ i} = O\left(h^2 + \frac{\log N_m}{n^2_m h^{2d_{\zeta}}} \right)$ with probability at least $1-\delta n_m^{-1/2}$. 

        \bigskip
        
        Consider the following approximation of $(\hat{G}_mW_m)' \hat{G}_mW_m  - (G_mW_m)' G_mW_m$\footnote{
            By the bias, the variance, and the probability bound of $(\hat{G}_mW_m - G_mW_m)_i$, higher order expansion does not (asymptotically) contribute the bias and variance. 
        }, Then, 
        \begin{align*}
            (\hat{G}_mW_m)' \hat{G}_mW_m  - (G_mW_m)' G_mW_m  \approx 2 (G_mW_m)'(\hat{G}_mW_m - G_mW_m). 
        \end{align*}
        It follows that, with probability at least $1-\delta n_m^{-1/2}$, 
        \begin{align*}
            \mb{E}\left((\hat{G}_mW_m)' \hat{G}_mW_m  - (G_mW_m)' G_mW_m \mid \{\zeta_{m, i}, W_{m, i}\}_{i=1}^{N_m}\right) = N_m O\left(  h^2 + \frac{\log N_m}{n^2_m h^{2d_{\zeta}}} \right), \\
              \mr{Var}\left((\hat{G}_mW_m)' \hat{G}_mW_m  - (G_mW_m)' G_mW_m \mid \{\zeta_{m, i}, W_{m, i}\}_{i=1}^{N_m}\right)  = N_m^2 O\left(  \delta_{N, n}^2 h^2 + \frac{ \log N_m }{n_m^2 h^{2d_{\zeta}}} \right),
        \end{align*}
        and $ (\hat{G}_mW_m)' \hat{G}_mW_m  - (G_mW_m)' G_mW_m   = O\left(h^2 + \frac{\log N_m}{n^2_m h^{2d_{\zeta}}}\right)$ with probability at least $1-\delta n_m^{-1/2}$. Define 
        \begin{align*}
            \Delta_{m, 1}:= (\hat{V}_m' \hat{Z}_m/ N_m) \Sigma_M (\hat{Z}_m'\hat{V}_m/ N_m)  - (V_m' Z_m /N_m) \Sigma_M (Z_m'V_m/N_m), \\
            \Delta_{m, 2}:=(\hat{V}_m' \hat{Z}_m/N_m) \Sigma_M (\hat{Z}_m'Y_m/N_m)  - (V_m' Z_m/N-m) \Sigma_M (Z_m'Y_m/N-m). 
        \end{align*}
        It is straightforward to extend the above results to show that, for $k\in \{1, 2\}$, with probability at least $1-\delta n_m^{-1/2}$, 
        \begin{equation}\label{eq:thm:linear_in_mean_delta}
          \begin{aligned}
             \mb{E}\left(\Delta_{m, k} \mid \{\zeta_{m, i}, W_{m, i}\}_{i=1}^{N_m}\right)  = & O\left(  h^2 + \frac{\log N_m}{n^2_m h^{2d_{\zeta}}} \right), \\
            \mr{Var}\left(\Delta_{m, k}\mid \{\zeta_{m, i}, W_{m, i}\}_{i=1}^{N_m}\right) = &  O\left(  \delta_{N, n}^2 h^2 + \frac{ \log N_m }{n_m^2 h^{2d_{\zeta}}} \right), \\
              \Delta_{m, k} = & O\left(h^2 + \frac{\log N_m}{n^2_m h^{2d_{\zeta}}}\right). 
        \end{aligned}  
        \end{equation}
        For notational simplicity, denote $\Delta_1:= \sum_{m=1}^{M} \Delta_{m, 1}$ and $\Delta_2:= \sum_{m=1}^{M} \Delta_{m, 2}$. Since each network is i.i.d., by~\eqref{eq:thm:linear_in_mean_delta}, for $k\in \{1, 2\}$, with probability at least $1-\delta n_m^{-1/2}$, we have 
        \begin{equation}\label{eq:thm:linear_in_mean_delta_big}
          \begin{aligned}
            \mb{E}\left(\Delta_{ k} \mid \{\zeta_{m, i}, W_{m, i}\}_{i=1}^{N_m}\right) = & N_m O\left(  h^2 + \frac{\log N_m}{n^2_m h^{2d_{\zeta}}} \right), \\
             \mr{Var}\left(\Delta_{ k}\mid \{\zeta_{m, i}, W_{m, i}\}_{i=1}^{N_m}\right) = & N_m O\left(  \delta_{N, n}^2 h^2 + \frac{ \log N_m }{n_m^2 h^{2d_{\zeta}}} \right). 
        \end{aligned}  
        \end{equation}
        The GMM estimator  $\hat{\alpha}$ can be approximated by\footnote{
            It is straightforward to verify that, by the bias, the variance, and the probability bound of $\Delta_{m, k}$, higher order expansion does not (asymptotically) contribute the bias and variance of $\hat{\alpha}$.  
        } 
        \begin{align*}
            \hat{\alpha} = & \left( \sum_{m=1}^{M}(\hat{V}_m' \hat{Z}_m / N_m) \Sigma_m (\hat{Z}_m'\hat{V}_m/N_m) \right)^{-1} \left( \sum_{m=1}^{M}(\hat{V}_m' \hat{Z}_m /N_m)\Sigma_m (\hat{Z}_m'Y_m/N_m) \right) \\
            = & \left( \sum_{m=1}^{M}(V_m' Z_m/N_m) \Sigma_m (Z_m'V_m/N_m) + \Delta_1 \right)^{-1} \left(\sum_{m=1}^{M}(V_m' Z_m/N_m) \Sigma_m (Z_m'Y_m/N_m) + \Delta_2 \right) \\
            \approx & \left(\sum_{m=1}^{M}V_m' Z_m \Sigma_m Z_m'V_m /N_m^2 \right)^{-1}\left(I + \Delta_1 \left(\sum_{m=1}^{M}V_m' Z_m \Sigma_m Z_m'V_m/N_m^2\right)^{-1}\right)\left(\sum_{m=1}^{M}V_m' Z_m \Sigma_m Z_m'Y_m/N_m^2 + \Delta_2 \right) \\
            \approx & \underbrace{\left(\sum_{m=1}^{M}V_m' Z_m \Sigma_m Z_m'V_m/N_m^2 \right)^{-1}\sum_{m=1}^{M}V_m' Z_m \Sigma_m Z_m'Y_m/N_m^2}_{U_1} + \underbrace{\left(\sum_{m=1}^{M}V_m' Z_m \Sigma_m Z_m'V_m/N_m^2 \right)^{-1}\Delta_2}_{U_2}  \\
            & - \underbrace{\left(\sum_{m=1}^{M}V_m' Z_m \Sigma_m Z_m'V_m/N_m^2 \right)^{-1}\Delta_1 \left(\sum_{m=1}^{M}V_m' Z_m \Sigma_m Z_m'V_m/N_m^2 \right)^{-1}\left(\sum_{m=1}^{M}V_m' Z_m \Sigma_m Z_m'Y_m /N_m^2 \right)}_{U_3}. 
        \end{align*}
        Here,  $\frac{1}{\sqrt{M}}(U_1 - \alpha) \cd N(0, \Omega)$ by Central Limit Theorem, where $\Omega$ is the asymptotic variance when using complete network data.  
        Let $N:= \frac{1}{M}\sum_{m=1}^{M}N_m$ and $n:=\frac{1}{M}\sum_{m=1}^{M}n_m$. By~\eqref{eq:thm:linear_in_mean_delta_big},  Assumption~\ref{assumption:linear_in_mean_peer_effect}\ref{item:linear_in_mean_peer_effect_sampling}, Assumption~\ref{assumption:linear_in_mean_peer_effect}\ref{item:linear_in_mean_peer_effect_asymptotics}, and Assumption~\ref{assumption:linear_in_mean_peer_effect}\ref{item:linear_in_mean_peer_effect_non_degenerate}, with probability at least $1-\delta n^{-1/2}$, we obtain 
        \begin{align*}
             \mb{E}\left(U_2 + U_3 \right)  = & O\left(h^2 + \frac{\log N_m}{n^2_m h^{2d_{\zeta}}} \right), \\
            \mr{Var} \left(U_2 + U_3 \right) = & \frac{1}{M}O\left(h^2\delta_{N, n}^2 + \frac{\log N_m}{n^2_m h^{2d_{\zeta}}} \right). 
        \end{align*}
        This completes the proof. 
    \end{proofthm}

    \begin{lemma}\label{lemma:sum_G_ij}
        Under conditions in Theorem~\ref{thm:bias_variance} and Assumption~\ref{assumption:linear_in_mean_peer_effect}, for any $i\in \mc{S}^c$, with probability at least $1-\delta n^{-1/2}$, 

        \begin{equation}\label{eq:lemma_sum_G_ij_bias}
        \begin{aligned}
             \mb{E}\left(\sum_{j = 1, \ldots, N}(A_{ij} - \hat{A}_{ij})\mid \{\zeta_i, W_i\}_{i=1}^{N}\right) = O(N h^2), \quad  \mb{E}\left(\sum_{j = 1, \ldots, N}(A_{ij} - \hat{A}_{ij})W_j\mid \{\zeta_i, W_i\}_{i=1}^{N}\right) = O(N h^2), 
        \end{aligned}   
        \end{equation}  
        and 
        \begin{equation}\label{eq:lemma_sum_G_ij_variance}
        \begin{aligned}
            \mr{Var}\left(\sum_{j = 1, \ldots, N} (\hat{A}_{ij} - A_{ij})  \mid \{\zeta_i, W_i\}_{i=1}^{N} \right) = O\left(N^2 \left(\delta_{N, n}^2 h^2 + \frac{\log N }{n^2 h^{2d_{\zeta}}}  \right)\right), \\
            \mr{Var}\left(\sum_{j = 1, \ldots, N} (\hat{A}_{ij} - A_{ij}) W_j  \mid \{\zeta_i, W_i\}_{i=1}^{N} \right) = O\left(N^2 \left(\delta_{N, n}^2 h^2 + \frac{\log N }{n^2 h^{2d_{\zeta}}}  \right)\right). 
        \end{aligned}   
        \end{equation}
        In addition, for any $\delta >0$, there exist constant $E_1, E_2 \geq 0$, such that 
        \begin{equation}\label{eq:lemma_sum_G_ij_bound}
        \begin{aligned}
            \mb{P}\left(\left| \sum_{ j = 1, \ldots, N } (\hat{A}_{ij} - A_{ij})\right|  \geq E_1 N \left( h^2 + \frac{\log N}{nh^{d_{\zeta}}} \right)\right) \leq \delta n^{-1/2}. 
        \end{aligned}   
        \end{equation}

    \end{lemma}

    \begin{prooflmm}{lemma:sum_G_ij}
        We only present the proof of~\eqref{eq:lemma_sum_G_ij_bias} and~\eqref{eq:lemma_sum_G_ij_variance}, as~\eqref{eq:lemma_sum_G_ij_bound} follows directly by using concentration inequalities in Lemma~\ref{lemma:numerator_1} and Lemma~\ref{lemma:numerator_2}. 
        For any $i, j\in \mc{S}^c$ and $i', j'\in \mc{S}^c$, define $Q^{(ij)}_2$ and $Q^{(ij)}_3$ as in~\eqref{eq:TWFE_solution}, and $\hat{R}^{(ij)}_3$ as in~\eqref{eq:definition_R_hat}, define $\hat{R}^{(ij)}_1$, $\hat{R}^{(ij)}_2$, and $\hat{R}^{(ij)}_3$ as in~\eqref{eq:definition_R_hat}, define $R^{(ij)}_1$, $R^{(ij)}_2$, and $R^{(ij)}_3$ as in~\eqref{eq:definition_R}, and define $H^{(ij)}_{i'}$, $H^{(ij)}_{j'}$, and $H^{(ij)}_{i'j'}$ as in~\eqref{eq:definition_H}. Note that 
        \begin{align*}
            \sum_{j = 1, \ldots, N} (\hat{A}_{ij} - A_{ij}) =  \sum_{j \in \mc{S}^c } \sum_{k=1, 2, 3}R ^{(ij)}_k  + \sum_{j \in \mc{S}^c } \sum_{k=1, 2, 3}(\hat{R}^{(ij)}_k - R^{(ij)}_k )+  \sum_{j \in \mc{S}^c } Q^{(ij)}_2 + \sum_{j \in \mc{S}^c } Q^{(ij)}_3  -  \sum_{j \in \mc{S}^c } \epsilon_{ij}. 
        \end{align*}
        \paragraph{Bias} 
        By~\eqref{eq:thm:bias_variance:bound_Q1},  
        \eqref{eq:thm:bias_variance:bound_Q2}, Assumption~\ref{assumption:regularity}\ref{item:consistency_empirics_assumption_independence}, and Assumption\ref{assumption:linear_in_mean_peer_effect}\ref{item:linear_in_mean_peer_effect_exogeneity}, we have, with probability at least $1-\delta n^{-1/2}$,  
        \begin{align*}
            \mb{E}\left(\sum_{j \in \mc{S}^c } \sum_{k=1, 2, 3}\hat{R}^{(ij)}_k \mid \{\zeta_i, W_i\}_{i=1}^{N}\right) \leq 6 N \rho_{f''} \eta^2h^2, 
        \end{align*}
        and 
        \begin{align*}
            \mb{E}\left(\sum_{j \in \mc{S}^c } Q^{(ij)}_2  -  \sum_{j \in \mc{S}^c } \epsilon_{ij}  \mid \{\zeta_i, W_i\}_{i=1}^{N}\right) = 0. 
        \end{align*}
        In addition, with probability at least $1-\delta n^{-1/2}$,
        \begin{equation}\label{eq:lemma:sum_G_ij_Q_3}
            \begin{aligned}
            \sum_{j\in \mc{S}^c}  Q^{(ij)}_{3} = \sum_{j\in \mc{S}^c} \frac{\sum_{i'\in \mc{S}_2  }\hat{K}^{(ij)}_{h, i'i'} f_{i'i'} }{ \sum_{i', j'\in \mc{S}_2   } \hat{K}^{(ij)}_{h, i'j'} }  \leqtext{(i)} 
            & \frac{1}{\underline{D}_2 n^2 h^{2d_{\zeta}}} \sum_{j\in \mc{S}^c}  \left|\sum_{i'\in \mc{S}_2  }\hat{K}^{(ij)}_{h, i'i'} f_{i'i'} \right|  \\
            \leq & \frac{1}{\underline{D}_2 n^2 h^{2d_{\zeta}}}   \left| \sum_{i'\in \mc{S}_2  } \sum_{j\in \mc{S}^c} \hat{K}^{(ij)}_{h, i'i'} f_{i'i'} \right|  \\
            \leq & \frac{\rho^2_K \rho_f}{\underline{D}_2 n^2 h^{2d_{\zeta}}}  \left|\sum_{i'\in \mc{S}_2} \bs{1} (d_{ii'} \leq h)\sum_{j'\in \mc{S}^c  } \bs{1}(d_{ij} \leq h) \right|  \\
            \leqtext{(ii)} & \frac{\rho^2_K \rho_f}{\underline{D}_2 n^2 h^{2d_{\zeta}}} \bar{C}_1^2  n N h^{2 d_{\zeta}} \\
            = & \frac{\bar{C}_1^2 \rho^2_K \rho_f N }{\underline{D}_2 n} .   
        \end{aligned}
        \end{equation} 
        Here, (i) holds with probability at least $1-\delta n^{-1/2}$ because of~\eqref{eq:lemma_numerator_2_K_hat} in Lemma~\ref{eq:lemma_numerator_2_K}, (ii) follows from~\eqref{eq:lemma_numerator_1_K} by replacing $K(\cdot)$ with a uniform kernel. 
        Therefore, we conclude that, with probability at least $1 - \delta n^{-1/2}$, 
        \begin{align*}
            \mb{E}\left(\sum_{j = 1, \ldots, N  } (\hat{A}_{ij} - f_{ij})\mid \{\zeta_i, W_i\}_{i=1}^{N}\right) \leq O\left(Nh^2 + \frac{N}{n}\right) = O(Nh^2). 
        \end{align*}
        The last equality holds because of conditions in Theorem~\ref{thm:bias_variance} and $d_{\zeta} \geq 2$. Since $W_i$ is uniformly bounded (Assumption~\ref{assumption:linear_in_mean_peer_effect}\ref{item:linear_in_mean_peer_effect_compact}), it immediately follows that, with probability at least $1 - \delta n^{-1/2}$, 
        \begin{align*}
            \mb{E}\left(\sum_{ j = 1, \ldots, N } (\hat{A}_{ij} - f_{ij})W_j \mid \{\zeta_i, W_i\}_{i=1}^{N}\right) = O(Nh^2). 
        \end{align*}

        \paragraph{Variance}
        
        Note that 
        \begin{align*}
            & \mr{Var}\left(\sum_{j = 1, \ldots, N} (\hat{A}_{ij} - A_{ij})  \mid \{\zeta_i, W_i\}_{i=1}^{N} \right)\\
            =  & 4 \mr{Var}\left( \sum_{j\in \mc{S}^c} \sum_{k=1, 2, 3}(\hat{R}^{(ij)}_k - R^{(ij)}_k ) \mid \{\zeta_i, W_i\}_{i=1}^{N} \right) + 4 \mr{Var}\left( \sum_{j\in \mc{S}^c}   Q^{(ij)}_2 \mid \{\zeta_i, W_i\}_{i=1}^{N} \right) \\
            &+ 4 \mr{Var}\left( \sum_{j \in \mc{S}^c } Q^{(ij)}_3 \mid \{\zeta_i, W_i\}_{i=1}^{N} \right)  + 4 \mr{Var}\left(  \sum_{j \in \mc{S}^c } \epsilon_{ij} \mid \{\zeta_i, W_i\}_{i=1}^{N} \right). 
        \end{align*}

        \paragraph{Step 1: Bound on $\mr{Var}\left( \sum_{k=1, 2, 3}(\hat{R}^{(ij)}_k - R^{(ij)}_k ) \mid \{\zeta_i, W_i\}_{i=1}^{N} \right)$} 
        By~\eqref{eq:thm_bias_variance_R_R_hat}, we have, with probability at least $1- \delta n^{-1/2} $, 
        \begin{align*}
            & \mr{Var}\left(\sum_{j\in \mc{S}^c}\hat{R}^{(ij)}_1 - R^{(ij)}_1 \mid  \{\zeta_i, W_i\}_{i=1}^{N} \right) \leq N^2 \max_{i, j\in \mc{S}^c }\mr{Var}\left(\hat{R}^{(ij)}_1 - R^{(ij)}_1 \mid  \{\zeta_i, W_i\}_{i=1}^{N} \right)= O(N^2 \delta_{N, n}^2 h^2), \\
            & \mr{Var}\left(\sum_{j\in \mc{S}^c}\hat{R}^{(ij)}_2 - R^{(ij)}_2 \mid  \{\zeta_i, W_i\}_{i=1}^{N} \right) \leq N^2 \max_{i, j\in \mc{S}^c }\mr{Var}\left(\hat{R}^{(ij)}_2 - R^{(ij)}_2 \mid  \{\zeta_i, W_i\}_{i=1}^{N} \right)= O(N^2 \delta_{N, n}^2 h^2). 
        \end{align*}
        Using the same argument, we obtain that, with probability at least $1- \delta n^{-1/2} $, 
        \begin{align*}
            \mr{Var}\left(\sum_{j\in \mc{S}^c}\hat{R}^{(ij)}_3 - R^{(ij)}_3 \mid  \{\zeta_i, W_i\}_{i=1}^{N} \right) \leq N^2 \max_{i, j\in \mc{S}^c }\mr{Var}\left(\hat{R}^{(ij)}_3 - R^{(ij)}_3 \mid  \{\zeta_i, W_i\}_{i=1}^{N} \right)= O(N^2 \delta_{N, n}^2 h^2). 
        \end{align*}
        Thus,  with probability at least $1- \delta n^{-1/2} $, 
        \begin{align*}
            \mr{Var}\left( \sum_{k=1, 2, 3}(\hat{R}^{(ij)}_k - R^{(ij)}_k ) \mid \{\zeta_i, W_i\}_{i=1}^{N} \right) = O(N^2 \delta_{N, n}^2 h^2). 
        \end{align*}

        \paragraph{Step 2: Bound on $\mr{Var}\left( \sum_{j\in \mc{S}^c}   Q^{(ij)}_2 \mid \{\zeta_i, W_i\}_{i=1}^{N} \right)$} 
        Note that  
        \begin{align*}
            \mr{Var}\left(\sum_{j\in \mc{S}^c }\frac{\sum_{i'\in \mc{S}_2 }\hat{K}^{(i)}_{h, i'} \epsilon_{i'j}}{\sum_{i'\in \mc{S}_2  }\hat{K}^{(i)}_{h, i'} } \mid  \{\zeta_i, W_i\}_{i=1}^{N} \right)  =  & \mr{Var}\left(\frac{\sum_{i'\in \mc{S}_2 }\hat{K}^{(i)}_{h, i'} \sum_{j\in \mc{S}^c } \epsilon_{i'j}}{\sum_{i'\in \mc{S}_2  }\hat{K}^{(i)}_{h, i'} } \mid  \{\zeta_i, W_i \}_{i=1}^{N} \right). 
        \end{align*}
        Since $\sum_{j\in \mc{S}^c } \epsilon_{i'j}$ is independent across $i'\in \mc{S}_2$ and sub-Gaussian with 
        $\sigma\sim \sqrt{n}$, we employ the method used in proving~\eqref{eq:thm_bias_variance_bound_Bernstein} to show that, with probability at least $1- \delta n^{-1/2} $, 
        \begin{align*}
            \mr{Var}\left(\sum_{j\in \mc{S}^c }\frac{\sum_{i'\in \mc{S}_2 }\hat{K}^{(i)}_{h, i'} \epsilon_{i'j}}{\sum_{i'\in \mc{S}_2  }\hat{K}^{(i)}_{h, i'} } \mid  \{\zeta_i, W_i\}_{i=1}^{N} \right)  =  O\left(\frac{N \log N}{nh^{d_{\zeta}}}\right). 
        \end{align*}
        Also, with probability at least $1- \delta n^{-1/2} $, 
        \begin{align*}
            \mr{Var}\left(\sum_{j\in \mc{S}^c }\frac{\sum_{j'\in \mc{S}_2 }\hat{K}^{(j)}_{h, j'} \epsilon_{ij'}}{\sum_{j'\in \mc{S}_2  }\hat{K}^{(j)}_{h, j'} } \mid  \{\zeta_i, W_i\}_{i=1}^{N} \right)  = & \mr{Var}\left(\frac{\sum_{j'\in \mc{S}_2 }  \left(\sum_{j\in \mc{S}^c } \hat{K}^{(j)}_{h, j'}\right) \epsilon_{ij'}}{\sum_{j'\in \mc{S}_2  }\hat{K}^{(j)}_{h, j'} } \mid  \{\zeta_i, W_i\}_{i=1}^{N} \right) \\
            \leq  & \frac{\max_{j'\in \mc{S}_2} \left(\sum_{j\in \mc{S}^c } \hat{K}^{(j)}_{h, j'}\right)^2  }{\min_{j'\in \mc{S}_2}\left(\sum_{j\in \mc{S}^c   }\hat{K}^{(j)}_{h, j'}\right)^2 } \mb{E}\left(\left(\sum_{j'\in \mc{S}_2 }  \epsilon_{ij'}\right)^2\right)  \\
            \leqtext{(i)} & O\left(\frac{N^2 h^{2d_{\zeta}}}{n^2 h^{2d_{\zeta}}} n \log N\right) \\
            = & O\left(\frac{N^2}{n} \log N\right). 
        \end{align*}
        In addition, with probability at least $1- \delta n^{-1/2} $,
        \begin{align*}
            \sum_{j\in \mc{S}^c} \mr{Var}\left(\sum_{j\in \mc{S}^c} \frac{\sum_{i', j'\in \mc{S}_2 \ }\hat{K}^{(ij)}_{h, i'j'} \epsilon_{i'j'}}{\sum_{i', j'\in \mc{S}_2 }\hat{K}^{(ij)}_{h, i'j'} } \mid \{\zeta_i, W_i \}_{i=1}^{N}\right) \leq & N^2 \max_{i, j \in \mc{S}^c} \mr{Var}\left(\frac{\sum_{i', j'\in \mc{S}_2 \ }\hat{K}^{(ij)}_{h, i'j'} \epsilon_{i'j'}}{\sum_{i', j'\in \mc{S}_2 }\hat{K}^{(ij)}_{h, i'j'} } \mid \{\zeta_i, W_i \}_{i=1}^{N}\right) \\
            \eqtext{(i)} & O\left( N^2\frac{\log N}{n^2 h^{2d_{\zeta}}}\right) , 
        \end{align*}
        where (i) follows from~\eqref{eq:thm_bias_variance_var_product_1}. 
        Therefore, we conclude that, when $d_{\zeta}\geq 2$, with probability at least $1- \delta n^{-1/2} $, 
        \begin{align*}
            \mr{Var}\left( \sum_{j\in \mc{S}^c}   Q^{(ij)}_2 \mid \{\zeta_i, W_i\}_{i=1}^{N} \right) = & O\left( \frac{N^2 \log N}{N nh^{d_{\zeta}}} + \frac{N^2}{n} \log N + N^2 \frac{\log N }{n^2 h^{2 d_{\zeta}}} \right) \\
            = & O\left( N^2 \frac{\log N }{n^2 h^{2 d_{\zeta}}} \right). 
        \end{align*}

        \paragraph{Step 3: Bound on $\mr{Var}\left( \sum_{j\in \mc{S}^c}   Q^{(ij)}_3 \mid \{\zeta_i, W_i\}_{i=1}^{N} \right)$} By~\eqref{eq:lemma:sum_G_ij_Q_3}, with probability at least $1- \delta n^{-1/2} $, 
        \begin{align*}
            \mr{Var}\left( \sum_{j\in \mc{S}^c}   Q^{(ij)}_3 \mid \{\zeta_i, W_i\}_{i=1}^{N} \right) = O\left(\frac{N^2}{n^2} \right). 
        \end{align*}
        \paragraph{Step 4: Bound on $ \mr{Var}\left(  \sum_{j \in \mc{S}^c } \epsilon_{ij} \mid \{\zeta_i, W_i\}_{i=1}^{N} \right) $ } Since $\{\epsilon_{ij} \mid i<j, i, j\in \mc{S}_2 \cup \mc{S}^c\}$, it is straightforward to verify that, with probability at least $1- \delta n^{-1/2} $, 
        \begin{align*}
            \mr{Var}\left(  \sum_{j \in \mc{S}^c } \epsilon_{ij} \mid \{\zeta_i, W_i\}_{i=1}^{N} \right) = O\left(N \log N\right). 
        \end{align*}

        Combining these results above, we conclude that, with probability at least $1- \delta n^{-1/2} $,         
        \begin{align*}
            \mr{Var}\left(\sum_{ j = 1, \ldots, N } (\hat{A}_{ij} - A_{ij})  \mid \{\zeta_i, W_i\}_{i=1}^{N} \right) = & O\left(N^2 \left(\delta_{N, n}^2 h^2 + \frac{\log N }{n^2 h^{2d_{\zeta}}} + \frac{1}{n^2} + \frac{1}{N} \right)\right)\\
            = & O\left(N^2 \left(\delta_{N, n}^2 h^2 + \frac{\log N }{n^2 h^{2d_{\zeta}}}  \right)\right). 
        \end{align*}
        Since $W_i$ is uniformly bounded (Assumption~\ref{assumption:linear_in_mean_peer_effect}\ref{item:linear_in_mean_peer_effect_compact}), it immediately follows that, with probability at least $1 - \delta n^{-1/2}$, 
        \begin{align*}
            \mr{Var}\left(\sum_{j = 1, \ldots, N } (\hat{A}_{ij} - A_{ij}) W_j  \mid \{\zeta_i, W_i\}_{i=1}^{N} \right) = O\left(N^2 \left(\delta_{N, n}^2 h^2 + \frac{\log N }{n^2 h^{2d_{\zeta}}}  \right)\right). 
        \end{align*}
        This completes the proof. 
    \end{prooflmm}

\clearpage

    \section{Proofs for Informativeness Condition and Technical Proofs}\label{appendix:supplementary}

    \begin{prooflmm}{lemma:SBM_informativeness}
        Without loss of generality, suppose that $f(\bar{\zeta}_1, \zeta_{g'}) >0 $ for all $g'$. Then, for any $g, g'\in \{1, \ldots, G\}$, 
        \begin{align*}
            & \max_{\zeta\in \{\bar{\zeta}_1, \ldots, \bar{\zeta}_G\} } \left|\int f(\zeta, \tilde{\zeta})(f(\bar{\zeta}_{g}, \tilde{\zeta}) - f(\bar{\zeta}_{g'}, \tilde{\zeta})) \mr{d}\mb{P}(\tilde{\zeta})\right| \geq \left|\int f(\bar{\zeta}_{g}, \tilde{\zeta})(f(\bar{\zeta}_{g}, \tilde{\zeta}) - f(\bar{\zeta}_{g'}, \tilde{\zeta})) \mr{d}\mb{P}(\tilde{\zeta})\right|, \\
            & \max_{\zeta\in \{\bar{\zeta}_1, \ldots, \bar{\zeta}_G\} } \left|\int f(\zeta, \tilde{\zeta})(f(\bar{\zeta}_{g}, \tilde{\zeta}) - f(\bar{\zeta}_{g'}, \tilde{\zeta})) \mr{d}\mb{P}(\tilde{\zeta})\right| \geq \left|\int f(\bar{\zeta}_{g'}, \tilde{\zeta})(f(\bar{\zeta}_{g}, \tilde{\zeta}) - f(\bar{\zeta}_{g'}, \tilde{\zeta})) \mr{d}\mb{P}(\tilde{\zeta})\right|.  
        \end{align*}
        Thus, 
        \begin{align*}
            \max_{\zeta\in \{\bar{\zeta}_1, \ldots, \bar{\zeta}_G\} } \left|\int f(\zeta, \tilde{\zeta})(f(\bar{\zeta}_{g}, \tilde{\zeta}) - f(\bar{\zeta}_{g'}, \tilde{\zeta})) \mr{d}\mb{P}(\tilde{\zeta})\right| \geq \frac{1}{2} \int (f(\bar{\zeta}_{g}, \tilde{\zeta}) - f(\bar{\zeta}_{g'}, \tilde{\zeta}))^2  \mr{d}\mb{P}(\tilde{\zeta}). 
        \end{align*}
        Therefore, the informativeness condition fails if and only if 
        \begin{align}\label{eq:SBM_sufficient}
            f(\bar{\zeta}_g, \zeta) = f(\bar{\zeta}_{g'}, \zeta), \quad \forall \zeta \in \{\bar{\zeta}_1, \ldots, \bar{\zeta}_G\}. 
        \end{align}
        In addition, when~\eqref{eq:SBM_sufficient} does not hold, let $d^2: = \min_{ g\neq g'} \frac{1}{2} \int (f(\bar{\zeta}_{g}, \tilde{\zeta}) - f(\bar{\zeta}_{g'}, \tilde{\zeta}))^2  \mr{d}\mb{P}(\tilde{\zeta})) $ and $r_{\infty}: = \max\{|\bar{\zeta}_1|, \ldots, |\bar{\zeta}_G| \}$. It follows that 
        \begin{align*}
            \lambda  = \frac{d^2}{2 r_{\infty}}. 
        \end{align*}
    \end{prooflmm}

    \begin{proofprop}{proposition:verification}
        By~\eqref{eq:verification_1}, for any $\delta_1 >0$, the compactness of $\mr{supp}(\zeta)$ (Assumption~\ref{assumption:regularity}\ref{item:regularity_assumption_compact_support}) and the continuity of $f$ imply that there exists a constant $\lambda_1 (\delta_1)$ which only depends on $\delta_1$ such that 
        \begin{align*}
            \inf_{\|\zeta_i - \zeta_{i'} \| \geq \delta_1 } \int \left(f(\zeta_{i} , \tilde{\zeta}) - f(\zeta_{i'}, \tilde{\zeta})\right)^2 \mr{d}\mb{P}(\tilde{\zeta}) \geq \lambda_1 (\delta_1). 
        \end{align*}
        In addition, since for any $\zeta_i, \zeta_{i'} \in \mr{supp}(\zeta)$, 
        \begin{align*}
            d(\zeta_{i}, \zeta_{i'}) \geq \frac{1}{2}\int \left(f(\zeta_{i} , \tilde{\zeta}) - f(\zeta_{i'}, \tilde{\zeta})\right)^2 \mr{d}\mb{P}(\tilde{\zeta}), 
        \end{align*}
        we conclude that 
        \begin{align*}
            \inf_{ \|\zeta_i - \zeta_{i'}  \|\geq \delta_1} d(\zeta_{i}, \zeta_{i'})  \geq \lambda_1(\delta_1). 
        \end{align*}
        Define $D:= \sup_{\zeta_i, \zeta_{i'} \in \mr{supp}(\zeta)} \|\zeta_{i} - \zeta_{i'}\|$. Since $\mr{supp}(\zeta)$ is compact, $D <\infty$. Therefore, for any $\delta_1 >0$, we have 
        \begin{align}\label{eq:verification_bound_1}
             d(\zeta_i, \zeta_{i'}) \geq \frac{\lambda_1(\delta_1)}{D} \|\zeta_i - \zeta_{i'}\|, \quad  \quad \forall \|\zeta_i - \zeta_{i'}  \|\geq \delta_1. 
        \end{align}

        Let $\rho_{f'} := \sup_{\zeta_1, \zeta_2\in \mr{supp}(\zeta)} \|\nabla f(\zeta_1, \zeta_2)\|_{\max}$, where $\|\cdot\|_{\max}$ denotes the largest absolute entry of a vector. Let $\rho_{f''} := \sup_{\zeta_1, \zeta_2\in \mr{supp}(\zeta)}\lambda_{\max}(\nabla_{\zeta_1\zeta_1}^2f(\zeta_1, \zeta_2))$, where $\lambda_{\max}(\cdot)$ denotes the largest singular value of a matrix. By Assumption~\ref{assumption:regularity}\ref{item:regularity_assumption_compact_support} and \ref{item:regularity_assumption_smoothness}, we have $\rho_{f'}, \rho_{f''} < \infty$.  Also, by Taylor expansion, we have, for any $\zeta_{i}, \zeta_{i'}\in  \mr{supp}(\zeta)$, 
        \begin{align*}
            d(\zeta_{i}, \zeta_{i'})\geq & \frac{1}{2}  \sup_{\zeta_1, \zeta_2 \in \mr{supp}(\zeta) } \left|  \int \left(f(\zeta_1, \tilde{\zeta}) -  f(\zeta_2, \tilde{\zeta})\right) \left(f(\zeta_{i}, \tilde{\zeta}) -  f(\zeta_{i'}, \tilde{\zeta})\right) \mr{d}\mb{P}(\tilde{\zeta})  \right| \\
            \geq & \frac{1}{2}  \sup_{\zeta_1 \in \mr{supp}(\zeta) } \left|  \int \left(f(\zeta_1, \tilde{\zeta}) -  f(\zeta_{i}, \tilde{\zeta})\right) \left(f(\zeta_{i}, \tilde{\zeta}) -  f(\zeta_{i'}, \tilde{\zeta})\right) \mr{d}\mb{P}(\tilde{\zeta})  \right| \\
            \geq & \frac{1}{2} \sup_{\zeta_1 \in \mr{supp}(\zeta)} \bigg\{\left|  \int \nabla'f(\zeta_i, \tilde{\zeta})(\zeta_1 - \zeta_i ) \nabla'f(\zeta_i, \tilde{\zeta})(\zeta_i - \zeta_{i'} )  \mr{d}\mb{P}(\tilde{\zeta})  \right|  \\
            & - \frac{1}{2}\sqrt{d_{\zeta}} \rho_{f'} \rho_{f''}\left(\|\zeta_1 - \zeta_i \|^2 \|\zeta_{i} - \zeta_{i'}\| + \|\zeta_1 - \zeta_i \| \|\zeta_{i} - \zeta_{i'}\|^2 \right) - \frac{1}{4} \rho_{f''}^2 \|\zeta_1 - \zeta_i \|^2 \|\zeta_{i} - \zeta_{i'}\|^2 \bigg\}\\
            \geq & \frac{1}{2} \sup_{\zeta_1 \in \mr{supp}(\zeta) \cap \mc{B}(\zeta_i, \delta_2)} \bigg\{\left|  \int \nabla'f(\zeta_i, \tilde{\zeta})(\zeta_1 - \zeta_i ) \nabla'f(\zeta_i, \tilde{\zeta})(\zeta_i - \zeta_{i'} )  \mr{d}\mb{P}(\tilde{\zeta})  \right|  \\
            & - \frac{1}{2}\sqrt{d_{\zeta}} \rho_{f'} \rho_{f''}\left(\|\zeta_1 - \zeta_i \|^2 \|\zeta_{i} - \zeta_{i'}\| + \|\zeta_1 - \zeta_i \| \|\zeta_{i} - \zeta_{i'}\|^2 \right) - \frac{1}{4} \rho_{f''}^2 \|\zeta_1 - \zeta_i \|^2 \|\zeta_{i} - \zeta_{i'}\|^2 \bigg\}\\ 
            \geqtext{(i)} & \frac{1}{2} \underline{\lambda} \min\left\{\frac{1}{2}, \frac{\delta_2}{2 D} \right\}\overline{\rho} \|\zeta_i - \zeta_{i'} \| - \frac{1}{4}\sqrt{d_{\zeta}} \rho_{f'} \rho_{f''} \|\zeta_i - \zeta_{i'}\|^2 \delta_2 \\
            & -  \left(\frac{1}{4} \sqrt{d_{\zeta}} \rho_{f'} \rho_{f''}\|\zeta_i - \zeta_{i'}\| +\frac{1}{8}\rho^2_{f''} \|\zeta_i - \zeta_{i'}\|^2 \right) \delta_2^2. 
        \end{align*}
        Here, inequality (i) follows from~\eqref{eq:verification_2} and Lemma~\ref{lemma:ball}, where $D:= \sup_{\zeta_i, \zeta_{i'} \in \mr{supp}(\zeta)} \|\zeta_{i} - \zeta_{i'}\| <\infty$, and $\overline{\rho} : = \sup_{\zeta \in \mr{supp}(\zeta)}\left\{\rho \ge 0:\ B(\zeta,\rho)\subset \mr{supp}(\zeta) \right\}$.

        We can choose a sufficiently small $\delta_2^* >0$ such that 
        \begin{align*}
            \underline{\lambda} \min\left\{\frac{1}{2}, \frac{\delta_2^{*}}{2 D} \right\}\overline{\rho} \geq  \left(\sqrt{d_{\zeta}} \rho_{f'} \rho_{f''}  + \frac{1}{2}\rho^2_{f''} D \right)\delta_2^{*2}, 
        \end{align*}
        and consequently, 
        \begin{align*}
            d(\zeta_{i}, \zeta_{i'})\geq \frac{1}{4} \underline{\lambda} \min\left\{\frac{1}{2}, \frac{\delta^*_2}{2 D} \right\}\overline{\rho} \|\zeta_i - \zeta_{i'} \| - \frac{1}{4}\sqrt{d_{\zeta}} \rho_{f'} \rho_{f''} \|\zeta_i - \zeta_{i'}\|^2 \delta_2^{*} , \quad \forall \zeta_{i}, \zeta_{i'} \in \mr{supp}(\zeta). 
        \end{align*}
        Also, we choose a sufficiently small $\delta_1^* >0$ such that $\delta_1^* \leq \frac{ \underline{\lambda} \min\left\{\frac{1}{2}, \frac{\delta^*_2}{2 D} \right\}\overline{\rho}}{2 \sqrt{d_{\zeta}} \rho_{f'} \rho_{f''}\delta_2^{*}}$. It is straightforward to verify that 
        \begin{align}\label{eq:verification_bound_2}
            d(\zeta_{i}, \zeta_{i'})\geq \frac{1}{8} \underline{\lambda} \min\left\{\frac{1}{2}, \frac{\delta^*_2}{2 D} \right\}\overline{\rho} \|\zeta_i - \zeta_{i'} \|, \quad \forall \|\zeta_{i}, \zeta_{i'} \| \leq \delta_1^*. 
        \end{align}
        Combining~\eqref{eq:verification_bound_1} and~\eqref{eq:verification_bound_2}, we have 
        \begin{align*}
            d(\zeta_i, \zeta_{i'}) \geq \min\left\{\frac{\lambda_1(\delta_1^*)}{D} , \frac{1}{8} \underline{\lambda} \min\left\{\frac{1}{2}, \frac{\delta^*_2}{2 D} \right\}\overline{\rho}\right\}\|\zeta_i - \zeta_{i'}\|, \quad \zeta_i, \zeta_{i'} \in \mr{supp}(\zeta). 
        \end{align*}
        Let $\lambda:=\min\left\{\frac{\lambda_1(\delta_1^*)}{D} , \frac{1}{8} \underline{\lambda} \min\left\{\frac{1}{2}, \frac{\delta^*_2}{2 D} \right\}\overline{\rho}\right\}$. This completes the proof.
    \end{proofprop}

    \begin{prooflmm}{lemma:verification_sigle_index_quadratic}
        We employ Proposition~\eqref{proposition:verification} to verify the informativeness condition. Since it is straightforward to verify that~\eqref{eq:verification_1} holds, it suffices to show~\eqref{eq:verification_2}. 
        Define $\zeta_{-1}' := \left(\zeta_2, \ldots, \zeta_{d_{\zeta}}\right)$ and $\beta_{-1}' := \left(\beta_{2}, \ldots, \beta_{d_{\zeta}}\right)$. 
        Since $g(\cdot)$ is strictly monotone and $\mr{supp}(\zeta)$ is compact, there exists a constant $\rho_{g'}>0$ such that
        \begin{align}\label{eq:verification_logit_quadratic_monotonic}
            \inf_{\zeta_i, \zeta_{i'} \in \mr{supp}(\zeta) } \left| g^{(1)}\left(\zeta_{i1} + \zeta_{i'1} + \sum_{d=2}^{d_{\zeta}} \beta_d \left(\zeta_{id} - \zeta_{i'd}\right)^2\right) \right|  \geq \rho_{g'},  
        \end{align}
        where $g^({i})(\cdot)$ denotes the first-order derivative of $g(\cdot)$. 
        Also, note that 
        \begin{align*}
            & \inf_{\zeta \in \mr{supp}(\zeta)}   \lambda_{\min} \left(\int   \nabla f(\zeta, \tilde{\zeta}) \nabla' f(\zeta, \tilde{\zeta}) \mr{d}\mb{P}(\tilde{\zeta})\right) \\
            = & \inf_{\zeta \in \mr{supp}(\zeta)}   \lambda_{\min} \bigg(4 \int  g^{(1)2}\left(\zeta_{1} + \tilde{\zeta}_{1} + \sum_{d=2}^{d_{\zeta}} \beta_d \left(\zeta_{d} - \tilde{\zeta}_{d}\right)^2\right) \\
            & \quad\quad \quad 
            \begin{pmatrix}
                1 & 0 \\
                0 & \mr{diag}(\beta_{-1})
            \end{pmatrix}
            \begin{pmatrix}
                1 & \zeta_{-1}' - \tilde{\zeta}_{-1}' \\
                \zeta_{-1} - \tilde{\zeta}_{-1} & (\zeta_{-1} - \tilde{\zeta}_{-1})(\zeta_{-1} - \tilde{\zeta}_{-1})'
            \end{pmatrix}
            \begin{pmatrix}
                1 & 0 \\
                0 & \mr{diag}(\beta_{-1})
            \end{pmatrix}
             \mr{d}\mb{P}(\tilde{\zeta})\bigg) \\
            \geqtext{(i)} & 4 \rho_{g'}^2 \left(\min\{\beta_{2}^2, \ldots, \beta_{d_{\zeta}}^2 \} \right) \inf_{\zeta \in \mr{supp}(\zeta)} \lambda_{\min} \left( 
                \begin{pmatrix}
                    1 & \zeta_{-1}' - \mb{E}(\tilde{\zeta}_{-1}') \\
                \zeta_{-1} -  \mb{E}(\tilde{\zeta}_{-1}) &  \mb{E}((\zeta_{-1} - \tilde{\zeta}_{-1})(\zeta_{-1} - \tilde{\zeta}_{-1})')
                \end{pmatrix}
            \right),
        \end{align*}
        where inequality (i) follows from~\eqref{eq:verification_logit_quadratic_monotonic}. It follows from Schur's decomposition that 
        \begin{align*}
            \begin{pmatrix}
                    1 & \zeta_{-1}' - \mb{E}(\tilde{\zeta}_{-1}') \\
                \zeta_{-1} -  \mb{E}(\tilde{\zeta}_{-1}) &  \mb{E}((\zeta_{-1} - \tilde{\zeta}_{-1})(\zeta_{-1} - \tilde{\zeta}_{-1})')
            \end{pmatrix}
        \end{align*}
        is positive definite if and only if 
        $\mr{Var}(\zeta_{-1} - \tilde{\zeta}_{-1})$ is positive definite. Also, for fixed $\zeta$, $\mr{Var}(\zeta_{-1} - \tilde{\zeta}_{-1})  = \mr{Var}(\tilde{\zeta}_{-1})$, which is positive definite by assumption. By the compactness of $\mr{supp}(\zeta)$ and continuity of $\lambda_{\min}(\cdot)$, it follows that  
        \begin{align*}
            \inf_{\zeta \in \mr{supp}(\zeta)} \lambda_{\min} \left( 
                \begin{pmatrix}
                    1 & \zeta_{-1}' - \mb{E}(\tilde{\zeta}_{-1}') \\
                \zeta_{-1} -  \mb{E}(\tilde{\zeta}_{-1}) &  \mb{E}((\zeta_{-1} - \tilde{\zeta}_{-1})(\zeta_{-1} - \tilde{\zeta}_{-1})'). 
                \end{pmatrix}
            \right) >0. 
        \end{align*}
        This verifies~\eqref{eq:verification_2} and completes the proof. 
    \end{prooflmm}

    \begin{prooflmm}{lemma:verification_absolute_logit}
        For each $\zeta_{i}, \zeta_{i'}\in [0, 1]$, without loss of generality, we assume $\zeta_{i} <  \zeta_{i'}$. Recall   
        \begin{align*}
            f(\zeta_{i}, \zeta_{i'}) = \frac{\exp\left(-|\zeta_{i} - \zeta_{i'}| \right)}{1 + \exp\left(-|\zeta_i - \zeta_{i'}| \right) }.
        \end{align*}
        Define 
        \begin{align*}
                h_{0}(\zeta) = \int  f(0, \tilde{\zeta})f(\zeta, \tilde{\zeta})  \mr{d}\tilde{\zeta}, \quad h_{1}(\zeta) = \int  f(1, \tilde{\zeta})f(\zeta, \tilde{\zeta})  \mr{d}\tilde{\zeta}. 
            \end{align*}
        It follows that 
        \begin{align*}
            d(\zeta_i, \zeta_{i'}) =   \sup_{\zeta \in [0, 1]} \left|\int   f(\zeta, \tilde{\zeta})(f(\zeta_i, \tilde{\zeta}) - f(\zeta_{i'}, \tilde{\zeta}))  \mr{d}\tilde{\zeta} \right| \geq \max \left\{ |h_{0}(\zeta_1) - h_{0}(\zeta_2)|, |h_{1}(\zeta_1) - h_{1}(\zeta_2)| \right\}.
        \end{align*}
        Thus, it suffices to prove that there exists a constant $\lambda >0$ such that for any $\zeta_i, \zeta_{i'} \in [0, 1]$, 
        \begin{align*}
            \max \left\{ |h_{0}(\zeta_1) - h_{0}(\zeta_2)|, |h_{1}(\zeta_1) - h_{1}(\zeta_2)| \right\} \geq \lambda |\zeta_1 - \zeta_2 |. 
        \end{align*}
        Note that 
        \begin{gather*}
            h'_0(\zeta) = - \int f(0, \tilde{\zeta}) \mr{sign}(\zeta - \tilde{\zeta}) f(\zeta, \tilde{\zeta}) (1 - f(\zeta, \tilde{\zeta}))  \mr{d}\tilde{\zeta}, \\
            h_0(\zeta) = h_1(1 - \zeta), \quad h'_1(\zeta) = -h'_0(1 - \zeta). 
        \end{gather*}
        Also, $h_0(\cdot)$  achieves its maximum at $\zeta_0^* <1/2$ and $h_1(\cdot)$ achieves its maximum at $\zeta_1^* = 1- \zeta_0^* > 1/2$. It is straightforward to verify that $h_0(\zeta)$ is a strictly increasing function in $\zeta$ for  $\zeta < \zeta_0^*$, and is a strictly decreasing function in $\zeta$ for $\zeta > \zeta_0^*$. 
         
        \paragraph{Step 1 $\zeta_i \in [\zeta_0^*, \zeta_1^*] \vee \zeta_{i'}\in [\zeta_0^*, \zeta_1^*] $} Fix an arbitrary positive constant $\delta < 1/2 - \zeta_0^*$, define 
        \begin{align*}
            \lambda_0 =  \inf_{\zeta\in [\zeta_0^* + \delta, 1] } |h'_0(\zeta) |. 
        \end{align*}
        One can check that $\lambda_0 >0 $ because $h'_0(\cdot)$ is strictly negative on $[\zeta_0^* + \delta, 1]$.  Then, 
        \begin{enumerate}[label=(\roman*)]
            \item when $\zeta_i \in  [0, \zeta_0^*]$ and $\zeta_{i'} \in [\zeta_0^*, \zeta_1^* - \delta ] $, we have $|h_1(\zeta_i) - h_1(\zeta_{i'})|\geq \lambda_0 | \zeta_i - \zeta_{i'} |$ by the mean value theorem;
            \item  when $\zeta_i \in  [0, \zeta_0^*]$ and $\zeta_{i'} \in [\zeta_1^* - \delta, \zeta_1^* ] $, $|h_1(\zeta_i) - h_1(\zeta_{i'})|\geq \lambda_0 (\zeta_1^* - \zeta_0^* - \delta)$ and thus, $|h_1(\zeta_i) - h_1(\zeta_{i'})|\geq \lambda_0 (\zeta_1^* - \zeta_0^* - \delta) |\zeta_{i} - \zeta_{i'}|$ because $|\zeta_{i} - \zeta_{i'}|\leq 1$;
            \item when $\zeta_i, \zeta_{i'} \in [\zeta_0^*, \zeta_1^*]$, since $\delta < 1/2 - \zeta_0^*$, it follows that $$\max \left\{|h_0(\zeta_i) - h_0(\zeta_{i'})|, |h_1(\zeta_i) - h_1(\zeta_{i'})|  \right\} \geq   \lambda_0 |\zeta_i - \zeta_{i'} |; $$
            \item when $\zeta_i \in  [\zeta_0^*,    \zeta_0^* + \delta]  $ and $\zeta_{i'} \in [\zeta_1^*, 1] $, $|h_0(\zeta_i) - h_0(\zeta_{i'})|\geq \lambda_0 (\zeta_1^* - \zeta_0^* - \delta)$ and thus, $|h_0(\zeta_i) - h_0(\zeta_{i'})|\geq \lambda_0 (\zeta_1^* - \zeta_0^* - \delta) |\zeta_{i} - \zeta_{i'}|$ because $|\zeta_{i} - \zeta_{i'}|\leq 1$;
            \item when $\zeta_{i} \in  [ \zeta_0^* + \delta, \zeta_1^*]$ and $\zeta_{i'} \in [\zeta_1^*, 1 ] $, we have $|h_0(\zeta_{i}) - h_0(\zeta_{i'})|\geq \lambda_0 | \zeta_{i} - \zeta_{i'} |$ by the mean value theorem. 
        \end{enumerate}
        Therefore, we obtain 
        \begin{align*}
            \max \left\{ |h_{0}(\zeta_{i}) - h_{0}(\zeta_{i'})|, |h_{1}(\zeta_{i}) - h_{1}(\zeta_{i'})| \right\} \geq \lambda_0 (\zeta_1^* - \zeta_0^* - \delta) |\zeta_{i} - \zeta_{i'} |.  
        \end{align*}

        \paragraph{Step 2 $\zeta_{i} \in [0, \zeta_0^*] \wedge \zeta_{i'}\in [\zeta_1^*, 1] $} Define 
        \begin{align*}
            \lambda_1 =  \inf_{\zeta\in [0, \zeta_0^* ] } |h_0(\zeta) - h_1(\zeta) |. 
        \end{align*}
        It follows that $\lambda_1 >0$ because $h_0(\zeta) > h_1(\zeta)$ when $\zeta\in [0, \zeta_0^*]$. Thus, for any $\zeta_{i} \in [0, \zeta_0^*] \wedge \zeta_{i'}\in [\zeta_1^*, 1] $, we have 
        \begin{enumerate}[label=(\roman*)]
            \item when $|h_0(\zeta_{i}) - h_0(\zeta_{i'}) | \leq \lambda_1$, we have $|h_1(\zeta_{i}) - h_1(\zeta_{i'}) | \geq 2\lambda_1 - \lambda_1 = \lambda_1$;
            \item  when $|h_1(\zeta_{i}) - h_1(\zeta_{i'}) | \leq \lambda_1$, we have $|h_0(\zeta_{i}) - h_0(\zeta_{i'}) | \geq 2\lambda_1 - \lambda_1 = \lambda_1$.    
        \end{enumerate}
        We obtain $\max\{|h_0(\zeta_{i}) - h_0(\zeta_{i'}) |,  |h_1(\zeta_{i}) - h_1(\zeta_{i'}) |\} \geq \lambda_1 \geq \lambda_1 |\zeta_{i} - \zeta_{i'}|$ when $\zeta_{i}\in [0, \zeta_0^*] \wedge \zeta_{i'}\in [\zeta_1^*, 1] $. 
        
        \paragraph{Step 3 $\zeta_{i}, \zeta_{i'} \in [0, \zeta_0^*] $ and $\zeta_{i}, \zeta_{i'} \in [\zeta_1^*, 1] $} It is straightforward to verify that, since $h$ is monotonic on $[0, \zeta_0^*]$ and $[\zeta_1^*, 1]$, 
        \begin{align*}
            \max \left\{|h_0(\zeta_i) - h_0(\zeta_{i'})|, |h_1(\zeta_i) - h_1(\zeta_{i'})|  \right\} \geq   \lambda_0 |\zeta_i - \zeta_{i'} |. 
        \end{align*}

        Combining the results from the three steps yields
        \begin{align*}
            d(\zeta_{i}, \zeta_{i'})\geq \max\{|h_0(\zeta_{i}) - h_0(\zeta_{i'}) |,  |h_1(\zeta_{i}) - h_1(\zeta_{i'}) |\}  \geq \min\{\lambda_0 (\zeta_1^* - \zeta_0^* - \delta) | , \lambda_1 \} |\zeta_{i'} - \zeta_{i}|, \quad \forall \zeta_{i}, \zeta_{i'} \in [0, 1]. 
        \end{align*}
        Let $\lambda = \min\{\lambda_0 (\zeta_1^* - \zeta_0^* - \delta) | , \lambda_1 \} $. This completes the proof. 
    \end{prooflmm}

    \begin{prooflmm}{lemma:verification_impossible}
        For any $t \in [0, 1]$, let $\zeta_{i} = (t, 0, 0, \ldots, 0)'$ and $\zeta_{i'} = ( 0, t, 0, \ldots, 0)$.  It is straightforward for verify that for any $\tilde{\zeta} \in [t, 1]^2 \times [0, 1]^{d_{\zeta} - 2}$, we have
        \begin{align*}
            \sum_{d=1}^{d_{\zeta}} | \zeta_{id} - \tilde{\zeta}_{d}| = \sum_{d=1}^{d_{\zeta}} \tilde{\zeta}_d - t, \\
            \sum_{d=1}^{d_{\zeta}} | \zeta_{i'd} - \tilde{\zeta}_{d}| = \sum_{d=1}^{d_{\zeta}} \tilde{\zeta}_d - t. 
        \end{align*}
        Therefore, 
        \begin{align}\label{eq:verification_impossible_bound_1}
            f(\zeta_i, \tilde{\zeta}) - f(\zeta_{i'}, \tilde{\zeta}) = 0, \quad \forall \tilde{\zeta} \in [t, 1]^2 \times [0, 1]^{d_{\zeta} - 2}. 
        \end{align}
        In addition, when $\tilde{\zeta} \in [0, t] \times [t, 1]\times  [0, 1]^{d_{\zeta} - 2}$
        \begin{align*}
            \sum_{d=1}^{d_{\zeta}} | \zeta_{id} - \tilde{\zeta}_{d}| = \sum_{d=2}^{d_{\zeta}} \tilde{\zeta}_2 + (t - \tilde{\zeta}_1), \\
            \sum_{d=1}^{d_{\zeta}} | \zeta_{i'd} - \tilde{\zeta}_{d}| = \sum_{d=2}^{d_{\zeta}} \tilde{\zeta}_d - (t - \tilde{\zeta}_1). 
        \end{align*} 
        Since the derivative of the logistic link function $\frac{\exp(\cdot)}{1 + \exp(\cdot)}$ is positive and uniformly bounded by $1/4$,  
        \begin{align}\label{eq:verification_impossible_bound_2}
            \left|f(\zeta_i, \tilde{\zeta}) - f(\zeta_{i'}, \tilde{\zeta})\right| \leq \frac{1}{4} \left|2 (t - \tilde{\zeta}_1) \right| \leq \frac{t}{2}, \quad \forall \tilde{\zeta} \in [0, t] \times [t, 1]\times  [0, 1]^{d_{\zeta} - 2}. 
        \end{align}
        By the same argument, we have 
        \begin{align}\label{eq:verification_impossible_bound_3}
            \left|f(\zeta_i, \tilde{\zeta}) - f(\zeta_{i'}, \tilde{\zeta})\right| \leq \frac{t}{2}, \quad \forall \tilde{\zeta} \in [t, 1] \times [0, t]\times  [0, 1]^{d_{\zeta} - 2}. 
        \end{align}
        When $\tilde{\zeta} \in [0, t]^2 \times  [0, 1]^{d_{\zeta} - 2}$ it is straightforward to verify that,  
        \begin{align}\label{eq:verification_impossible_bound_4}
            \left|f(\zeta_i, \tilde{\zeta}) - f(\zeta_{i'}, \tilde{\zeta})\right| \leq \frac{1}{4}\left|2(\tilde{\zeta}_1 -\tilde{\zeta}_2) \right| \leq \frac{t}{2}, \quad \forall \tilde{\zeta} \in [0, t]^2 \times  [0, 1]^{d_{\zeta} - 2}. 
        \end{align}
        Combining~\eqref{eq:verification_impossible_bound_1},~\eqref{eq:verification_impossible_bound_2},~\eqref{eq:verification_impossible_bound_3}, and~\eqref{eq:verification_impossible_bound_4},  we obtain 
        \begin{align*}
            &\sup_{\zeta\in [0, 1]^{d_{\zeta}}} \left| \int  f(\zeta, \tilde{\zeta}) \left(f(\zeta_i, \tilde{\zeta}) - f(\zeta_{i'}, \tilde{\zeta})\right) \mr{d}\mb{P}(\tilde{\zeta})\right|  \\
            = & \sup_{\zeta\in [0, 1]^{d_{\zeta}}} \left| \int \bs{1}(\tilde{\zeta} \notin [t, 1]^2 \times [0, 1]^{d_{\zeta} - 2})  f(\zeta, \tilde{\zeta}) \left(f(\zeta_i, \tilde{\zeta}) - f(\zeta_{i'}, \tilde{\zeta})\right) \mr{d}\mb{P}(\tilde{\zeta})\right| \\
            \leq & \sup_{\zeta\in [0, 1]^{d_{\zeta}}}  \int \bs{1}(\tilde{\zeta} \notin [t, 1]^2 \times [0, 1]^{d_{\zeta} - 2})  \left|f(\zeta, \tilde{\zeta}) \left(f(\zeta_i, \tilde{\zeta}) - f(\zeta_{i'}, \tilde{\zeta})\right)\right| \mr{d}\mb{P}(\tilde{\zeta}) \\
            \leq & \frac{t}{2} \mb{P}\left(\tilde{\zeta} \notin [t, 1]^2 \times [0, 1]^{d_{\zeta} - 2}\right) \\
            \leq & \frac{t(2t - t^2)}{2}. 
        \end{align*} 
        Since $\|\zeta_{i} - \zeta_{i'}\| = \sqrt{2}t$, we have $\frac{d(\zeta_{i}, \zeta_{i'})}{\|\zeta_{i} - \zeta_{i'}\| } \leq \frac{t(2t - t^2)}{2\sqrt{2}t}$ and 
        \begin{align*}
            \lim_{t\rightarrow 0}\frac{d(\zeta_{i}, \zeta_{i'})}{\|\zeta_{i} - \zeta_{i'}\| } = 0. 
        \end{align*}
        Therefore, there does not exist a constant $\lambda>0$ such that $d(\zeta_i, \zeta_{i'})
        \geq \lambda \|\zeta_{i} -\zeta_{i'}\|$ for all $\zeta_i, \zeta_{i'} \in [0, 1]^{d_{\zeta}}$.  Hence,  Assumption~\ref{assumption:informativeness} fails.
    \end{prooflmm}

    \begin{lemma}\label{lemma:ball}
        Suppose $\mr{supp}(\zeta)$ is compact, convex, and $\mr{affine}(\mr{supp}(\zeta)) = \mb{R}^{d_{\zeta}}$. Given a  symmetric positive definite matrix $\Omega \in \mb{R}^{d_{\zeta}\times d_{\zeta}}$ such that the minimum singular value of $\Omega$ satisfies $\lambda_{\min}(\Omega) > \underline{\lambda} >0$. Then, for any $\zeta_{i}, \zeta_{i'} \in \mr{supp}(\zeta)$ and for any given $\delta>0$, 
        \begin{align*}
            \sup_{\zeta_1 \in \mr{supp}(\zeta) \cap \mc{B}(\zeta_i, \delta)} \left| (\zeta_1 - \zeta_i)'\Omega (\zeta_i - \zeta_{i'})\right| \geq \min\left\{\frac{1}{2}, \frac{\delta}{2 D} \right\}\overline{\rho} \underline{\lambda}  \|\zeta_i - \zeta_{i'} \|, 
        \end{align*} 
        where $D:= \sup_{\zeta_i, \zeta_{i'} \in \mr{supp}(\zeta)} \|\zeta_{i} - \zeta_{i'}\| <\infty$, and $\overline{\rho} : = \sup_{\zeta \in \mr{supp}(\zeta)}\left\{\rho \ge 0:\ B(\zeta,\rho)\subset \mr{supp}(\zeta) \right\}$. 
    \end{lemma}
    \begin{prooflmm}{lemma:ball}
        By definition, $\overline{\rho}$ is the largest radius of a Euclidean ball that can be placed entirely inside $\mr{supp}(\zeta)$. Since $\mr{supp}(\zeta)$ is compact, convex, and $\mr{affine}(\mr{supp}(\zeta)) = \mb{R}^{d_{\zeta}}$, such ball exists and $0 < \overline{\rho} < \infty$. Pick one of these balls and suppose its center is $\zeta_0$ such that $\mc{B}(\zeta_0, \overline{\rho}) \subset \mr{supp}(\zeta)$. Define $\ell := \min\{1/2, \delta/2D\}$ and $\tilde{\zeta}_{i} = (1-\ell)\zeta_i + \ell \zeta_{0}$. In addition, define the ball as
        \begin{align*}
            \mc{B}(\tilde{\zeta}_i, \ell \overline{\rho} ):=\left\{  (1-\ell)\zeta_i + \ell \zeta \mid  \zeta \in \mc{B}(\zeta_0, \overline{\rho}) \right\}. 
        \end{align*}
        It is straightforward to verify that $\mc{B}(\tilde{\zeta}_i, \ell \overline{\rho} ) \subset\mr{supp}(\zeta) \cap \mc{B}(\zeta_i, \delta) $. Then, 
        \begin{align*}
            & \sup_{\zeta_1 \in \mr{supp}(\zeta) \cap \mc{B}(\zeta_i, \delta)} \left| (\zeta_1 - \zeta_i)'\Omega (\zeta_i - \zeta_{i'})\right| \\
            \geq &  \max\bigg \{\left| \ell (\zeta_i - \zeta_0)'\Omega (\zeta_{i'} - \zeta_i) + \frac{\ell \overline{\rho}}{\|\zeta_{i'} - \zeta_i\|} (\zeta_{i'} - \zeta_{i})'\Omega  (\zeta_{i'} - \zeta_{i})\right|, \\
            & \qquad   \qquad \left| \ell (\zeta_i - \zeta_0)'\Omega (\zeta_{i'} - \zeta_i) - \frac{\ell \overline{\rho}}{\|\zeta_{i'} - \zeta_i\|} (\zeta_{i'} - \zeta_{i})'\Omega  (\zeta_{i'} - \zeta_{i})\right| \bigg \} \\
            \geqtext{(i)} &  \ell\overline{\rho} \underline{\lambda} \|\zeta_{i'} - \zeta_i\| \\
            \geq & \min\left\{\frac{1}{2}, \frac{\delta}{2 D} \right\}\overline{\rho} \underline{\lambda}  \|\zeta_i - \zeta_{i'} \|. 
        \end{align*}
        Here, (i) follows from the fact that $\max\{|a+b|, |a-b|\} \geq |b|$. This completes the proof. 
    \end{prooflmm}

    \begin{lemma}\label{lemma:Bernstein}
        Let $\{X_i\}_{i=1}^{N}$ be a collection of independent random
        variables, and let $\{y_i\}_{i=1}^{N}$ be a sequence of
         constants. Assume that there exist
        $B_X,B_Y,\sigma_X>0$ such that, for every $1\leq i\leq N$,
        \begin{align*}
            \mb{E}(X_i)=0,\qquad
            |X_i|\leq B_X,
            \qquad |y_i|\leq B_Y,
        \end{align*}
        and $\sup_{1\leq i\leq N}\mr{Var}(X_i)\leq \sigma_X^2$. 
        Then, for any $\epsilon>0$,
        \begin{align*}
            \mb{P}\left(
                \left|\frac{1}{N}\sum_{i=1}^{N}
                 X_iy_i  \right|
                \geq \epsilon
            \right)
            \leq
            2\exp\left(
                -\frac{N\epsilon^2}
                {2 \sigma_X^2B_Y^2+ \frac{2}{3}B_XB_Y\epsilon }
            \right).
        \end{align*}
    \end{lemma}

\begin{prooflmm}{lemma:Bernstein}
    Since $y_i$ is deterministic and 
    \begin{align*}
        |X_iy_i|&\leq B_XB_Y, \quad 
        \mr{Var} (X_iy_i) =y_i^2\mr{Var}(X_i)
        \leq \sigma_X^2B_Y^2.
    \end{align*}
    Therefore, applying Bernstein's inequality (see \citet[Proposition~2.14]{wainwright2019high}) to the
    independent, mean-zero random variables
    $\{X_iy_i\}_{i=1}^{N}$ directly yields
    \begin{align*}
            \mb{P}\left(
                \left|\frac{1}{N}\sum_{i=1}^{N}
                 X_iy_i  \right|
                \geq \epsilon
            \right)
            \leq
            2\exp\left(
                -\frac{N\epsilon^2}
                {2 \sigma_X^2B_Y^2+ \frac{2}{3}B_XB_Y\epsilon }
            \right).
        \end{align*}
    This completes the proof.
\end{prooflmm}

\end{document}